\documentclass[preprint]{aastex}
\usepackage{graphicx}        
\usepackage{natbib}         
\usepackage{amssymb}        
\usepackage{amsmath}
\usepackage{color}           
\usepackage{url}             
\usepackage{epstopdf}
\usepackage{soul}
\usepackage{threeparttable}
\usepackage{ulem}

\providecommand\bnabla{\boldsymbol{\nabla}}

\newcommand\bu{{\boldsymbol{u}}}

\newcommand\be{{\boldsymbol{e}}}
\newcommand\bA{{\boldsymbol{A}}}
\newcommand\bB{{\boldsymbol{B}}}
\newcommand\bF{{\boldsymbol{F}}}

\renewcommand\Re{{\rm Re}}

\newcommand\Ri{{\rm Ri}}
\newcommand\Pe{{\rm Pe}}
\renewcommand\Pr{{\rm Pr}}

\begin{document}



\title{Turbulent transport by diffusive stratified shear flows: from local to global models. Part II: Limitations of local models}

\author{Damien Gagnier, \\
IRAP, Universit\'e de Toulouse, CNRS, UPS, CNES, 14 avenue \'Edouard Belin, F-31400 Toulouse, France.\\
Pascale Garaud, \\
Department of Applied Mathematics and Statistics, Baskin School of Engineering, \\
University of California at Santa Cruz, 1156 High Street, Santa Cruz CA 95064.\\
}



\vspace{1cm}
\begin{abstract}

This paper continues the systematic investigation of diffusive shear instabilities initiated in Part I of this series. In this work, we primarily focus on quantifying the impact of non-local mixing, which is not taken into account in Zahn's mixing model \citep{Zahn92}. We present the results of direct numerical simulations in a new model setup designed to contain coexisting laminar and turbulent shear layers. As in Part I, we use the Low P\'eclet Number approximation of \citet{Lign1999} to model the evolution of the perturbations. Our main findings are twofold. First, turbulence is not necessarily generated whenever Zahn's nonlinear criterion  \citep{Zahn1974} $J{\rm Pr} < (J{\rm Pr})_c$ is satisfied, where $J=N^2/S^2$ is the local gradient Richardson number, ${\rm Pr} = \nu/ \kappa_T$ is the Prandtl number, and $(J{\rm Pr})_c \simeq 0.007$. We have demonstrated that the presence or absence of turbulent mixing in this limit hysteretically depends on the history of the shear layer. Second, Zahn's nonlinear instability criterion only approximately locates the edge of the turbulent layer, and mixing beyond the region where $J{\rm Pr} < (J{\rm Pr})_c$ can also take place in a manner analogous to convective overshoot. We found that the turbulent kinetic energy decays roughly exponentially beyond the edge of the shear-unstable region, on a lengthscale $\delta$ that is directly proportional to the scale of the turbulent eddies, which are themselves of the order of the Zahn scale (see Part I). Our results suggest that mixing by diffusive shear instabilities should be modeled with more care than is currently standard in stellar evolution codes.

\end{abstract}

\section{Introduction}
\label{sec:intro}

The diffusive shear instability was first discussed in the context of stellar astrophysics by \cite{Zahn1974} as a potential source of mixing in stably stratified radiation zones which extracts its energy from the star's rotational shear. Zahn indeed noted that since the Prandtl number $\Pr = \nu/\kappa_T$ (where $\nu$ is the kinematic viscosity and $\kappa_T$ is the thermal diffusivity) is very small in stars, $\nu \ll \kappa_T$, and one can always find a scale $l$ such that the Reynolds and P\'eclet numbers based on that scale satisfy 
\begin{equation}
\Pe_l \ll 1 \ll \Re_l, \mbox{ where } \Pe_l = Sl^2/\kappa_T, \quad \Re_l = Sl^2/\nu \ ,
\end{equation}
and $S$ is the local shearing rate. On that scale, fluid flows are both strongly diffusive, which implies that they barely feel the thermal stratification, but also very weakly viscous, a necessary condition for shear instability to take place. \citet{Zahn1974} concluded that such diffusive shear instabilities should be fairly ubiquitous in stars and argued that an appropriate instability criterion must be of the form 
\begin{equation}
J \Pr < (J \Pr)_c \ ,
\label{eq:zahnstab}
\end{equation}
where $J = N^2 / S^2$ (where $N$ is the Brunt-V\"ais\"al\"a frequency) is the local gradient Richardson number, and $(J \Pr)_c$ is a universal constant. He later proposed a simple model for mixing by diffusive shear instabilities, where the turbulent viscosity $\nu_{\rm turb}$ and the turbulent diffusivity of a passive scalar $D_{\rm turb}$ are given by 
\begin{equation}
D_{\rm turb} \simeq \nu_{\rm turb} \simeq C\frac{\kappa_T}{J} \ ,
\label{eq:zahnmodel}
\end{equation}
where $C$ is another universal constant.
The arguments leading to the derivation of (\ref{eq:zahnstab}) and (\ref{eq:zahnmodel}) are reviewed for instance in \citet{Garaudal17} (Paper I hereafter). 

Significant progress has been made in recent years toward testing Zahn's model using three-dimensional (3D) Direct Numerical Simulations (DNSs) with encouraging results, although much remains to be done. \citet{PratLignieres13,PratLignieres14}, \cite{Pratal2016} and Paper I studied uniform shear flows (i.e. flows where both the background shear and the Brunt-V\"ais\"al\"a frequency are constant) using different forcing methods. \citet{Garaudal15} and \citet{Garaud2016} studied stratified Kolmogorov flows, where the background Brunt-V\"ais\"al\"a frequency is constant and the background shear flow is driven by a spatially sinusoidal body-force. In all cases the Reynolds numbers achieved were sufficiently high to ensure that the viscous scale was much smaller than the domain scale. 

The findings reported in these papers are remarkably consistent with one another (given the differences in model setups), and show that whenever the P\'eclet number based on the turbulent eddy scale $l_e$ is smaller than one {\it and} $l_e$ is both significantly larger than the viscous scale but also significantly smaller than the domain scale (i.e. whenever the turbulent dynamics only depend on the local shear) then:
\begin{itemize}
\item Stratified diffusive shear instabilities are only excited provided \citep{Pratal2016,Garaud2016,Garaudal17}
\begin{equation}
J \Pr < (J \Pr)_c \simeq 0.007 \ ,
\label{eq:zahnstab2}
\end{equation}
which not only recovers the criterion proposed by \citet{Zahn1974} but also estimates the previously unknown constant $(J \Pr)_c$.
\item Both the turbulent viscosity $\nu_{\rm turb}$ and the turbulent diffusivity $D_{\rm turb}$ are roughly equal to one another, and can be approximated by 
\begin{equation}
D_{\rm turb} \simeq \nu_{\rm turb} \simeq C\frac{\kappa_T}{J} \ ,
\label{eq:zahnmodel2}
\end{equation}
which recovers the model proposed by \citet{Zahn92}. Note that $C \simeq 0.03$ for \citet{Pratal2016}, and $C \simeq 0.08$ for Paper I, the latter value being more reliable as it covers a much wider range of simulations. 
\end{itemize}

A comparative study of all the works listed above also reveals interesting discrepancies (with one another, and with the Zahn model) when $l_e$ approaches the domain scale (or the scale of the shear, in the sinusoidal case), revealing limitations of a theory that is local by construction. Larger eddy sizes are expected as the stratification decreases, which is a situation that can occur in the vicinity of a convection zone. Meanwhile the amplitude of the shear can vary significantly with depth depending on what forces it, so the shear lengthscale could sometimes be fairly small (e.g. the solar tachocline). This raises the issue of how should one modify Zahn's model to account for non-local effects. In particular, two questions arise: (i) does the stability criterion given in equation (\ref{eq:zahnstab}) correctly predict the  location of the boundary between turbulent and laminar regions if the shear is localized, and more generally (ii)  is the local turbulent mixing model adequate if the shear varies rapidly with radius, or in the vicinity of the edge of the turbulent region (where the size of the eddies becomes comparable with the shear lengthscale and/or with the distance to the edge of the turbulent shear layer)? 

To answer  these questions, which are a natural follow-up of our work from Paper I, we consider here a new model setup where the shear layers are specifically designed to contain both laminar and turbulent regions. This setup is presented in Section \ref{sec:mod}. For completeness, the linear stability properties of this model are presented in Section \ref{sec:stab}. In Section \ref{sec:num} we present the results of a few selected numerical simulations, illustrating the salient properties of turbulent shear layers adjacent to stable regions. In particular, we demonstrate the failure of Zahn's model to account for mixing beyond the edge of the theoretically unstable shear layer. In Section~\ref{sec:extension}, we propose a simple exponential prescription for overshoot in the vicinity of a turbulent shear flow, and conclude in Section~\ref{sec:ccl} by summarizing our results, and raising prospects for modeling diffusive shear instabilities beyond Zahn's simple models.  

\section{The model}
\label{sec:mod}

In this paper we consider a relatively small region of a stellar radiative zone which we assume to be located around a radius $r_0$ and whose vertical extent $L_z$ is substantially smaller than the local pressure, temperature and density scaleheights. In this region, we can define a local background density profile and a local background temperature profile, respectively $\bar{\rho}(r) \simeq \rho_0 + (r-r_0)d\bar{\rho}/dr + ...$ and $\bar{T}(r) \simeq T_0 + (r-r_0)d\bar{T}/dr + ...$ where $\rho_0$ and $T_0$ are the mean density and the mean temperature in the region considered. This area is also characterized by a background adiabatic temperature gradient $dT_{ad}/dr \simeq (T_0/H_p)\nabla_{ad}$ where $H_p$ is the pressure scaleheight at $r=r_0$. This small region is then modeled using a Cartesian coordinate system where gravity defines the $z$-direction, namely ${\bf g} = -g {\bf e}_z$. In what follows, we use the Boussinesq approximation in which fluctuations of density are neglected everywhere except in the buoyancy term. They are related to the temperature fluctuations $T$ away from the mean $T_0$ through a linearized equation of state, i.e.:
\begin{equation}
\frac{\rho}{\rho_0} = - \alpha T \ ,
\end{equation}
where $\alpha$ is the coefficient of thermal expansion $\alpha=-\rho^{-1}_0(\partial{\rho}/\partial{T})_p$. As in \citet{Garaudal15} and \citet{Garaud2016}, we assume that the shear is created by a steady triply-periodic body-force ${\bF}$ applied in the $x$-direction, whose amplitude varies in the $z$-direction.

The system we have described can be represented by the following dimensional Boussinesq equations (where we neglect the role of compositional stratification for now):

\begin{equation}\label{eq1}
\frac{\partial \bu}{\partial t}+ \bu \cdot \bnabla \bu=-\frac{1}{\rho_{0}}\bnabla p + \alpha g T \be_z + \nu \nabla^2 \bu + \frac{1}{\rho_{0}}\bF \ ,
\end{equation}
\begin{equation}\label{eq2}
\bnabla \cdot \bu = 0 \ ,
\end{equation}
\begin{equation}\label{eq3}
\frac{\partial T}{\partial t}+ \bu \cdot \bnabla T + w\left(\frac{d\bar{T}}{dr}-\frac{dT_{ad}}{dr}\right)=\kappa_{T} \nabla^2 T \ ,
\end{equation}
where $p$ is the pressure perturbation away from hydrostatic equilibrium, and all the perturbations and the velocity field $\bu = (u,v,w)$ are considered triply-periodic. The quantities $\rho_0$, $\alpha$, $g$, $\nu$, $\kappa_T$, $d\bar{T}/dr$ and $dT_{ad}/dr$ are all taken as constant.


In previous papers by \citet{Garaudal15}, and \citet{Garaud2016} the force $\bF$  was chosen to be a sinusoidal function of $z$, namely ${\bF} = F_0 \sin(kz) {\bf e}_x$. In this paper, the force $\bF=F(z){\bf e}_x$ is chosen to generate the vertically periodic and horizontally invariant flow profile:
\begin{equation}\label{eq:laminar}
U(z)= U_0 \frac{\tanh(a \cdot \sin(2 \pi z / L_z))}{\tanh(a)} \ ,
\end{equation}
where the functions $U(z)$ and $F(z)$ are related by the equation:

\begin{equation}
\nu \nabla^2 U(z) + \frac{1}{\rho_0} F(z)=0 \ .
\end{equation}

\begin{figure}[t]
  \centerline{\includegraphics[width=0.7\textwidth]{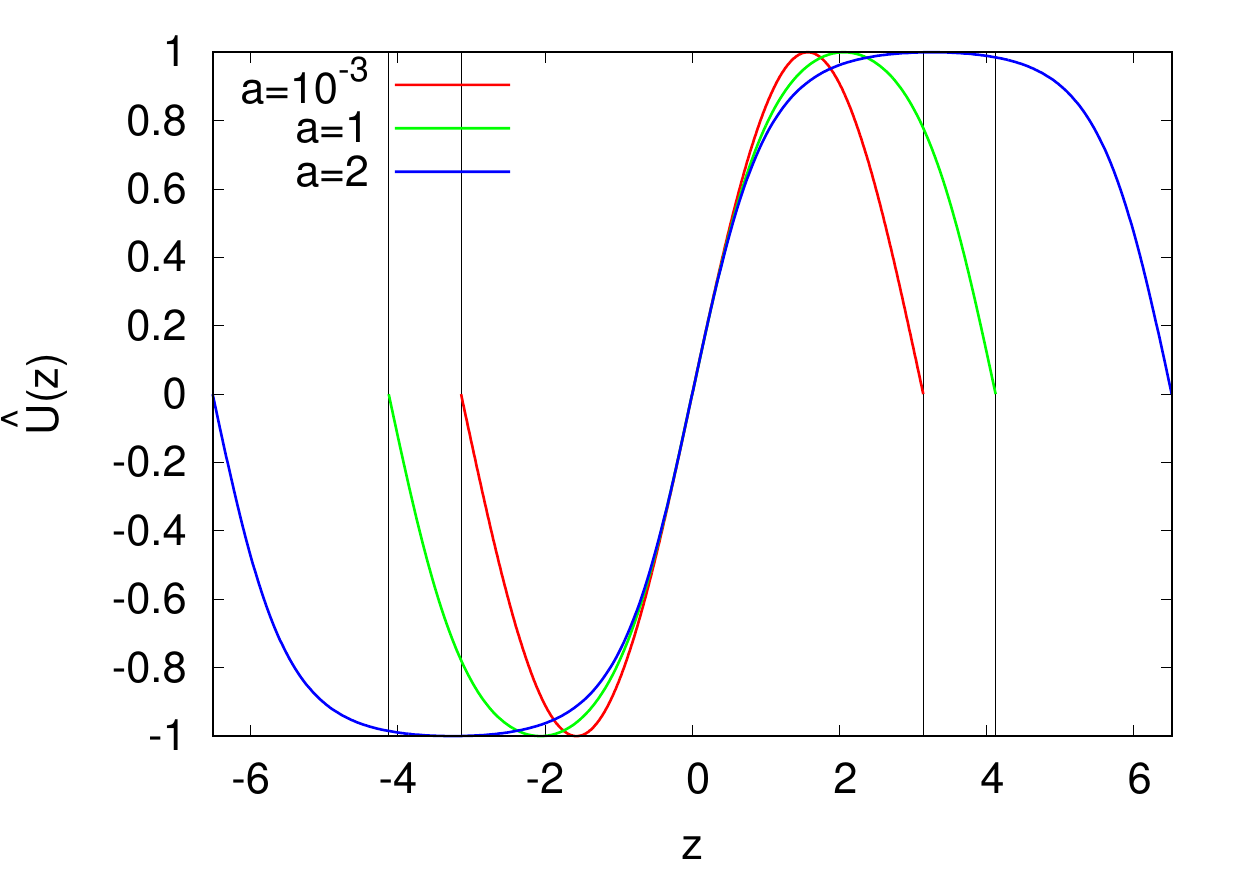}}
\caption{Shape of the non-dimensional laminar shear flow $\hat{U}(z)$ for $a=10^{-3}, \ 1 \ \text{and} \ 2$. The vertical lines represent the limits of the vertical domains for $a=10^{-3}$ and $a=1$. Peak velocities, and shearing rates at $z=0$, are equal to $1$ in all cases.}
\label{fig:shear}
\end{figure}

This profile is such that $U(z)$ varies from $-U_0$ to $+U_0$ over a shear layer of width $L_0 = \tanh(a)L_z / (2 \pi a)$. 
At fixed $L_z$, changing the value of $a$ therefore varies $L_0$. In what follows, however, we fix the shear layer scaleheight $L_0$, and vary the overall domain size $L_z$ with $a$, as in $L_z = 2 \pi a L_0 / \tanh(a)$. With this choice, we can compare the behavior of different laminar flows with an identical central shear and peak velocity but different laminar region sizes, only by varying the parameter $a$ (see Figure~\ref{fig:shear}).  Setting the parameter $a$ to a very small value recovers the sinusoidal laminar profile studied by \citet{Garaudal15} and \citet{Garaud2016}, while using very large values of $a$ asymptotically recovers the hyperbolic tangent profile of \citet{Lignal1999}. 

The amplitude $U_0$ and lengthscale $L_0$ of the laminar flow profile can be used to define a new unit system where
\begin{align}
\begin{split}
[u]&= U_0 \textrm{ is the unit velocity,}\\
[l]&= L_0 \textrm{ is the unit length,}\\
[t]&= \frac{[l]}{[u]} = \frac{L_0}{U_0} \textrm{ is the unit time,}\\
[T]&= L_0\left(\frac{d\bar{T}}{dr}-\frac{dT_{ad}}{dr}\right) \textrm{ is the unit temperature.}
\end{split}
\end{align}
In this non-dimensionalization, $L_z$ is now the domain size in units of the shear layer size $L_0$. Equations (\ref{eq1}), (\ref{eq2}) and (\ref{eq3}) can be re-written:

\begin{equation}\label{eq:1}
\frac{\partial \hat{\bu}}{\partial t}+ \hat{\bu} \cdot \bnabla \hat{\bu}=-\bnabla \hat{p} + {\rm Ri}\hat{T} \be_z + \frac{1}{\rm Re} \nabla^2 (\hat{\bu} - \hat{U}(z) \be_x),
\end{equation}
\begin{equation}\label{eq:2}
\bnabla \cdot \hat{\bu} = 0,
\end{equation}
\begin{equation}\label{eq:3}
\frac{\partial \hat{T}}{\partial t}+ \hat{\bu} \cdot \bnabla \hat{T} + \hat{w}=\frac{1}{\Pe}\nabla^2 \hat{T},
\end{equation}
where $\hat{\bu}$, $\hat{p}$ and $\hat{T}$ are the non-dimensional velocity, pressure and temperature fields. The
differential operators as well as the independent variables have also been implicitly non-dimensionalized. This non-dimensionalization results in the appearance of three dimensionless parameters, 

\begin{align}
\begin{split}
{\rm Re}&= \frac{U_0 L_0}{\nu},\\
{\rm Pe}&= \frac{U_0 L_0}{\kappa_T},\\
{\rm Ri}&= \frac{\alpha g \left(\frac{d\bar{T}}{dr}-\frac{dT_{ad}}{dr}\right)L_0^2}{U_0^2}.
\end{split}
\end{align}

The Reynolds number $\rm Re$ can be seen as the ratio of viscous to convective timescales, the P\'eclet number ${\rm Pe}$ as the ratio of thermal diffusion to convective timescales and the Richardson number ${\rm Ri}$ is the typical ratio of potential energy lost to kinetic energy gained by turbulent eddies.

In this work, we will focus on low P\'eclet number flows, i.e. flows for which thermal diffusion is
significant as in Zahn's original paper \citep{Zahn1974}. \citet{Lign1999} and \citet{Lignal1999} showed that in this limit, temperature fluctuations are entirely slaved
to the vertical velocity. The flow dynamics can be modeled using his proposed low-P\'eclet number (LPN) equations which were validated against Direct Numerical Simulations of fully nonlinear low P\'eclet number flows by \citet{Garaud2016} \citep[see also][]{PratLignieres13}:

\begin{equation}\label{LPN1.1}
\frac{\partial \hat{\bu}}{\partial t}+ \hat{\bu} \cdot \bnabla \hat{\bu}=-\bnabla \hat{p} + {\rm RiPe} \nabla^{-2}\hat{w} \be_z + \frac{1}{\rm Re} \nabla^2( \hat{\bu} - \hat{U}(z) \be_x) \ ,
\end{equation}
\begin{equation}\label{LPN1.2}
\bnabla \cdot \hat{\bu} = 0 \ .
\end{equation}

Using the LPN equations has several advantages. They only depend on two control parameters: the Reynolds number and the product of the Richardson and P\'eclet numbers called the “Richardson-P\'eclet number hereafter. Moreover, their numerical integration is not limited by the necessity of capturing the very rapid thermal diffusion timescale  and  therefore can be a lot faster. \citet{Garaud2016} showed that the low-P\'eclet-number limit could be relevant in the envelopes of very massive stars where thermal diffusivity exceeds $10^{14}$ cm\textsuperscript{2}/s, but does not apply for lower-mass stars, or deep within the interiors of massive stars where the thermal diffusivity is much smaller.

 We shall now present the linear stability properties of this particular shear flow as a function of the governing parameters $a$, $\rm Re$, and $\rm RiPe$. The reader who is only interested in the mixing properties of this shear layer can jump to Section \ref{sec:num}. 

\section{Linear stability of the periodic tanh profile}
\label{sec:stab}

Hydrodynamic stability theory consists in the study of the response of a fluid flow to perturbations of various amplitudes. If it returns to its original state one defines the flow as stable, however, if the amplitude of the disturbance grows and substantially modifies the original flow, then it can be considered unstable. In what follows, we will use the terminology ``linearly unstable'' to refer to flows that are unstable to infinitesimal perturbations, and ``nonlinearly unstable'' to refer to flows that are stable to infinitesimal perturbations but can be destabilized by appropriate finite amplitude perturbations. Homogeneous diffusive shear flows for instance are known to be linearly stable but are also nonlinearly unstable provided they satisfy Zahn's nonlinear instability criterion~(\ref{eq:zahnstab}) \citep[see][]{PratLignieres13, PratLignieres14}. Sinusoidally forced diffusive shear flows on the other hand can either be linearly unstable or nonlinearly unstable depending on the value of ${\rm RiPe}$ \citep{Garaudal15}. In this section, we now study the linear stability of the selected periodic tanh profile.

As discussed in the previous section, our model is such that our target flow profiles $U(z)$ all have nearly identical shear layers regardless of the selected value of $a$. Naively, we would therefore expect the results of the linear stability analyses to be fairly independent of $a$ as well. We also expect the linear eigenmodes of instability to be mostly localized in the shearing region, and have vanishingly small amplitude far from it. However, as we shall demonstrate, neither of these naive assumptions are correct in this case. 

\begin{figure}[h]
	\includegraphics[width=.5\textwidth]{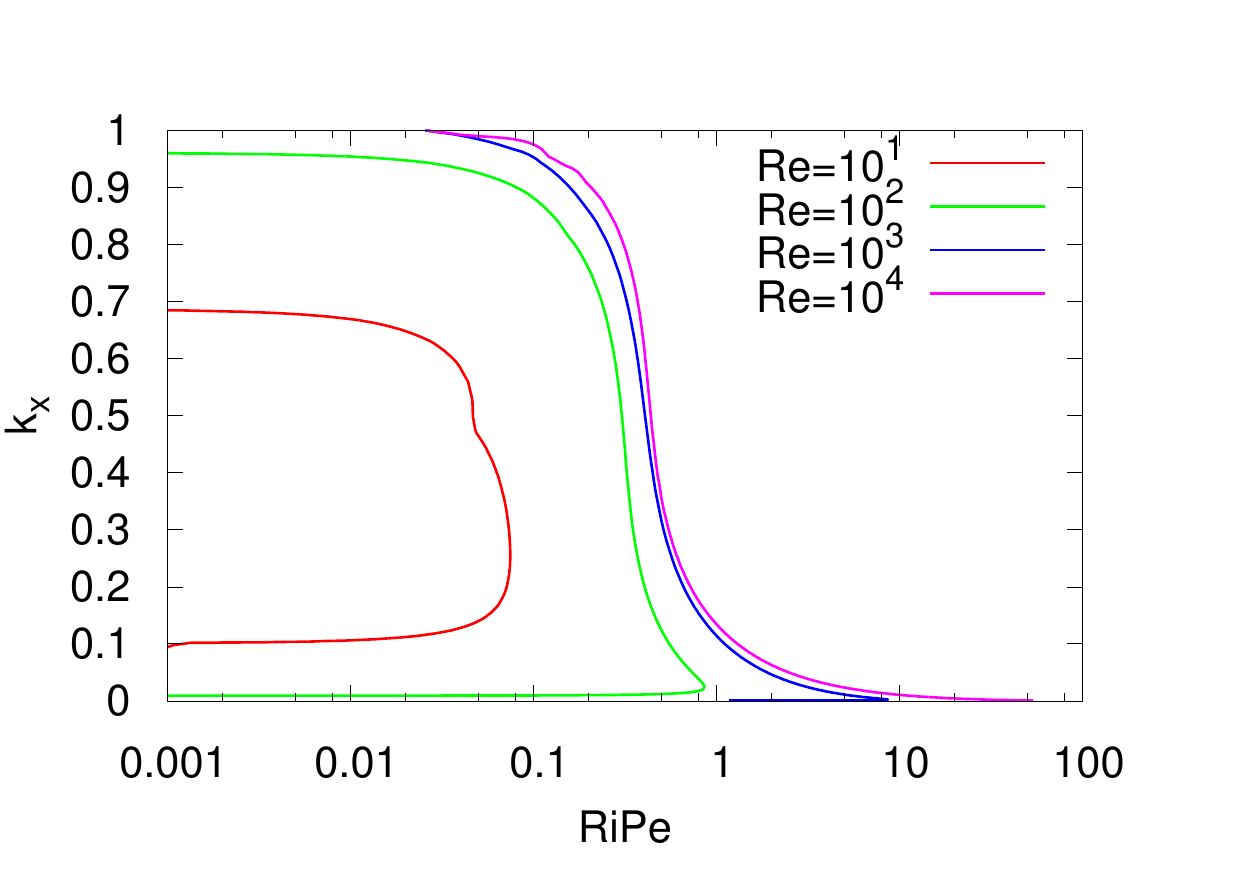}\hfill
	\includegraphics[width=.5\textwidth]{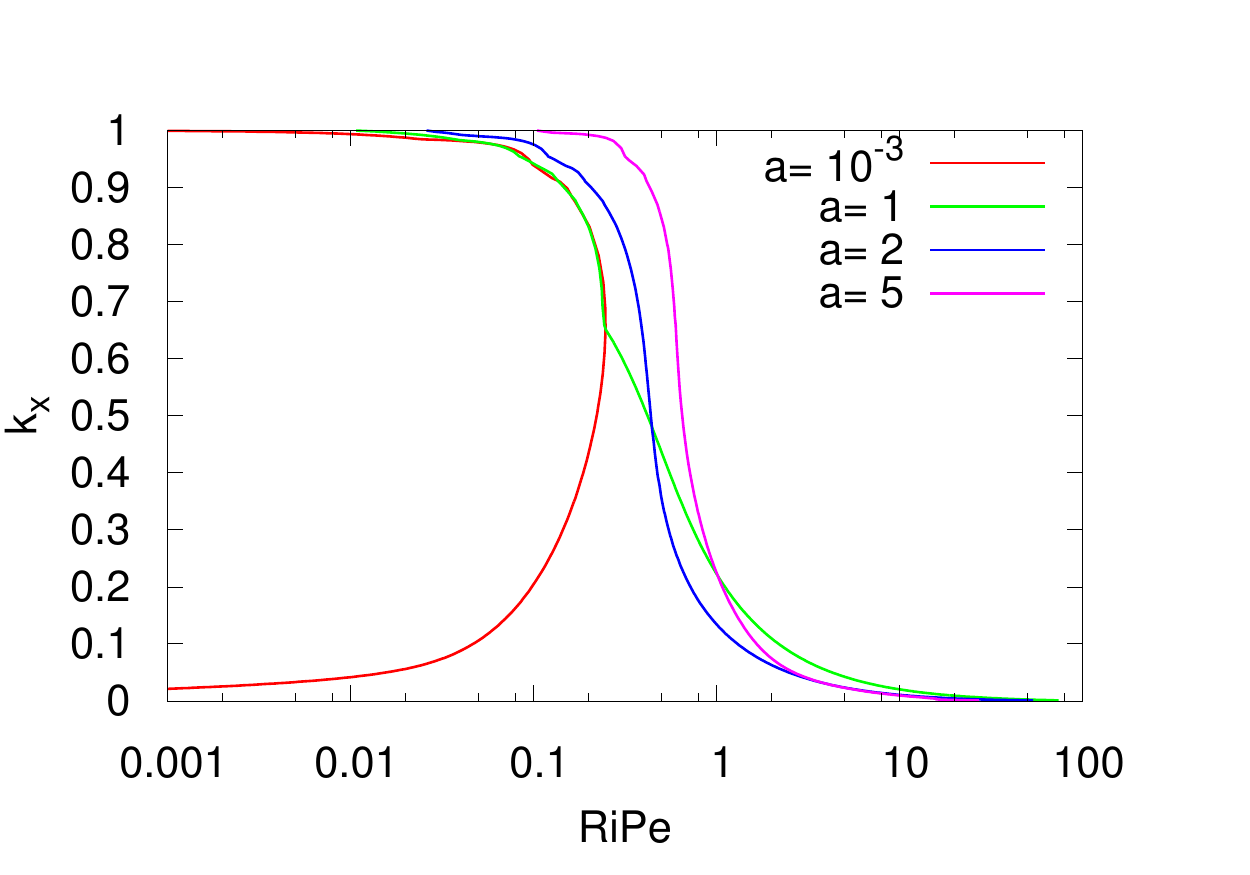}
  \caption{Left : Marginal stability curves at various $\Re$ and $a=2$. The system is unstable in the area to the left of each curves. Right : As on the left, but for varying $a$, at fixed  $\Re=10^4$.}
\label{fig:marginalstab}
\end{figure}

We study the linear stability of the stratified shear flow $\hat{U}(z)$ in the LPN limit. Details of the technique used to carry out the analysis are presented in Appendix A. We assume that perturbations are two-dimensional in the ($x,z$) plane, and take the form $\hat{q}(x,z,t) = \tilde{q}(z) \exp(i k_x x + \lambda t)$. Figure \ref{fig:marginalstab} shows the marginal stability curve for different values of ${\rm Re}$ and $a$. The system is unstable ($\operatorname{\mathbb{R}e}(\lambda) > 0$) on the left side of each curve and stable ($\operatorname{\mathbb{R}e}(\lambda) < 0$) on the right side.

In Figure \ref{fig:marginalstab}a, we hold the shape parameter $a$ constant with value $a=2$, and vary the Reynolds number ${\rm Re}$. For small ${\rm Re}$ the unstable region is quite small but it increases in size as ${\rm Re}$ increases as expected. For large ${\rm Re}$, the maximum value of ${\rm RiPe}$ for instability of mid-range $k_x$ values ($k_x \simeq 0.5$) asymptotes to $\sim 0.45$, but we also see the development of a tail of low $k_x$ unstable modes at larger ${\rm RiPe}$. As ${\rm Re}$ tends to infinity, the marginal stability curve converges to the curve one would obtain in the inviscid limit (infinite ${\rm Re}$). The shape of that inviscid curve is very reminiscent of the one presented by \citet{Lignal1999} for the pure hyperbolic tangent profile but is not exactly the same. Instead, the tail has a finite extent, and there exists a maximum value of ${\rm RiPe}$ for which unstable modes exist (see later for more on this topic).

In Figure \ref{fig:marginalstab}a, $a$ varies while ${\rm Re}$ is held constant at ${\rm Re}=10^4$. This specific value of ${\rm Re}$ has been chosen to ensure that viscous effects are not particularly important. As expected, for very small values of $a$, we recover the marginal stability curves from \citet{Garaudal15}, and for larger $a$ we recover marginal stability curves that look very similar to those from \citet{Lignal1999}. This is an important result because it shows that the sinusoidal profile and the hyperbolic tangent profile behave in fundamentally different ways. This difference does not come from the difference between periodic and non-periodic profiles nor from shear sign variability in the domain as proposed by \citet{Garaudal15}. The reason behind this substantial difference, i.e. why the hyperbolic tangent profile is unstable to low $k_x$ modes while the sinusoidal profile is not, can be attributed to the fact that there are two distinct modes of instability, one of which disappears when the size of the domain becomes too small (i.e. when $a \rightarrow 0$).

\begin{figure}[h]
    \includegraphics[width=.5\textwidth]{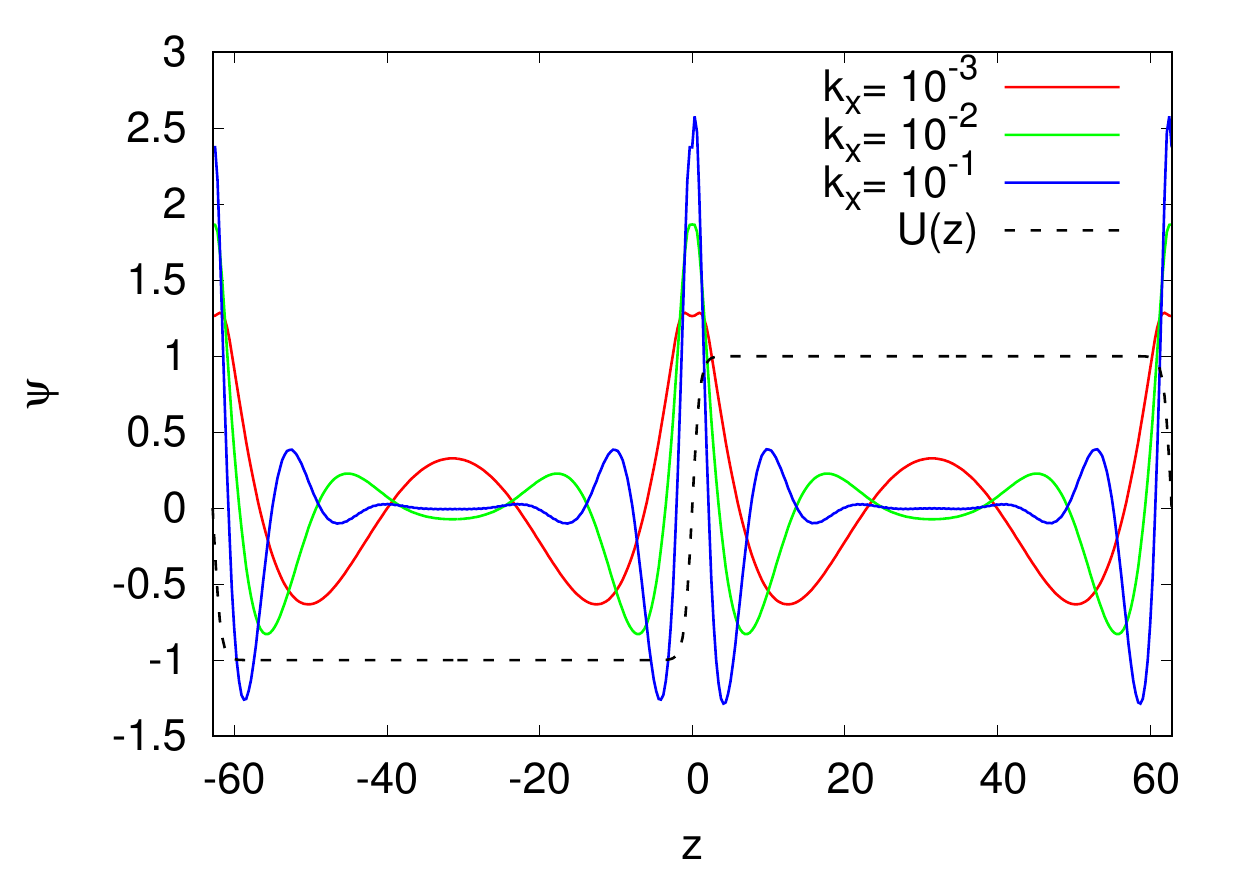}\hfill
  	\includegraphics[width=.5\textwidth]{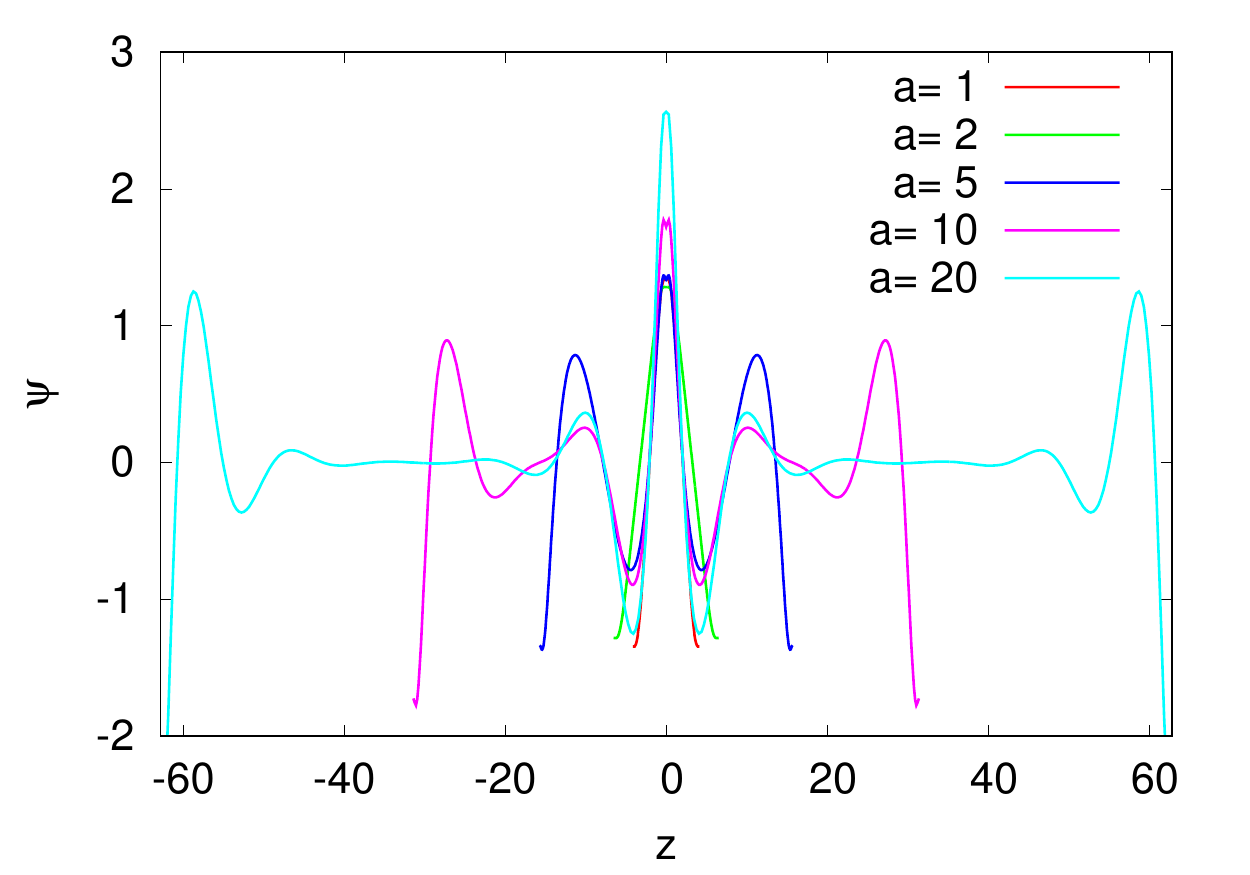}
  \caption{Streamfunction $\psi$ as a function of $z$. Left: ${\rm Re}= 10^4$, ${\rm RiPe= 1}$, $a= 20$ and $k_x$ varies. The smaller $k_x$, the further modes extend beyond the sheared region. Right: ${\rm Re}= 10^4$, ${\rm RiPe= 1}$, $k_x= 0.1$ and $a$ varies. The shape of $\psi$ in the vicinity of the sheared region is more-or-less independent of $a$. However, when $a$ drops below $0.45$, that mode is no longer unstable.}
\label{fig:varyk&a}
\end{figure}

 To see why more clearly,  we now look at a case with $a = 20$, ${\rm Re} = 10^4$ and ${\rm RiPe} = 1$. For such a large value of $a$, the flow mimics the hyperbolic tangent profile of \citet{Lignal1999}. Figure \ref{fig:varyk&a}a shows the shape of the stream function $\psi(z)$, for various values of $k_x$. Note that $\psi(z)$ (see Appendix A) is related to $\hat{u}$ and $\hat{w}$ via $i k_x \psi  = \hat{w}$ and $\partial \psi/ \partial z = -\hat{u}$, so $\hat{w} = 0$ whenever $\psi = 0$, while $\hat{u} = 0$ whenever $\partial \psi/ \partial z = 0$. In all cases, $\psi(z)$ oscillates about $0$, showing that a given mode is composed of several overturning cells stacked on top of each other. We see that varying the horizontal wavenumber affects the vertical structure of the mode: modes with larger $k_x$ are more concentrated towards the shear layer itself and decay exponentially rapidly away from it, while modes with smaller $k_x$ can extend vertically far beyond the sheared region. 

However, as shown in Figure \ref{fig:varyk&a}b, the small $k_x$ modes need enough space to exist. Indeed, for a fixed value of $k_x$ (here for example $k_x = 0.1$), the shape of the stream function in the vicinity of the sheared region is more-or-less independent of the overall domain size. As the domain size decreases with decreasing $a$, the mode has to adjust to the applied boundary conditions. The shrinking streamfunction has fewer and fewer nodes, thus reducing the number of vertically-stacked overturning cells it has. Below some critical value of $a$ ($\sim 0.45$ for this particular $k_x$), the mode no longer has enough room to exist, and is stabilized. 
For larger values of $k_x$, on the other hand, the vertical extent of the modes is much smaller, and they remain confined to the sheared region. These modes can therefore exist for much smaller values of $a$ (albeit for sufficiently small ${\rm RiPe}$). 

\begin{figure}[t]
  \includegraphics[width=0.5\textwidth]{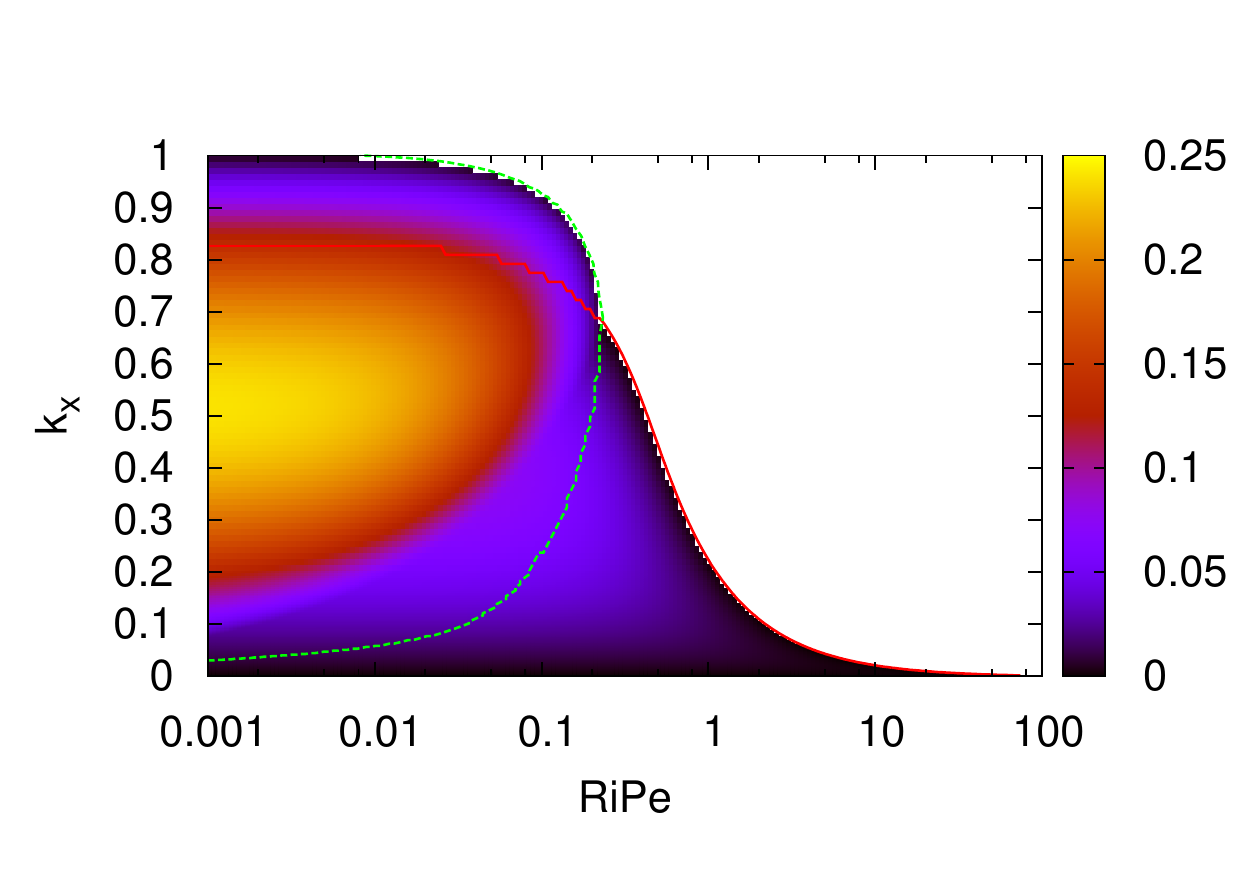}\hfill
  \includegraphics[width=0.5\textwidth]{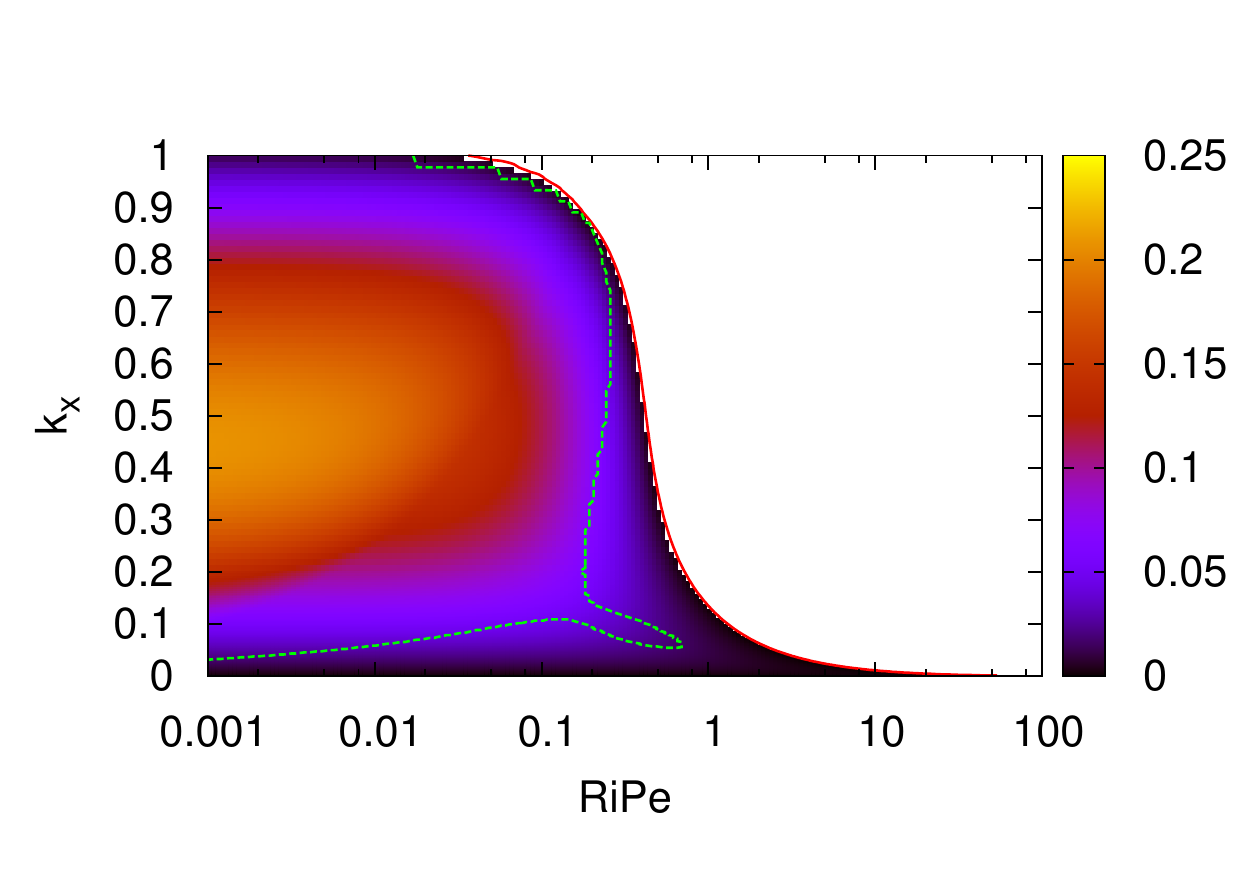}
  \caption{Marginal stability curves at ${\rm Re}=10^4$ and $a=1$ (left) and $a=2$ (right) for Mode S (green) and mode H (red), respectively. Each mode is unstable to the left of its corresponding curve. Colors show the actual growth rate of the fastest growing of the two modes.}
\label{fig:twomodes}
\end{figure}

Figure \ref{fig:twomodes} presents the marginal stability contour ($\operatorname{\mathbb{R}e}(\lambda) = 0$) for two distinct unstable modes of the linear equations for $a=1$ and $\Re= 10^4$. The mode labeled ``Mode S'' typically dominates at large $k_x$, and its marginal stability boundary has a shape that is reminiscent of the one obtained in the purely sinusoidal case \citep{Garaudal15}. The mode labeled ``Mode H'' on the other hand dominates at small $k_x$, and its marginal stability curve has a shape that is reminiscent of the one obtained by \citet{Lignal1999} for the hyperbolic tangent case.  The stability boundary presented earlier in Figure \ref{fig:marginalstab}b, for $a=  1$ and ${\rm Re} = 10^4$, is the envelope of these two curves.

\begin{figure}[t]
  \includegraphics[width=.5\textwidth]{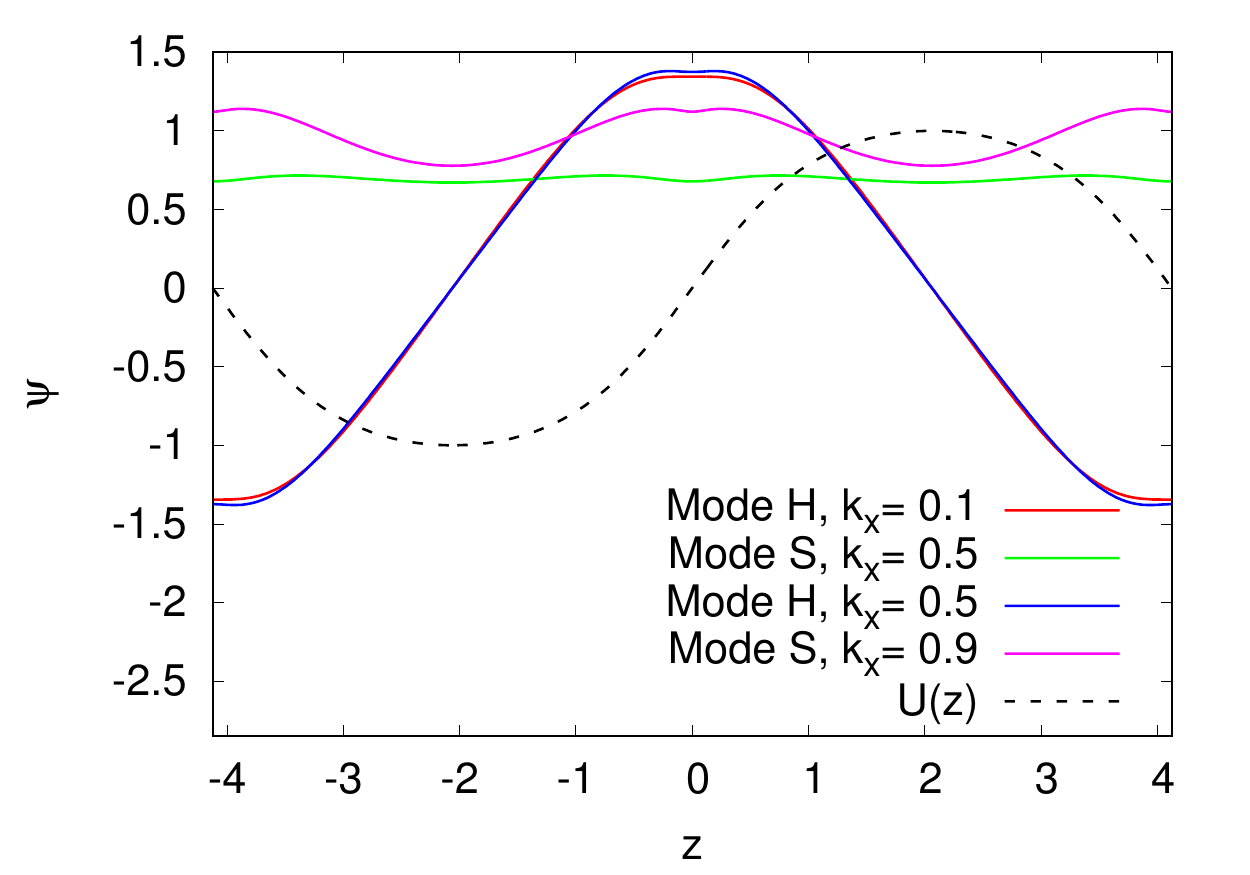}\hfill
  \includegraphics[width=.5\textwidth]{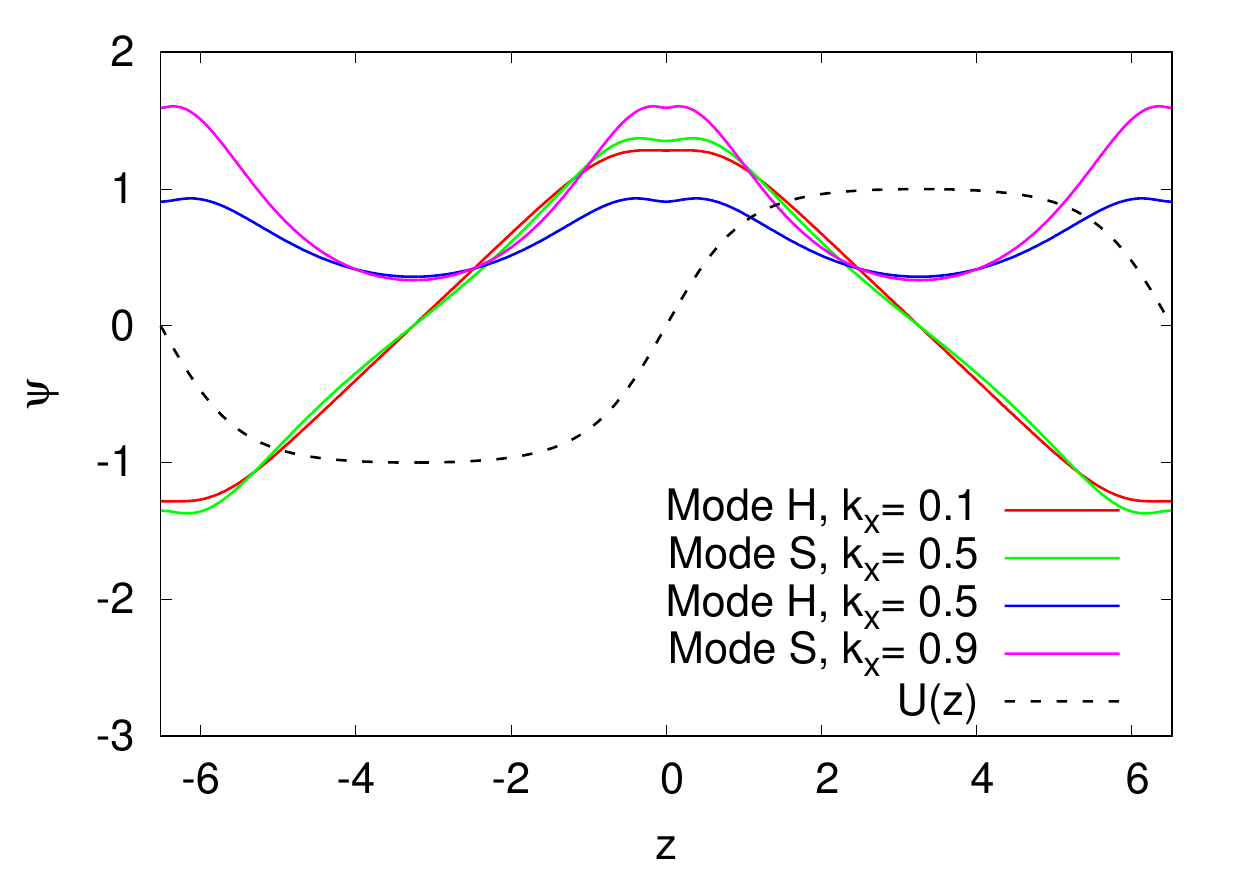} 
  \caption{The streamfunction $\psi(z)$ for $a=1$ (left) and $a=2$ (right) at ${\rm RiPe=0.1}$, for various values of $k_x$. Mode H is the only one unstable at $k_x=0.1$. Similarly, Mode S is the only one unstable for $k_x=0.9$. At $k_x=0.5$ both modes are unstable. Also shown in both cases is the laminar flow profile $\hat U(z)$, to illustrate the position and vertical extent of the shear layer. }
\label{fig:psi_z_ex}
\end{figure}

In order to gain some insight into the difference between the two types of modes, we compare their corresponding streamfunctions $\psi$ in Figure \ref{fig:psi_z_ex}. The parameters used here are $a = 1$, ${\rm Re} = 10^4$ and ${\rm RiPe} = 0.1$. We saw from Figure 4 that for $k_x=0.1$ and $k_x=0.9$ the only modes contributing to linear instability are respectively Mode H and Mode S. For $k_x=0.5$, both are contributing, Mode S being the faster growing one. Figure~\ref{fig:psi_z_ex} shows that Mode S has a streamfunction that never changes sign, which in turn implies that, for a given value of $x$ that is not at a node, the vertical velocity never changes sign (since $\hat{w}=  ik_x \psi$). By contrast, Mode H has a vertical velocity that is zero (and changes sign) when the shear is zero.

\begin{figure}[t]
  \centerline{\includegraphics[width=0.7\textwidth]{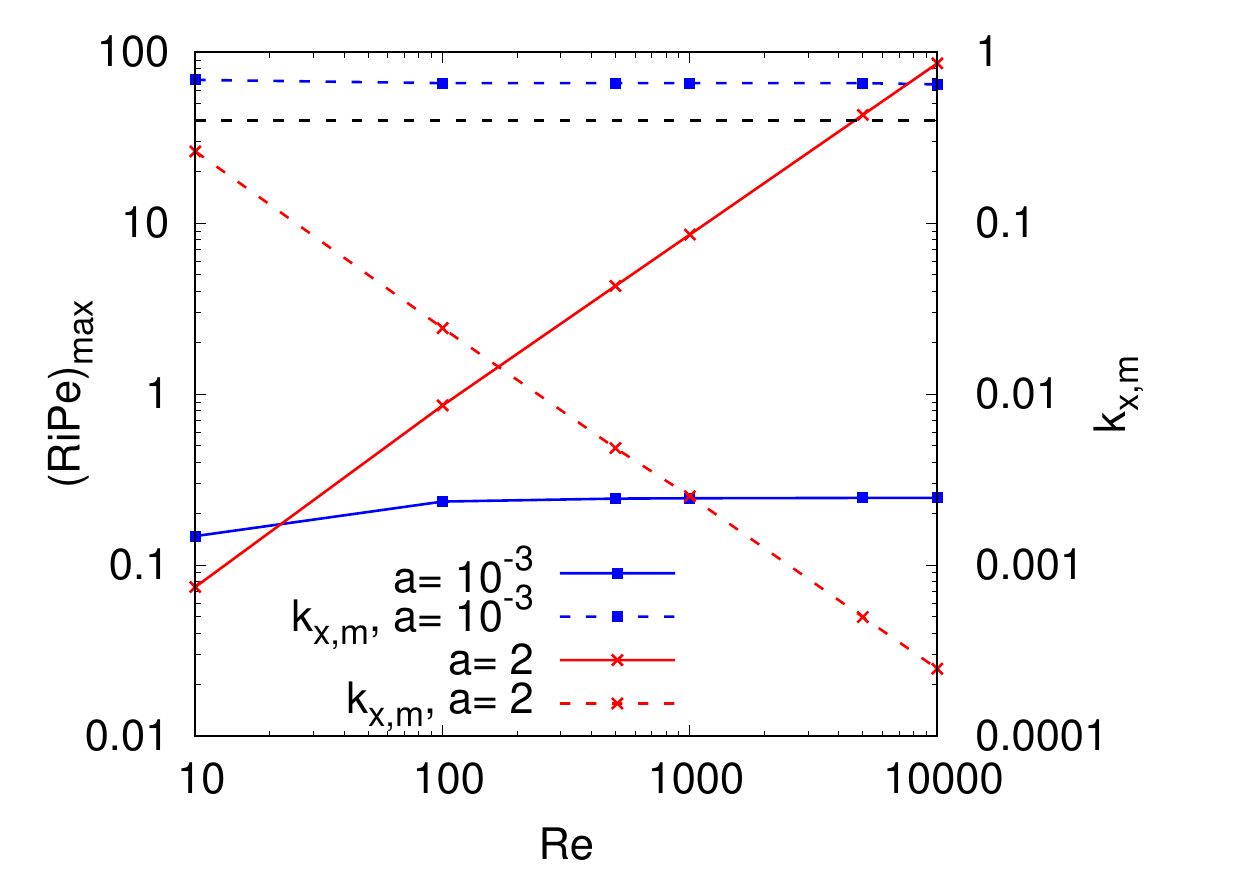}}
  \caption{Solid lines: Maximum value of ${\rm RiPe}$ for which the flow is subject to linear instability as a function of ${\rm Re}$ for the sinusoidal case with $a= 10^{-3}$ (blue) and for the periodic tanh case with $a=2$ (red).  $({\rm RiPe})_{\rm max} \simeq {\rm cste}$ for $a=10^{-3}$ whereas $({\rm RiPe})_{\rm max} \propto {\rm Re}$ for $a=2$. Dashed lines: Wavenumber of the marginally stable mode $k_{x,m}$ as a function of ${\rm Re}$ for $a= 10^{-3}$ and $2$. While $k_{x,m} \simeq {\rm cste}$ for $a= 10^{-3}$, $k_{x,m} \propto 1/{\rm Re}$ for $a= 2$. The black dashed line corresponds to the smallest possible value of $k_x$ achievable in DNS when the horizontal domain size is $L_x = 5\pi$ (see Section~\ref{sec:numsetup}).}
\label{fig:RiPemax_Re}
\end{figure}

In the numerical simulations presented later, we shall focus on one particular value of $a$, namely $a = 2$, which behaves a little like the hyperbolic tangent case. We now compare the stability domains of the $a = 2$ case to the sinusoidal case (approximated here by taking $a =0.001$). 
Figure \ref{fig:RiPemax_Re} shows (on the left axis) the maximum value of $ {\rm RiPe}$ that can be linearly unstable for a given ${\rm Re}$, for sinusoidal flows and for $a = 2$. We thus recover the results of \citet{Garaudal15} in the sinusoidal case ($a = 0.001$), who found that the largest linearly unstable $ {\rm RiPe}$ (called ${\rm RiPe}_{\rm max}$ here) is close to $0.25$ regardless of ${\rm Re}$. When $a = 2$ on the other hand, we find that ${\rm RiPe}_{\rm max}$ scales linearly with ${\rm Re}$, as for the pure hyperbolic tangent flow of \citet{Lignal1999}. Based on the results described earlier, we can now understand the difference between the two scalings as arising from the existence and behavior of Mode H, which dominates the instability at high $ {\rm RiPe}$ but disappears entirely when the vertical domain size is too small. 
Note, however, that the existence of linearly unstable modes for large ${\rm RiPe}$ in the case of $a = 2$ is also contingent on having a large horizontal domain, since they have very low $k_x$. This can be seen in Figure \ref{fig:marginalstab}, and is summarized in Figure \ref{fig:RiPemax_Re} (right axis), which shows the wavenumber of the marginally unstable mode $k_{x,m}$ for a given ${\rm Re}$, and ${\rm RiPe} = {\rm RiPe}_{\rm max}$. We see that $k_{x,m}$ is close to $0.45$ for the sinusoidal case, but decreases as $1/{\rm Re}$ for $a =2$ \citep{Lignal1999}. Hence, the horizontal extent of the domain must be larger than $2 \pi/k_{x,\rm {m}}$ for linear instability to be possible up to ${\rm RiPe} = {\rm RiPe}_{\rm max}$. This is not necessarily an issue in a star if the shear layer is thin compared with its radius, but does affect numerical simulations. Indeed, if the computational domain is smaller than $2\pi/k_{x,m}$, then the largest value of $\Ri\Pe$ for which linear instability can take place will be smaller than ${\rm RiPe}_{\rm max}$.

\section{Numerical simulations}
\label{sec:num}

\subsection{Numerical setup}\label{sec:numsetup}
In order to study the nonlinear evolution of the shear instability, we now solve the set of equations (\ref{LPN1.1}) and (\ref{LPN1.2}) in a triply periodic domain of size $(L_x,L_y,L_z)$. In all simulations we take $L_z = 2\pi a/\tanh(a) \simeq 13.04$ for the shape factor $a=2$. 

As discussed in Section~\ref{sec:mod}, the value of $a$ selected controls the width of the flat plateaus separating the periodically-spaced hyperbolic tangent shear layers. We originally wanted to explore a wide range of values of $a$,  but this turns out to be computationally prohibitive. Indeed, we are primarily interested in fully turbulent flows, which require a high Reynolds number and an equivalently high resolution. Then of course, the larger $a$, the larger $L_z$ has to be (e.g. $L_z \simeq 8.25$ for $a = 1$, $L_z \simeq 13.04$ for $a = 2$, $L_z \simeq 31.42$ for $a =5$, etc). Also, in order to be able to probe the linearly-unstable modes at high $\Ri\Pe$, i.e. in the very small $k_x$ regime, we would need a very large $L_x$ (see Section 3). These constraints, when combined, imply that it is in practice impossible to pick a value of $a$ larger than a few. We have therefore decided that the value $a = 2$ provides a good compromise for our purposes. Note how the corresponding flow $\hat U(z)$ (see Figure 1) contains two substantial regions with negligible shear. 

We use the PADDI pseudo-spectral code described in \citet{Traxleral2011a}. This code has been modified to include the effect of the body force $\bF$ and to solve the LPN momentum equation together with the continuity equation by \citet{Garaudal15}. All perturbations are assumed to be triply-periodic in the domain. 
From here on, we will work using the forcing-based non-dimensionalization introduced by \citet{Garaud2016} which assumes a balance between the forcing term and the nonlinear inertial terms. This implies that the new unit velocity is
\begin{equation}
[u]=U_F\equiv\left(\frac{F_0 L_0}{\rho_0}\right)^{1/2} = \frac{U_0}{\rm Re} \ ,
\end{equation}
since 
\begin{equation}
F_0= \frac{\rho_0 U_0 \nu}{L_0^2} \ .
\end{equation}
The unit length and temperature remain defined as before. Using $U_F$, one can define a new Richardson number, P\'eclet number and Reynolds number based on the forcing as

\begin{align}\label{forcingnondim}
\begin{split}
	  \Re_F&=\frac{U_F L_0}{\nu} ={\rm Re}^{1/2} \ , \\
      \Pe_F&=\frac{U_F L_0}{ \kappa_T} ={\rm Re }^{-1/2}{\rm Pe} \ , \\
      \Ri_F&=\frac{N^2 L_0^2}{U_F^2} = \frac{N^2 L_0\rho_0}{ F_0}={\rm ReRi}.
\end{split}
\end{align}

In this new system of units, the velocity, pressure and temperature fields are denoted as $\breve{\bu}$, $\breve{p}$ and $\breve{T}$, and the system of governing equations is

\begin{equation}\label{LPN1}
\frac{\partial \breve{\bu}}{\partial t}+ \breve{\bu} \cdot \bnabla \breve{\bu}=-\bnabla \breve{p} + \Ri_F\Pe_F \nabla^{-2}\breve{w}\be_z + \frac{1}{\Re_F} \nabla^2 \breve{\bu} + \breve{F}(z)\be_x \ ,
\end{equation}
\begin{equation}\label{LPN2}
\bnabla \cdot \breve{\bu} = 0 \ ,
\end{equation}
where 
\begin{equation}
\breve{F}(z)= \frac{d^2}{dz^2} \left(\frac{\tanh(a \sin(2 \pi z/ L_z))}{\tanh(a)}\right) \ .
\end{equation}
The interest of this non-dimensionalization compared to the one based on the amplitude of the laminar flow is that the new velocity scale $U_F$ is a reasonably good estimate for the r.m.s. velocity of the fully turbulent flow, so $\Re_F$ and $\Ri_F\Pe_F$ are correspondingly good estimates of its actual Reynolds and Richardson-P\'eclet numbers \citep{Garaud2016}. 


In what follows, we will present a series of DNS results from high Reynolds number simulations (with ${\rm Re}_F = 100$  or equivalently ${\rm Re} = 10^4$),  at $a = 2$ (close to a hyperbolic tangent case), and with $L_y = 2\pi$, $L_x = 5\pi$ and $L_z= 2 \pi a / \tanh(a) \simeq 13.04$. Our selection of the horizontal domain size $L_x$ was guided by the following consideration: since $L_x$ cannot be taken to be as large as $2\pi/k_{x,m}$ anyway (which would need to be of order $10^{4} \pi$ here, see Section 3 and Figure \ref{fig:RiPemax_Re}) there is no hope of being able to capture the linearly unstable modes that exist at very large Richardson-P\'eclet numbers. Instead, we have selected to take $L_x = 5\pi$, simply for computational feasibility. Having  $L_x = 5\pi$  implies that the smallest possible value of $k_x$ is about $0.4$. The spanwise domain size $L_y$ has been chosen to be $2\pi$ following \cite{Garaud2016}. We checked that this is large enough to avoid affecting the dynamics of the fully turbulent flow. In the moderately and strongly stratified regimes in particular, it is many times larger than the spanwise eddy scale. The largest value of the Richardson-P\'eclet number  for which our flow remains linearly unstable is now simply dictated by the domain size and is approximately equal to $\Ri\Pe = 0.45$ (see figure \ref{fig:marginalstab}b), corresponding to $\Ri_F\Pe_F=45$. While not extremely large, this is still larger than in the sinusoidal case (for which the maximum linearly unstable $\Ri_F\Pe_F$ is about $25$). Furthermore, the linearly unstable mode in the $a = 2$ simulation is predicted to be of type  H  for moderate and large ${\Ri}_F{\Pe}_F$, and is therefore expected to behave the same way as the modes that would be unstable at lower $k_x$ should the domain be larger. 
Table~\ref{table:0} summarizes the properties of each DNS performed, as well as the initial conditions used in each case. From here on the breve symbols are dropped for simplicity of notation, and all the results are reported in the non-dimensionalization based on the forcing unless specifically mentioned.

\subsection{Sample simulations}

In this section we present the various regimes observed in the simulations for $\Re_F= 100$ and $a=2$: a weakly stratified run ($\Ri_F\Pe_F=0.1$), a linearly unstable run that has significant stratification ($\Ri_F\Pe_F=10$), and a linearly stable run that exhibits a nonlinear instability ($\Ri_F\Pe_F=50$). In all that follows, $\langle . \rangle $ implies a volume average over the entire domain, while $\bar{.}$ implies a horizontal average. 

\subsubsection{Weak stratification regime}
\label{sec:weakstrat}

\begin{figure}[h]
	\includegraphics[width=.5\textwidth]{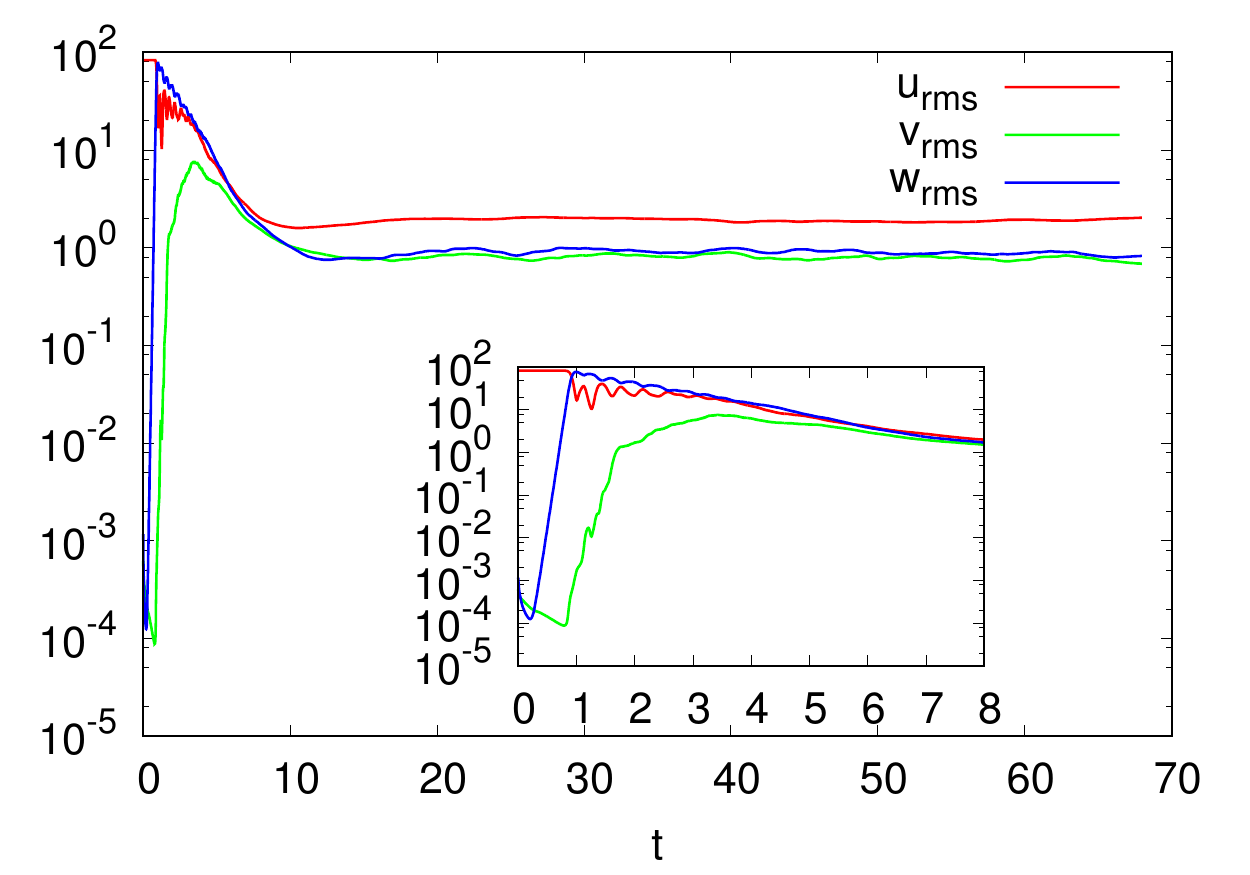}\hfill
	\includegraphics[width=.5\textwidth]{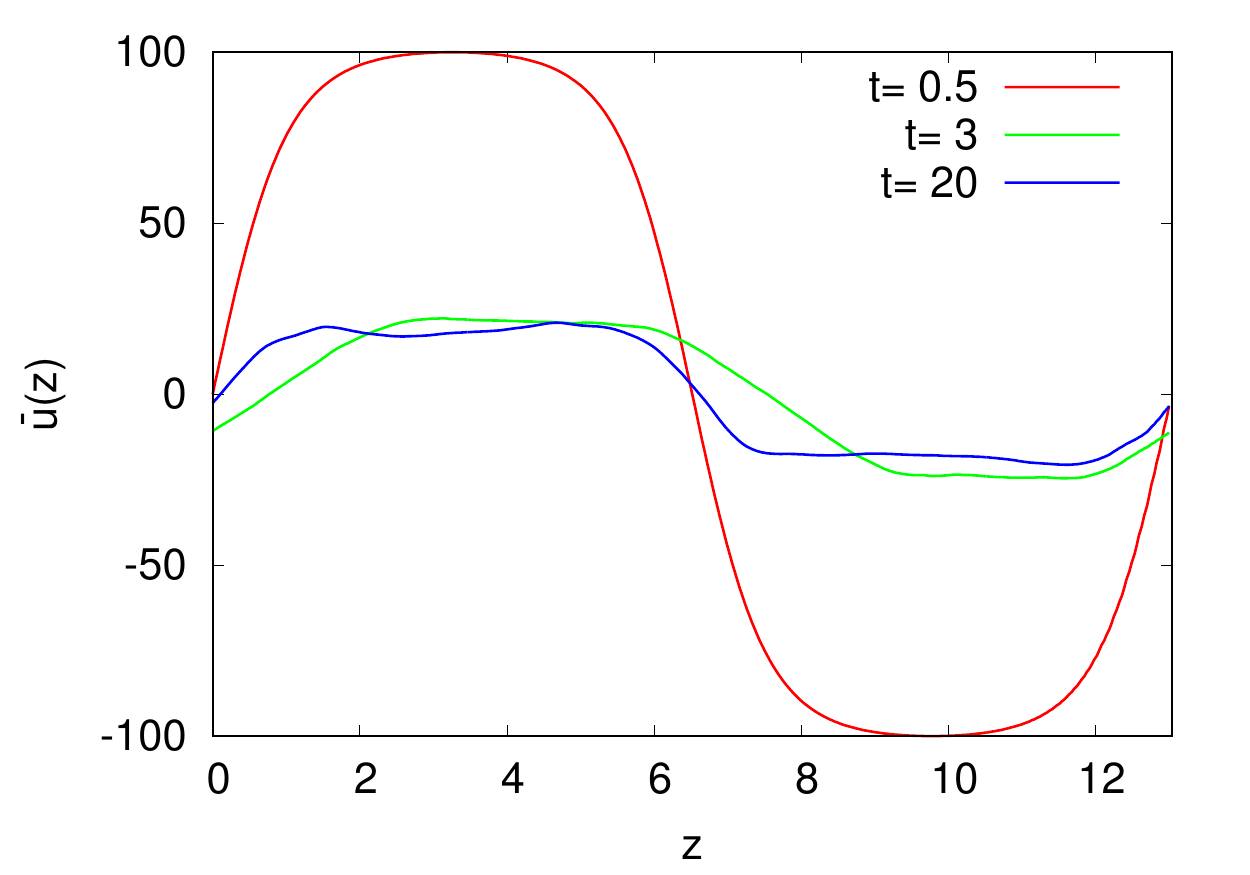}

  \caption{Left : $u_{\rm rms}$, $v_{\rm rms}$ and $w_{\rm rms}$ as a function of time for  a DNS with $\Re_F=100$, $\Ri_F\Pe_F=0.1$ and $a=2$. The statistically steady state is reached at around $t\sim 10$. Right : snapshots of the mean flow $\bar u(z)$ as a function of $z$ at different time steps. The red curve corresponds to the initial linearly unstable profile, the green curve corresponds to the short transient phase and the blue curve is shown  during the statistically stationary state. The $t=20$ velocity profile has been multiplied by a factor $10$ for visibility.}
\label{fig:u_z_u_t}
\end{figure}

\begin{figure}[h]
	\includegraphics[width=.33\textwidth]{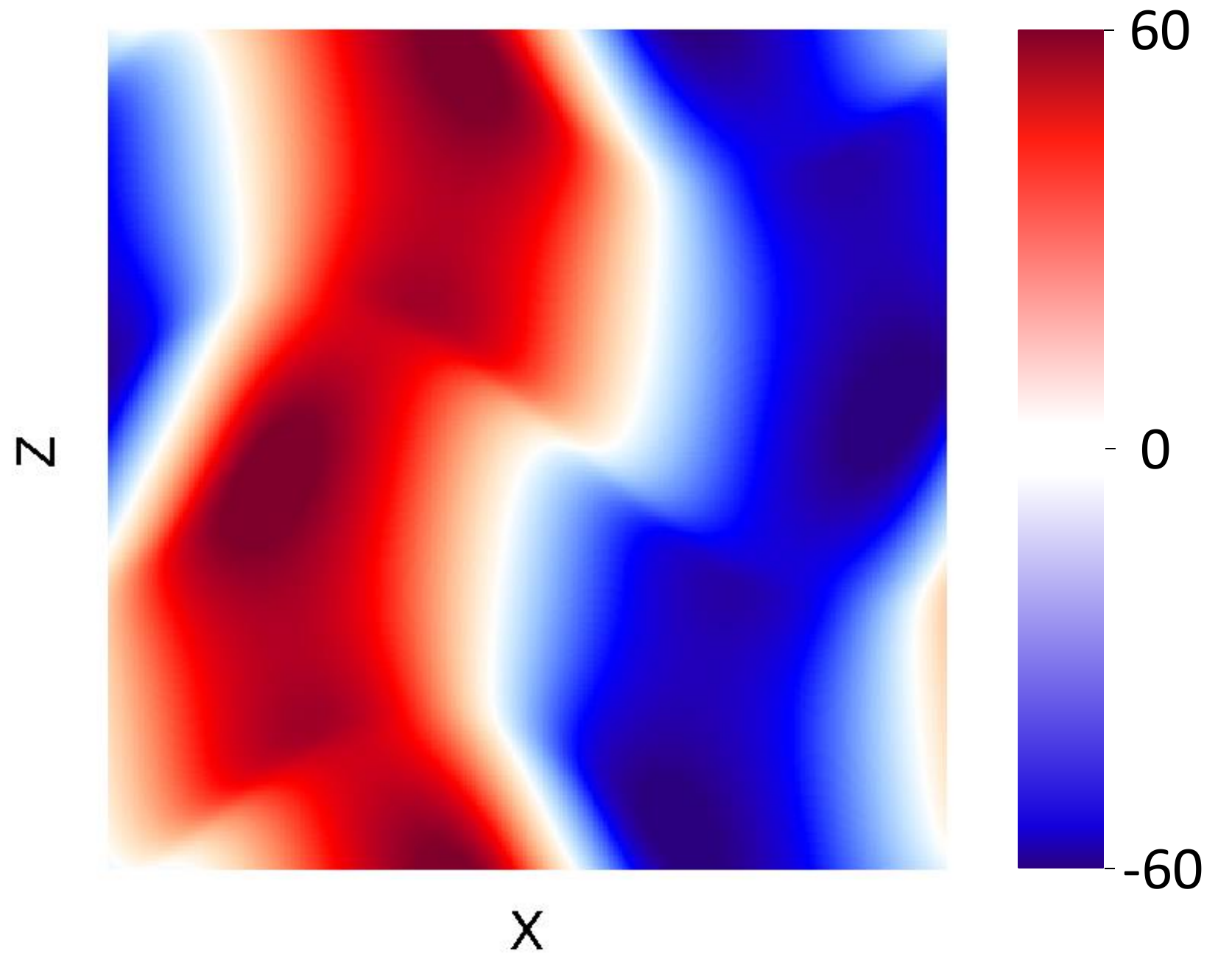}\hfill
	\includegraphics[width=.33\textwidth]{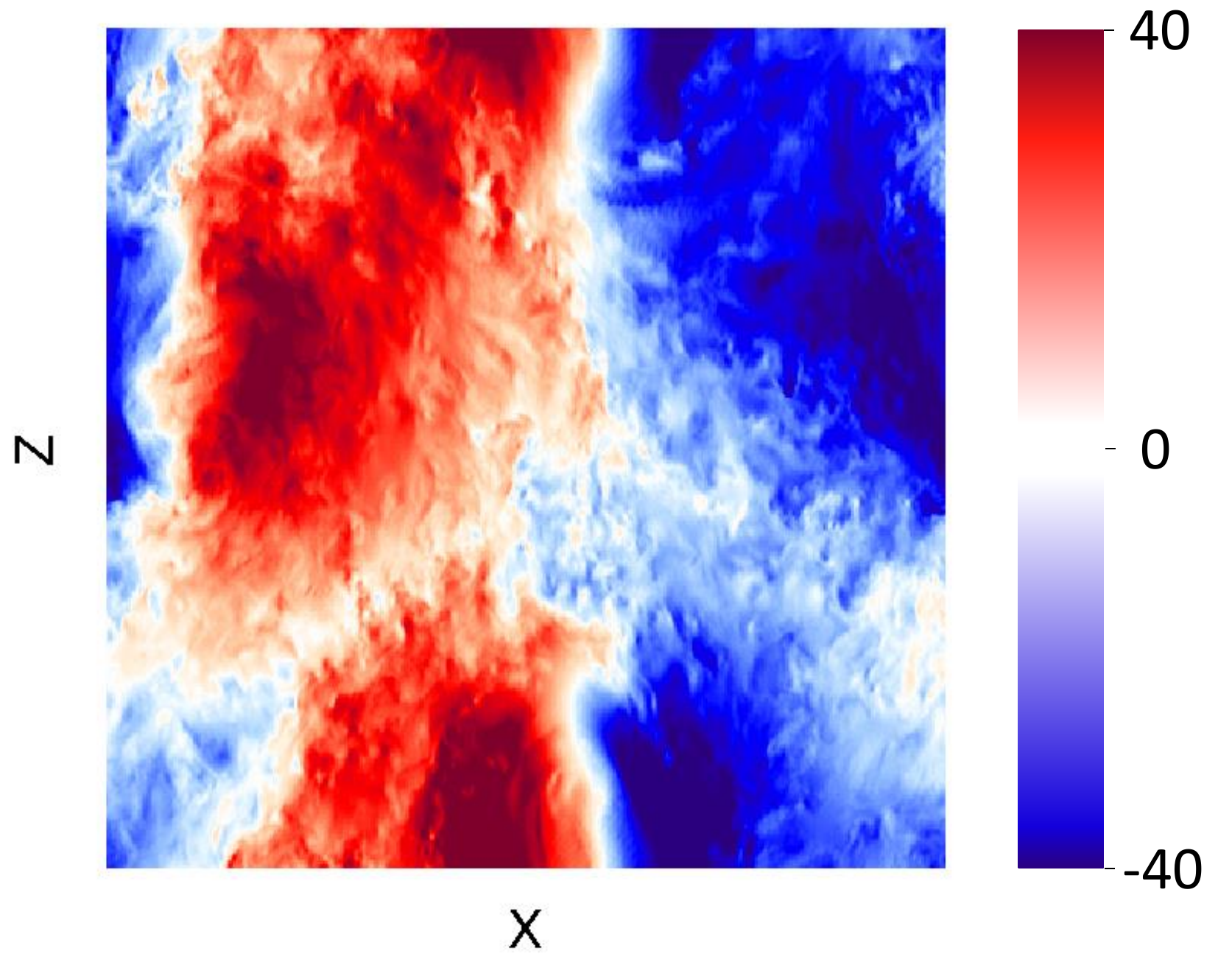}\hfill
	\includegraphics[width=.33\textwidth]{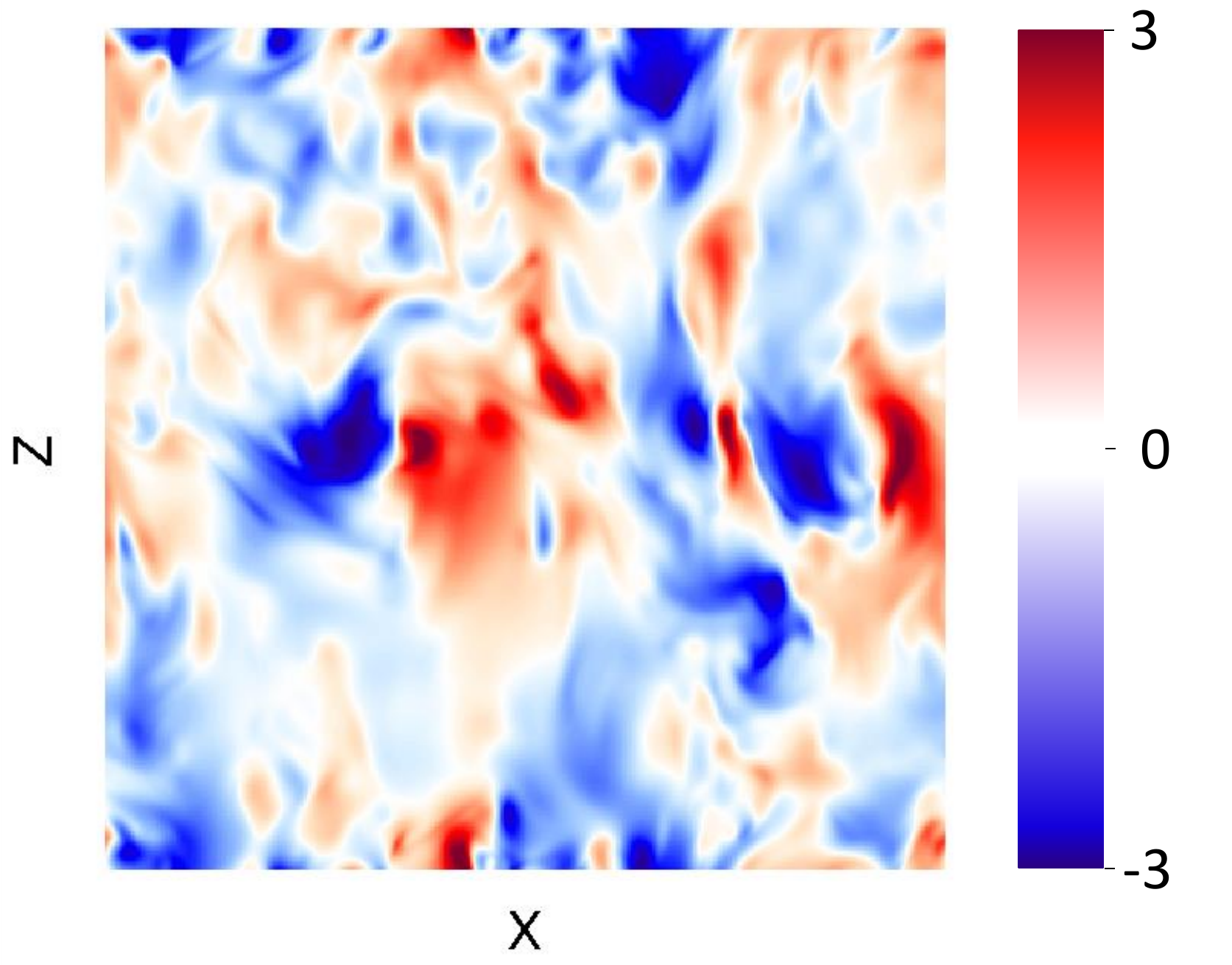}
  \caption{Snapshots of the vertical velocity field  $w(x,y=0,z,t)$ for $\Re_F=100$, $\Ri_F\Pe_F=0.1$ and $a=2$ at three different time-steps ($t= 0.5, \ 3, \ \rm{and} \ 20$). The simulation has been started from the laminar profile from equation (\ref{eq:laminar}) and} grows into a Mode S (left), which later becomes unstable to 3D perturbations (middle) and finally settles into a statistically steady state (right).

\label{fig:ev}
\end{figure}

We begin by looking at the evolution of a weakly stratified LPN simulation, at $\Re_F= 100$, $\Ri_F\Pe_F=0.1$ and $a=2$.   Figure \ref{fig:u_z_u_t}a shows the time dependence of the root mean squared velocities $u_{\rm rms}$, $v_{\rm rms}$ and $w_{\rm rms}$ (where $u_{\rm rms}  =  \langle u^2 \rangle^{1/2}$  and similarly for the other two components). Snapshots of the horizontally-averaged  mean flow profile $\bar{u}(z)$ at selected times are shown in Figure~\ref{fig:u_z_u_t}b  while snapshots of the vertical velocity field $w(x,y=0,z)$ at the same times are shown in Figure~\ref{fig:ev}. We see that the laminar flow profile rapidly becomes unstable to a 2D instability with motion in the $x$ and $z$ directions only. Inspection of the vertical velocity snapshot (Figure~\ref{fig:ev}a) shows that the linearly unstable mode is of type S (recall that in type S modes, the vertical velocity does not change sign with $z$, even when the shearing rate goes to zero).  This instability causes a first transfer of energy from the mean flow to  $y$-invariant fluid motions, up to $t \sim 1$.  
   At this point, the two-dimensional perturbation reaches a sufficient amplitude to become unstable to 3D perturbations and the energy in spanwise ($y-$) motions starts increasing. This is a well-known feature of the transition to turbulence in stratified shear flows \citep{Peltier2003}. Beyond this point, the system rapidly relaxes to a fully-turbulent statistically stationary state (see Figure~\ref{fig:ev}c), which it reaches around $t \sim 10$ (see Figure~\ref{fig:u_z_u_t}a).
   This final state is characterized  by a $u_{\rm rms}$ that is roughly twice as large as $v_{\rm rms}$ and $w_{\rm rms}$ and is of order unity, as expected from the selected non-dimensionalization. Figure \ref{fig:u_z_u_t}b shows the evolution of the mean flow from the initial laminar solution towards the ultimate statistically stationary state. We see that the amplitude of the latter is nearly two orders of magnitude smaller than the former. 
   
\begin{figure}[h]
	\includegraphics[width=.45\textwidth]{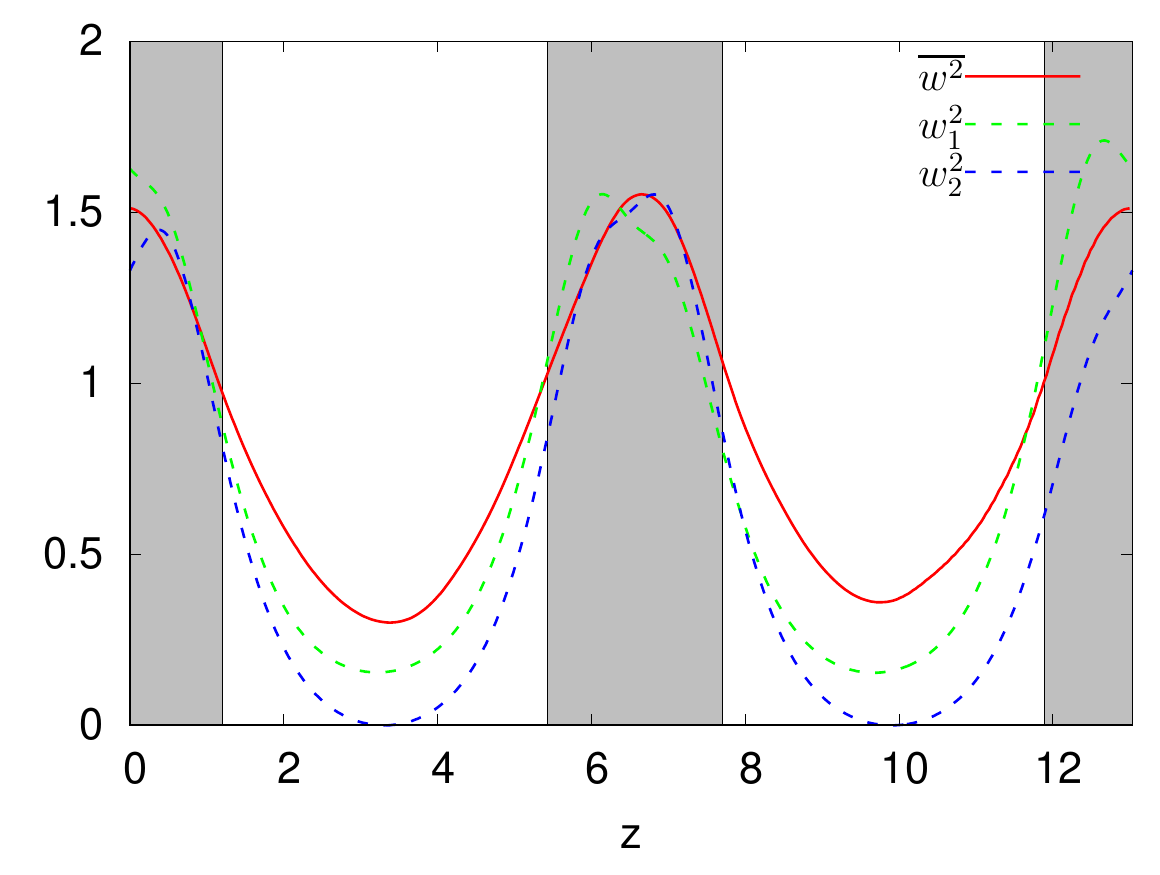}\hfill
	\includegraphics[width=.5\textwidth]{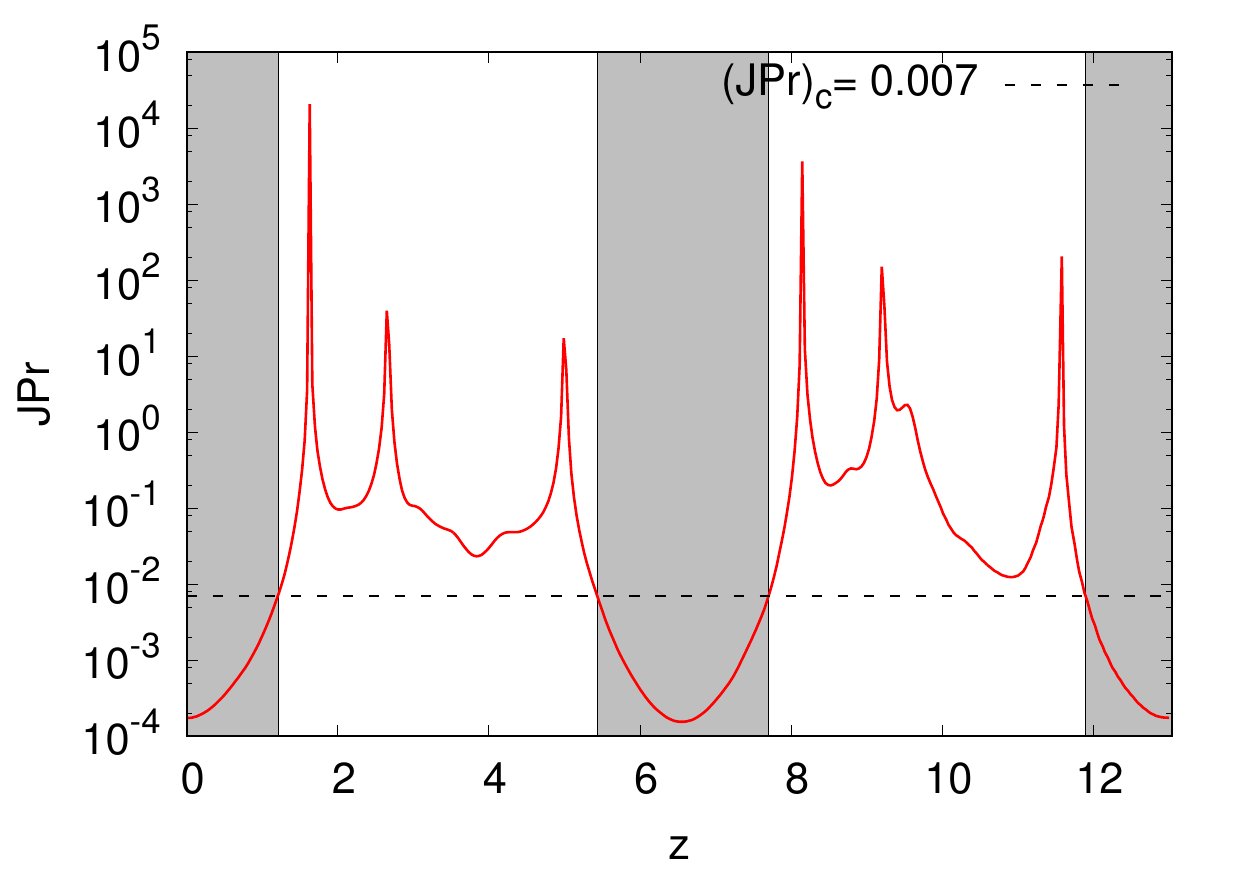}

  \caption{Left: Time and horizontally averaged vertical velocity $\overline{w^2}$  as a function of $z$, for $\Re_F=100$, $\Ri_F\Pe_F=0.1$ and $a=2$, once the statistically steady state is reached. The green and blue curves show  the $\overline{w^2}$ profiles for the fastest and the second fastest growing modes, $w_1^2$ and $w_2^2$, respectively (see text for detail). Right : Corresponding profile of $J\Pr$ as a function of $z$. The black dotted line shows $(J\Pr)_c=0.007$ below which the shear should be unstable according to  Zahn's nonlinear criterion for instability. The shaded areas in both plots highlight the regions that are unstable under this criterion.}
\label{fig:wrms}
\end{figure}

Zahn's nonlinear criterion for instability \citep{Zahn1974}, combined with the recent numerical results of Paper I and \citet{Pratal2016}, states than in order for the flow to be unstable $J\Pr$ has to drop below the critical value  $(J\Pr)_c=0.007$. We can now compare this prediction to the DNSs. Figure~\ref{fig:wrms}a shows the horizontally and time averaged vertical velocity squared $\overline{w^2}$ as a function of $z$. The time average is taken during the statistically stationary state. The quantity $\overline{w^2}$ measures the intensity of the turbulent fluid motions everywhere in the domain.  Figure~\ref{fig:wrms}b shows the time- and horizontally averaged value of $J\Pr$ in the statistically stationary state as a function of $z$, calculated as

\begin{equation}
J\Pr= \frac{\Ri_F\Pe_F}{\Re_F \bar{S}^2} \ ,
\end{equation}
where
\begin{equation}
\bar{S}=\frac{d \bar{u}}{dz} \ .
\end{equation}
We clearly see that the $\rm{r.m.s.}$ vertical velocity of the flow remains significant for all $z$ even in regions where $J{\rm Pr}$ largely exceeds the critical value 0.007. This shows that Zahn's mixing model fails in these regions of low shear: non-local effects from the nearby unstable shear regions drive turbulent mixing everywhere in the domain, including in the theoretically stable regions. One may naturally wonder whether this extension of the turbulent motions into the region of low shear can be captured by the spatial structure of the linearly unstable modes, which can sometimes be quite extended (see Section~\ref{sec:stab}). To check this, we performed a linear stability analysis of the mean flow profile once the latter has reached a statistically steady state. The real part of the growth rate $\lambda$ of the two fastest-growing modes as a function of the horizontal wavenumber $k_x$ of the perturbations, is shown in Figure~\ref{fig:lambda}. For this simulation, we find that the statistically stationary profile remains linearly unstable, and the growth rates peak around  $k_x \sim 0.68$, which corresponds to a horizontal wavelength of about 9.25-- much larger than the observed horizontal size of the turbulent eddies in Figure~\ref{fig:ev}c. Since both modes are of type $S$, their vertical structure extends throughout the full domain. The vertical profile of $w^2$ ($w_1^2$ and $w_2^2$ for the fastest and second fasted growing modes, respectively) for these linearly unstable modes is shown in Figure~\ref{fig:wrms}a (the arbitrary amplitude has been scaled to match that of the observed profile of $\overline w^2$ for ease of comparison). We see that it does provide a reasonably good match to the fully nonlinear flow, but tends to underestimate mixing somewhat in the stable regions.  

As such, it is interesting to see that using information from the linearly unstable mode for this flow can provide some qualitative information about the turbulent profile everywhere in the domain. We will see below, however, that this is no longer the case for more strongly stratified flows.

\begin{figure}[h]\label{fig:lambda}
\centerline{\includegraphics[width=.7\textwidth]{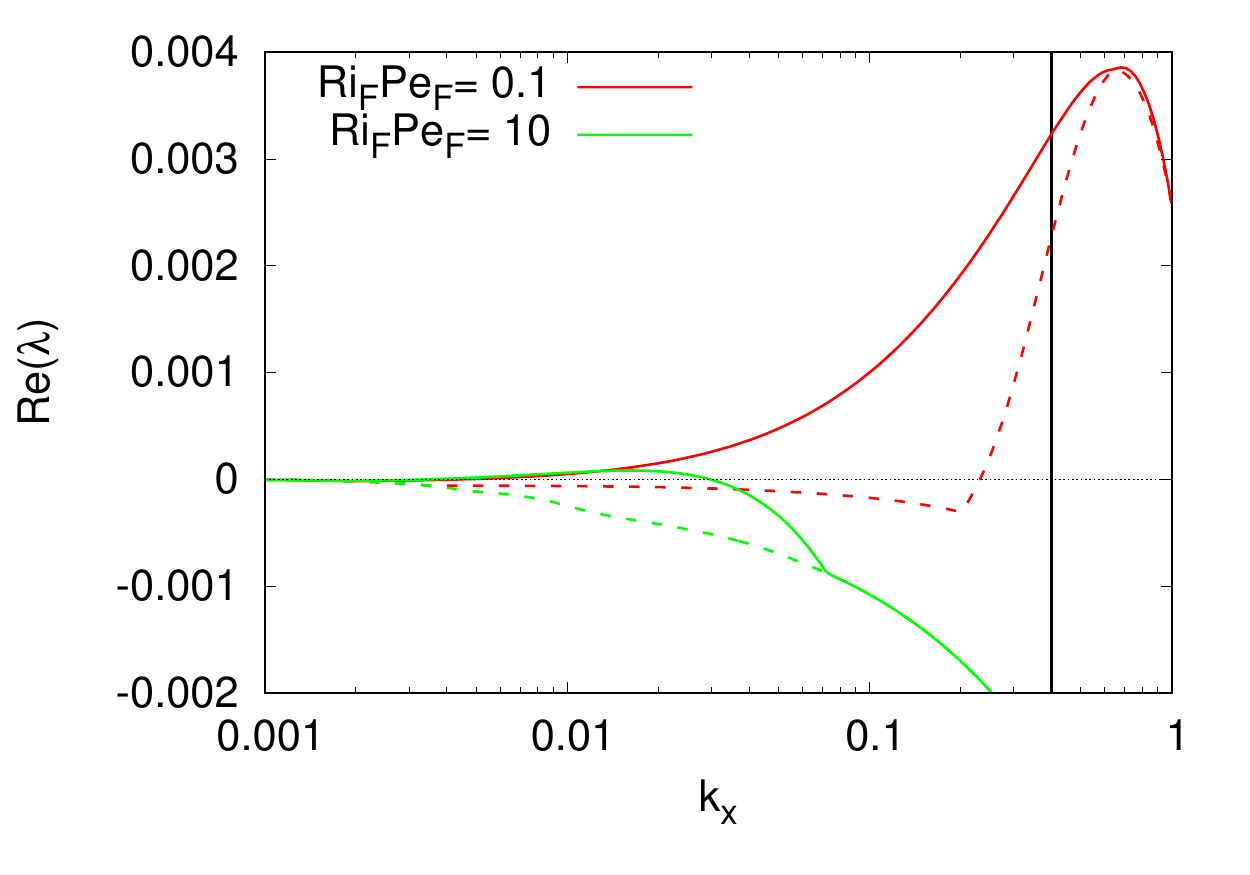}}
  \caption{Real part of the two largest growth rates $\lambda$ from linear stability theory of the mean flow at the statistically steady state as a function of the horizontal wavenumber $k_x$, for the weakly stratified regime with $\Ri_F\Pe_F= 0.1$ and the moderately stratified regime with $\Ri_F\Pe_F= 10$ (Simulation A). The largest and second largest growth rates are shown in solid and dashed lines, respectively. The black vertical line shows the smallest value of $k_x$, namely $2\pi/L_x = 0.4$, available in our domain. }
\label{fig:lambda}
\end{figure}

\subsubsection{Intermediate stratification regime}
\label{sec:intermstrat}
\begin{figure}[h]
	\includegraphics[width=.5\textwidth]{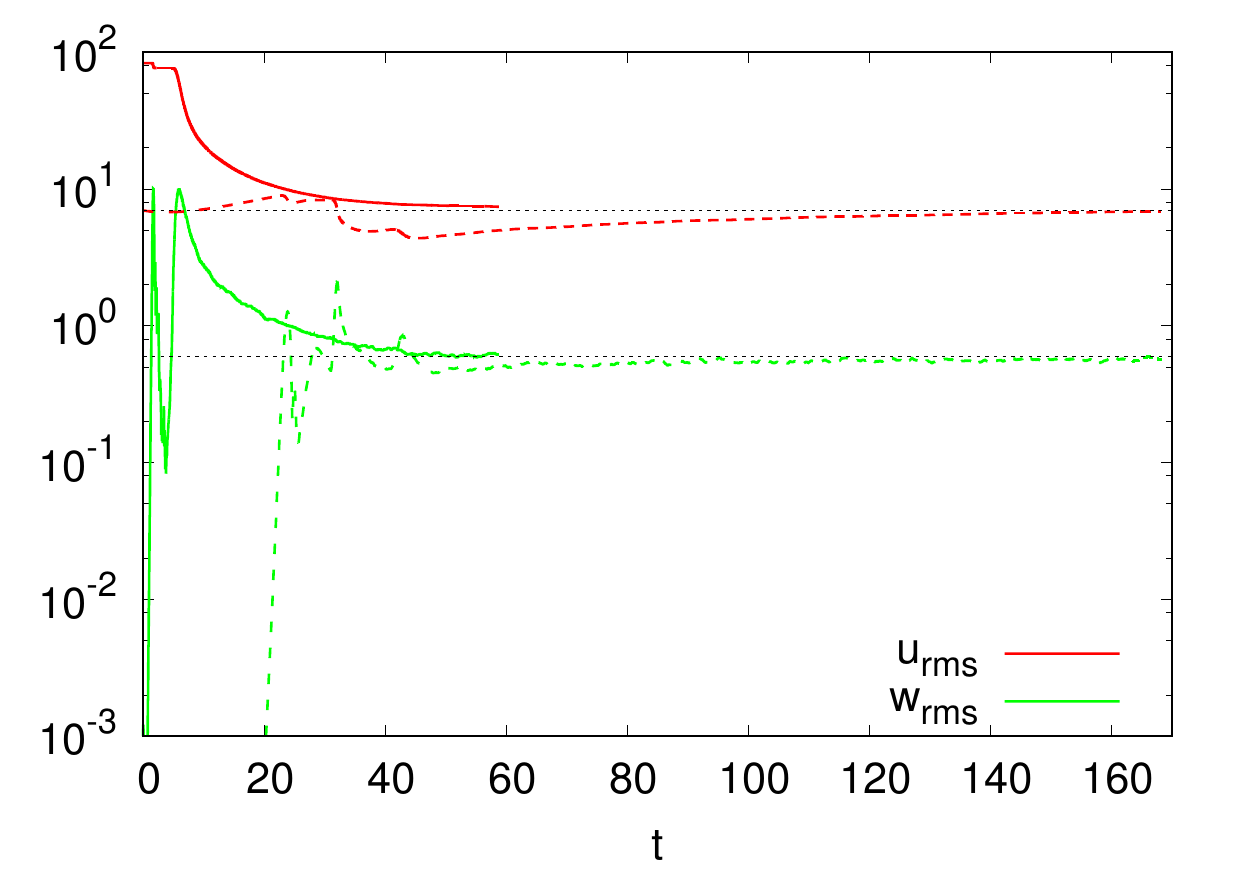}\hfill
	\includegraphics[width=.5\textwidth]{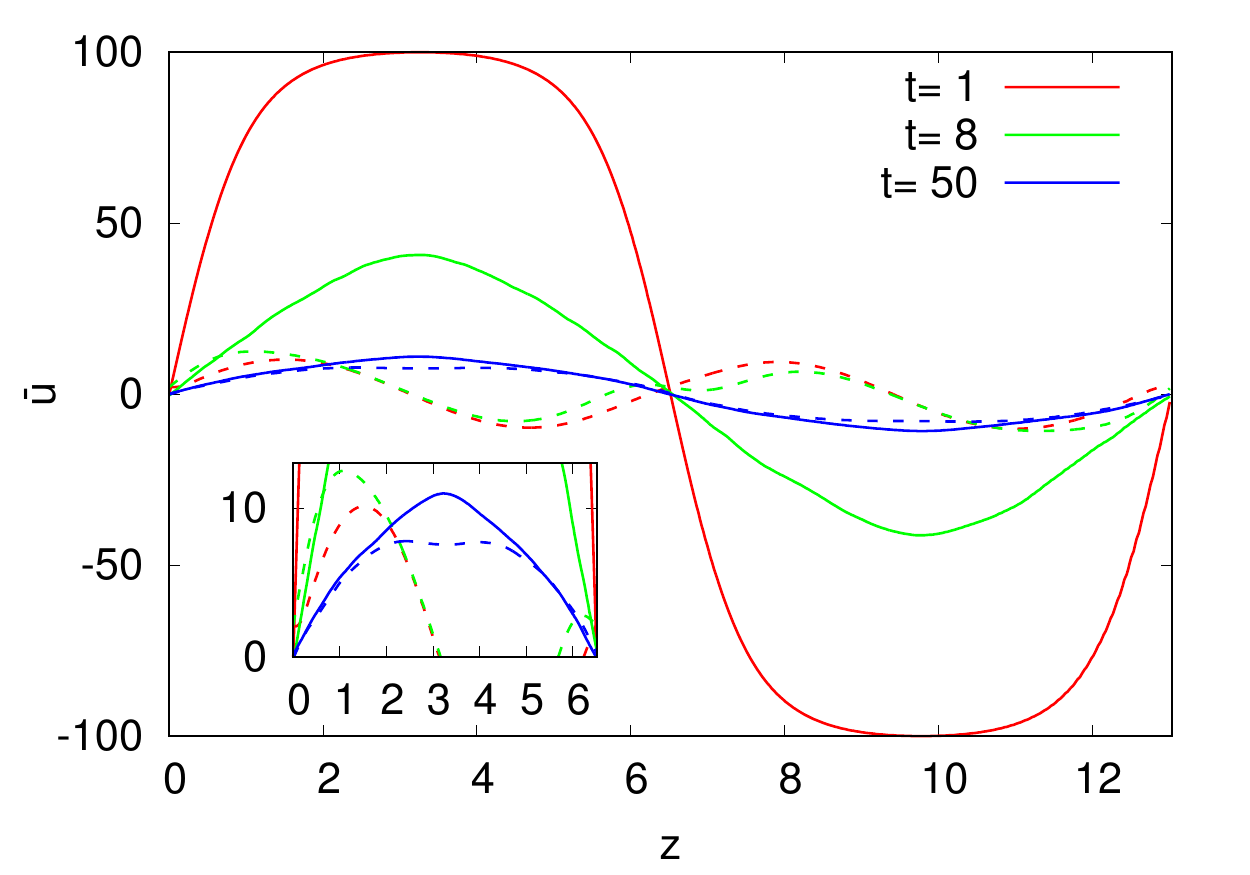}

  \caption{Left : $u_{\rm rms}$ and $w_{\rm rms}$ as a function of time for $\Re_F=100$, $\Ri_F\Pe_F=10$ and $a=2$ for Simulations A and B in solid and dashed lines, respectively. The statistically stationary state is reached around $t\sim 50$ in Simulation A and $t \sim 160$ in Simulation B. The black horizontal lines roughly mark the statistically stationary values. Right : Snapshots of the mean flow $\bar u(z)$ at different times for Simulations A and B in full and dashed lines, respectively. At $t= 50$, Simulation A has reached a statistically stationary state while Simulation B has not. The strongly sheared turbulent regions (e.g. near $z =0$) have equilibrated, but the stable regions are still evolving. }
\label{fig:u_t_u_z_2}
\end{figure}

\begin{figure}[h]
	\includegraphics[width=.33\textwidth]{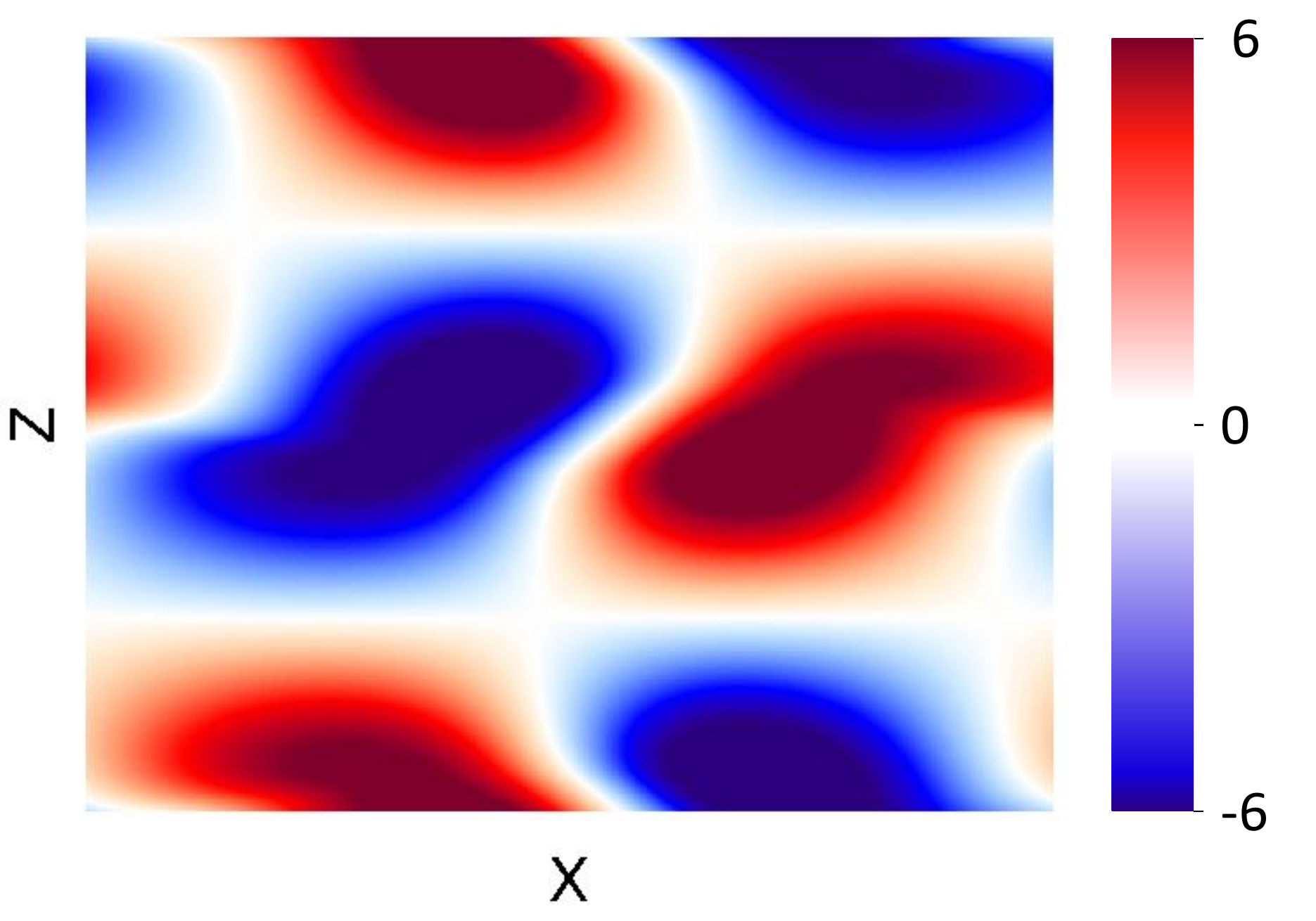}\hfill
		\includegraphics[width=.33\textwidth]{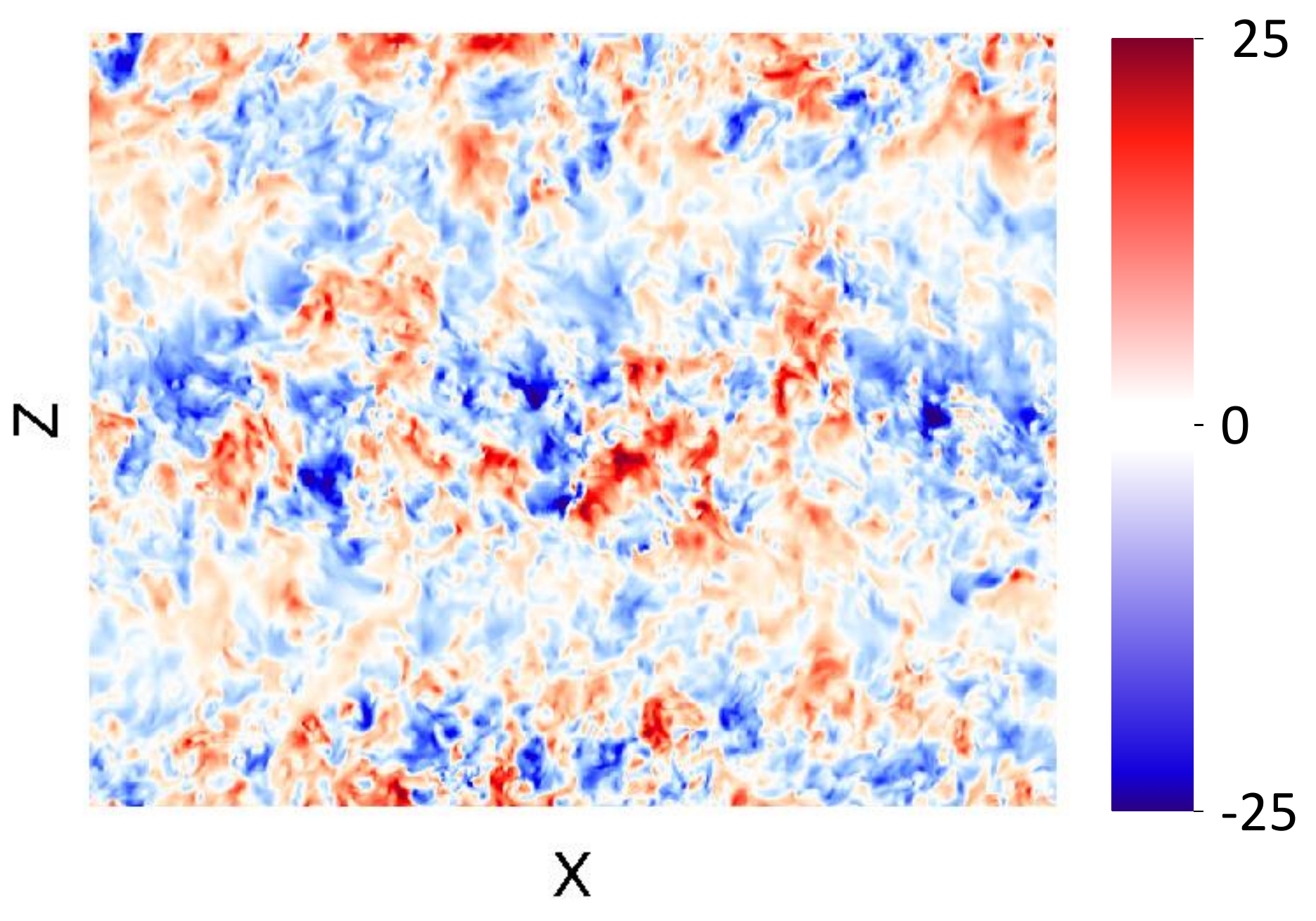}\hfill
	\includegraphics[width=.33\textwidth]{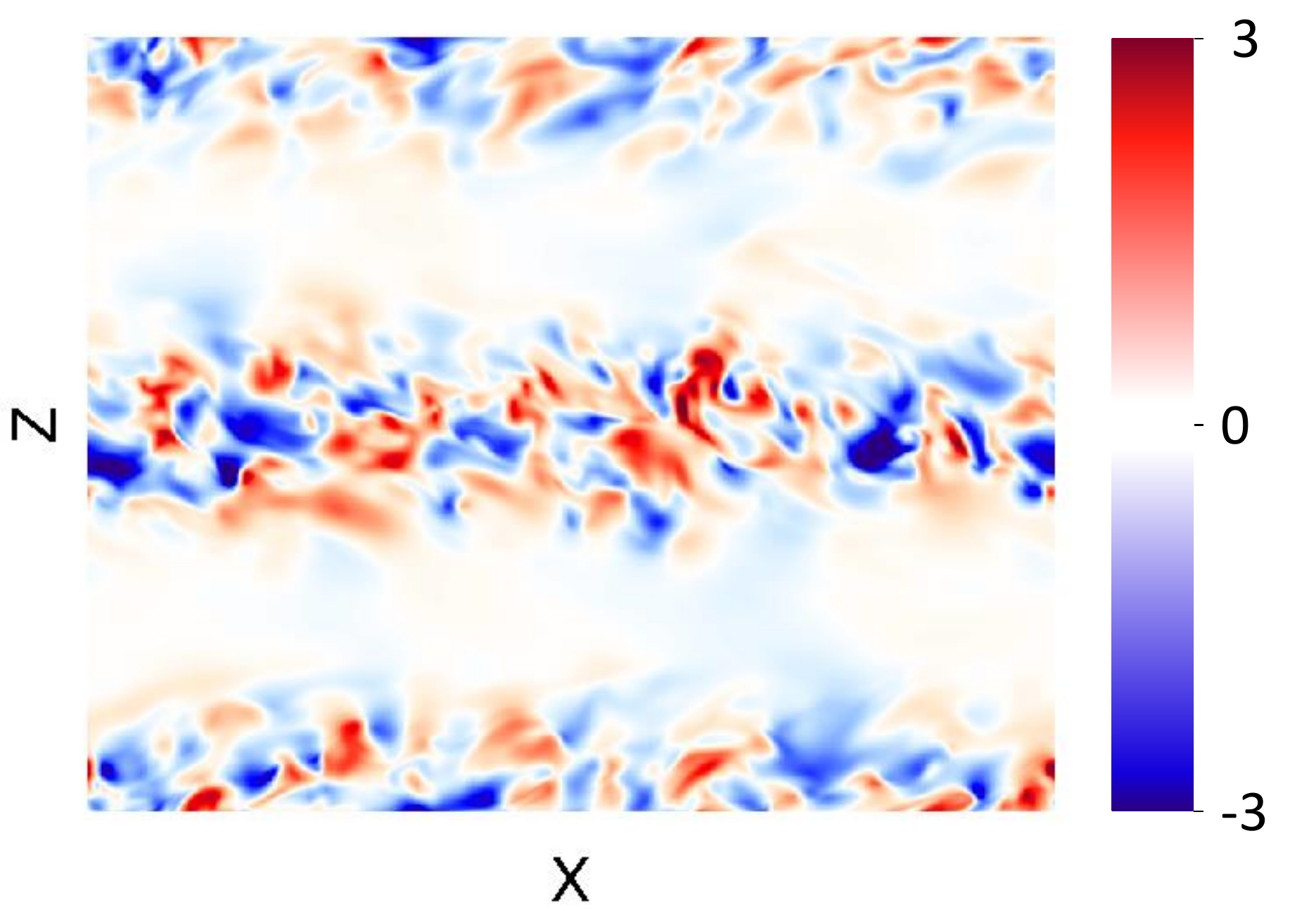}

  \caption{Snapshots of the vertical velocity field $w(x,y=0,z,t)$ for $\Re_F=100$, $\Ri_F\Pe_F=10$ and $a=2$ taken from Simulation A at the same three time-steps as Figure~\ref{fig:u_t_u_z_2}. The flow is subject to linear instability (left) and, after a transient phase (middle), it settles into a statistically stationary state where turbulence is localized in the regions of strongest shear (right).}
\label{fig:snap}
\end{figure}

We now look at LPN simulations with $\Re_F = 100$, $\Ri_F\Pe_F = 10$, and $a = 2$. With this choice of parameters, we are able to probe an intermediate regime  which is still linearly unstable, but where the effects of stratification become important.  
Again, we start by looking at the time dependence of $u_{\rm rms}$ and $w_{\rm rms}$ ($v_{\rm rms}$ is not shown for clarity), see Figure~\ref{fig:u_t_u_z_2}a. This time we present the results of two different simulations with two different initializations: in Simulation A, the mean flow is initialized with the laminar profile  from equation~(\ref{eq:laminar}) while Simulation B is initialized with a weak amplitude sinusoidal flow.  Figure~\ref{fig:u_t_u_z_2}b and Figure~\ref{fig:snap} show snapshots of the mean horizontal velocity profile $\bar{u}$  and of the vertical velocity field $w(x,y=0,z)$ respectively, from Simulation A, at selected times. Both simulations ultimately reach the same statistically stationary state, but do so through different routes: Simulation A rapidly becomes linearly unstable, while Simulation B is first linearly stable, and only becomes unstable once the mean flow has grown to sufficiently large amplitude.  

As before, the linear instability (in Simulation A in particular) is first purely two-dimensional, then later becomes three-dimensional. The linearly unstable mode is global, and spans the entire domain (see Figures~\ref{fig:psi_z_ex} and~\ref{fig:ev}). Looking at Figure~\ref{fig:snap}a, we see that this time the  dominant linear mode is Mode H.  Indeed, as mentioned in Section~\ref{sec:stab}, H modes are characterized by a vertical velocity that changes sign where the shear is zero. Beyond the linear instability phase, however, turbulence becomes strongly localized in the shearing regions once the statistically stationary  state is reached, while the regions on either side become much more quiescent. The turbulent eddy size in this regime is now significantly smaller than at lower $\Ri_F \Pe_F$. The amplitude of the shear is reduced by  one order of magnitude compared to the initial one, and the shape of the mean flow has changed. A linear stability analysis of the new mean flow profile (see Figure~\ref{fig:lambda}) now reveals the latter should be {\it stable} -- the  weak instability of the very long wavelength modes around $k_x \simeq 0.02$ being suppressed in our selected computational domain. The turbulence observed must therefore be nonlinearly driven. This is a good example of a system whose linear stability properties completely fail to give any insight into the nonlinear saturation of the instability, by contrast with our results from the more weakly stratified case discussed in Section~\ref{sec:weakstrat}.

An important point to notice is that in both cases A and B, the flow takes much longer to reach a statistically stationary state than in the more weakly stratified limit. This is particularly true of the simulation initialized with a weaker amplitude mean flow (Simulation B). 
This can be understood by noting that the mean flow evolves in response to the lack of balance between the imposed force, viscous stresses, and the divergence of the turbulent momentum flux. The turbulent flux is naturally large in the strongly sheared unstable regions, but drops to nearly zero in the weakly sheared nominally stable regions. As a result the stable regions evolve on a much slower timescale than the unstable ones. This can be seen in  Figure~\ref{fig:u_t_u_z_2}b (note the inset in particular), which shows that the velocity profile in the strongly sheared regions converges rapidly to a statistically stationary state, but that the more weakly sheared regions continue to evolve on a much longer timescale. Simulation B therefore illustrates a situation that is closer to  the reality of stellar interiors where large scale laminar flows evolve over millions of years while the turbulent regions adjust themselves to the evolving forcing conditions very rapidly.

\begin{figure}[h]
	\includegraphics[width=.475\textwidth]{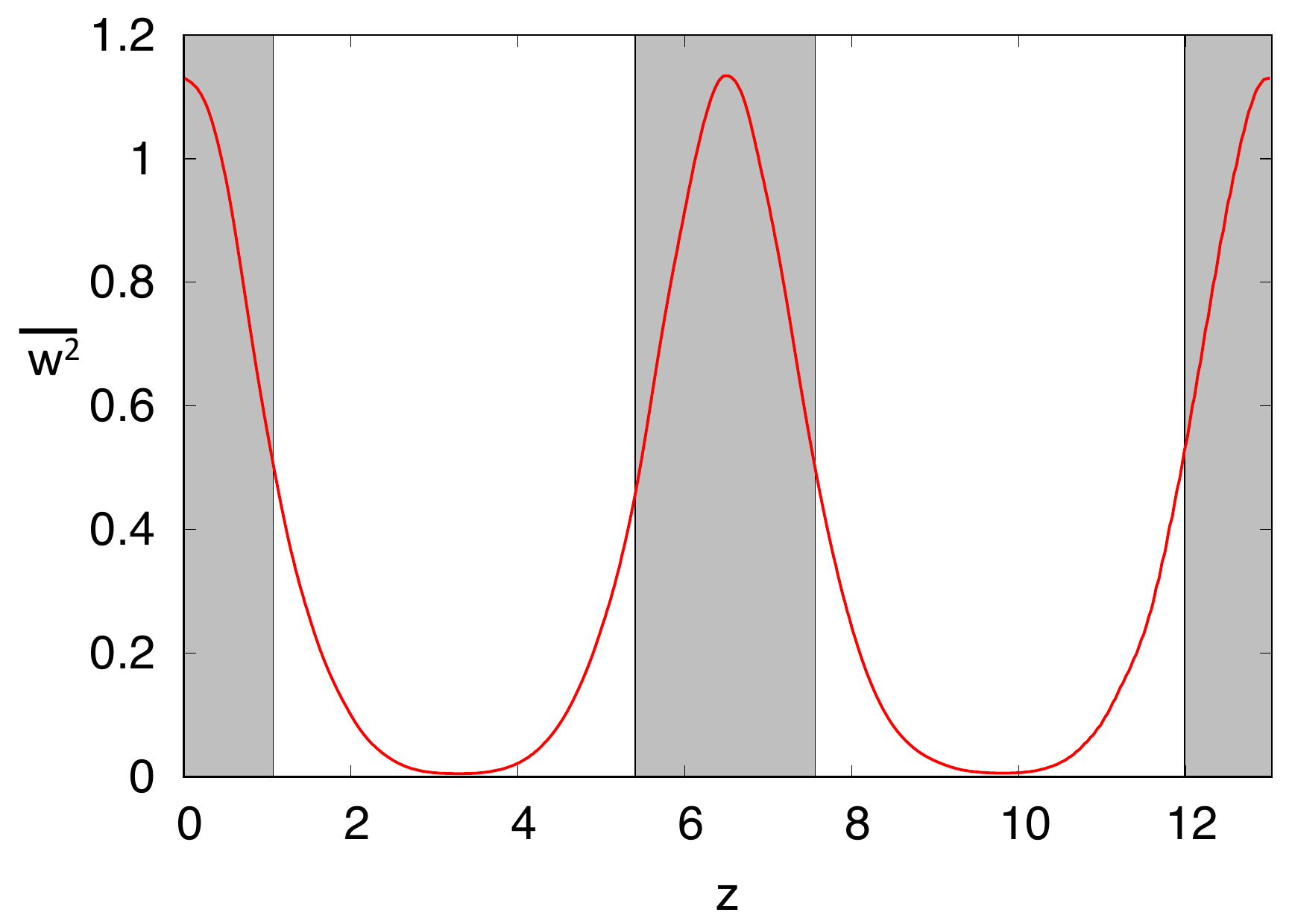}\hfill
	\includegraphics[width=.5\textwidth]{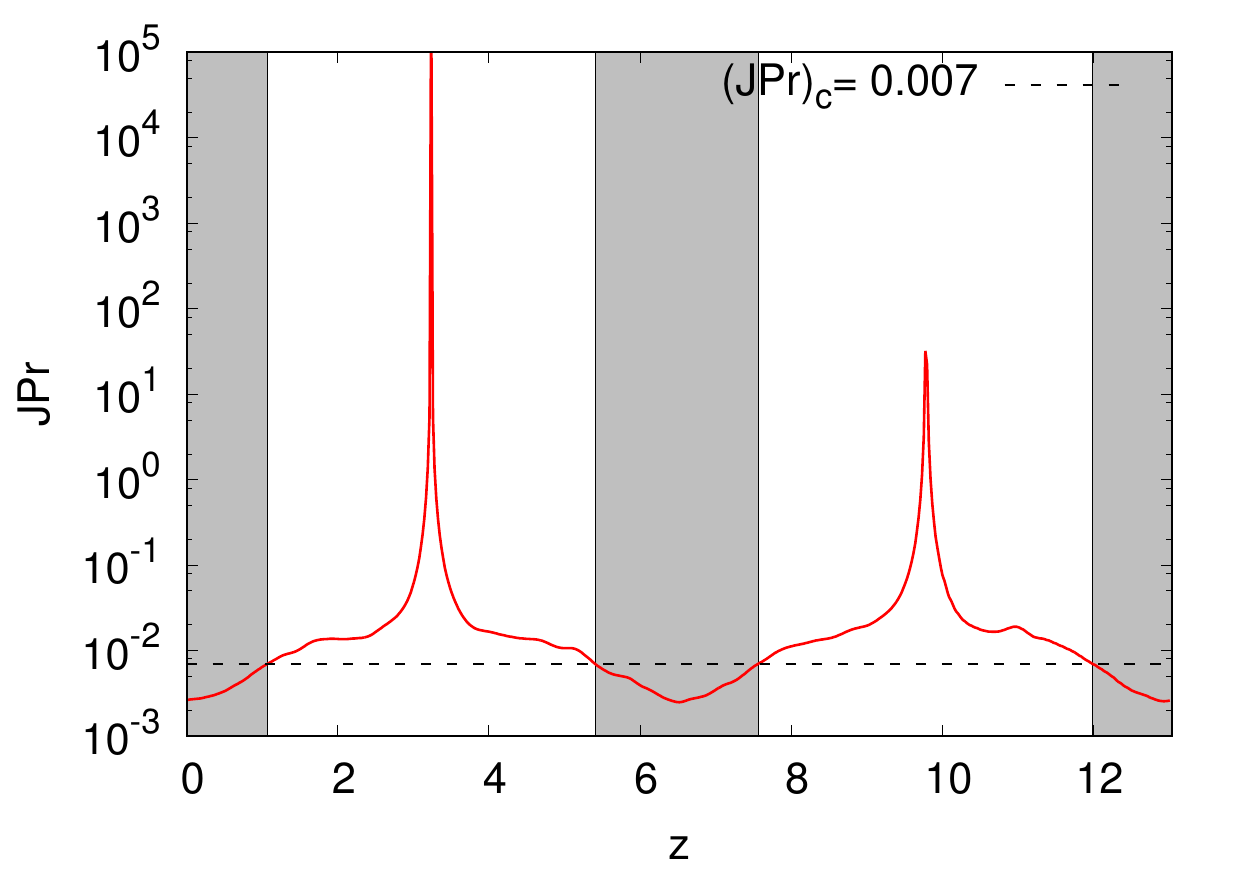}

  \caption{Left: time and horizontally averaged vertical velocity $\overline{w^2}$  as a function of $z$, for $\Re_F=100$, $\Ri_F\Pe_F=10$ and $a=2$ (Simulation A), once the statistically steady state is reached. Right: Corresponding profile of $J\Pr$ as a function of $z$. The black dotted line shows $(J\Pr)_c=0.007$ below which the shear should be unstable according to  Zahn's nonlinear criterion for instability. The shaded areas in both plots highlight the regions that are unstable under this criterion.}
\label{fig:wrms2}
\end{figure}

 As we have seen before in the case of very weak stratification, regions that are nominally stable to Zahn's local criterion for instability can still be mixed, or partially mixed, by the non-local influence of turbulence generated in nearby unstable shear layers. This can be seen in Figure~\ref{fig:wrms2} which shows $\overline{w^2}$ as a function of $z$ on the left, and $J\Pr$ as a function of $z$ on the right, for Simulation A. In both cases we are showing the horizontal and time average of these quantities once they have reached a statistically stationary state.
Mixing driven from strong shearing regions where $J\Pr < 0.007$ clearly extends into the theoretically stable regions, implying again that non-local mixing is important and needs to be taken into account. This time, however, it cannot be attributed to the spatially extended ``tail'' of linearly unstable mode as in Section 4.2.1 since the flow profile is now linearly stable. The extension of mixing into the theoretically stable regions will be investigated in Section~\ref{sec:extension} and in Part III of this series.

Now that we have seen the behavior of mixing in the linearly unstable case for weak and intermediate thermal stratification regimes, we complete our tour of parameter space by looking at a regime that is linearly stable but nonlinearly unstable.

\subsubsection{Linearly stable regime}

We now focus on two LPN simulations with $\Re_F= 100$, $\Ri_F\Pe_F= 50$ and $a=2$. According to the linear stability analysis of Section~\ref{sec:stab}, the laminar flow driven by the force $F(z)$ is linearly stable in this parameter regime. We used the endpoint of another simulation at lower $\Ri_F\Pe_F$ as initial condition in order to obtain a turbulent solution. We compare two cases, one which is restarted from a run at $\Ri_F\Pe_F= 10$ (namely, the Simulation A discussed in the previous section), and one which is restarted from a run at $\Ri_F\Pe_F= 30$ (see Table~\ref{table:0}). These will be referred to as Simulation I and Simulation II hereafter. In both cases, the preceding simulation was evolved until a statistically stationary state was reached before increasing $\Ri_F\Pe_F$ to 50. A snapshot of the corresponding vertical velocity field at the restart time can be seen in Figure~\ref{fig:sv0} (leftmost panels), and shows in both cases two well-defined turbulent shear layers (one in the middle of the domain, and one at the top/bottom edge of the domain).

\begin{figure}[h]
  \centerline{\includegraphics[width=0.7\textwidth]{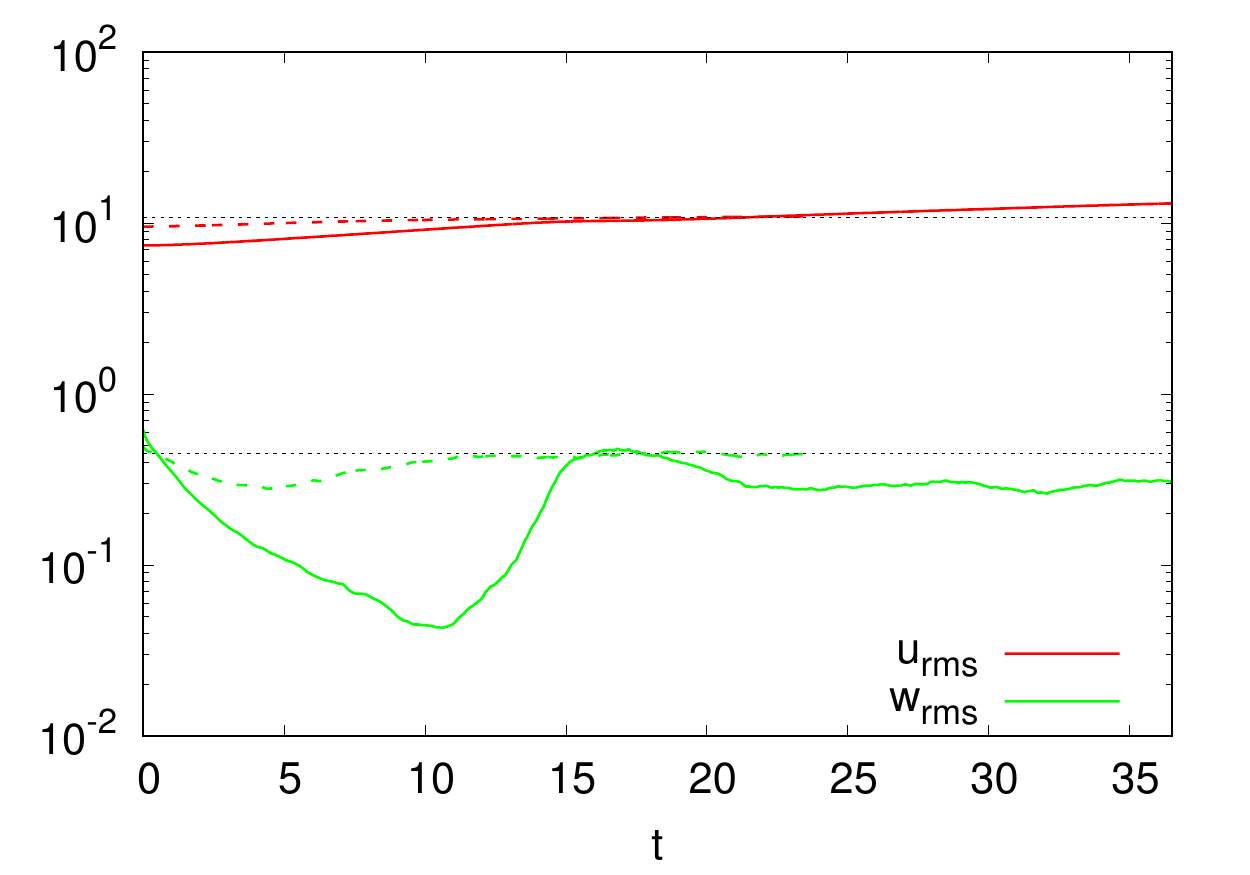}}
  \caption{ $u_{\rm rms}$ and $w_{\rm rms}$ as a function of time for $\Re_F = 100$, $\Ri_F\Pe_F = 50$ and $a = 2$, for Simulation I (solid line) and Simulation II (dashed line). Simulation II rapidly adjusts to the change in background stratification, and reaches a statistically stationary state around t $\sim$ 12. By contrast, Simulation I takes much longer to recover from the change in the background stratification. $w_{\rm rms}$ reaches a statistically stationary state around $t \sim 20$, and the value achieved is about 1/2 of that of Simulation II. The $u_{\rm rms}$ curve on the other hand continues to evolve on the slower viscous timescale for much longer, and has not reached a stationary state by the end of the simulation.}
  \label{fig:urmsRiPe50}
\end{figure}

\begin{figure}[h]
    \includegraphics[width=.2\textwidth]{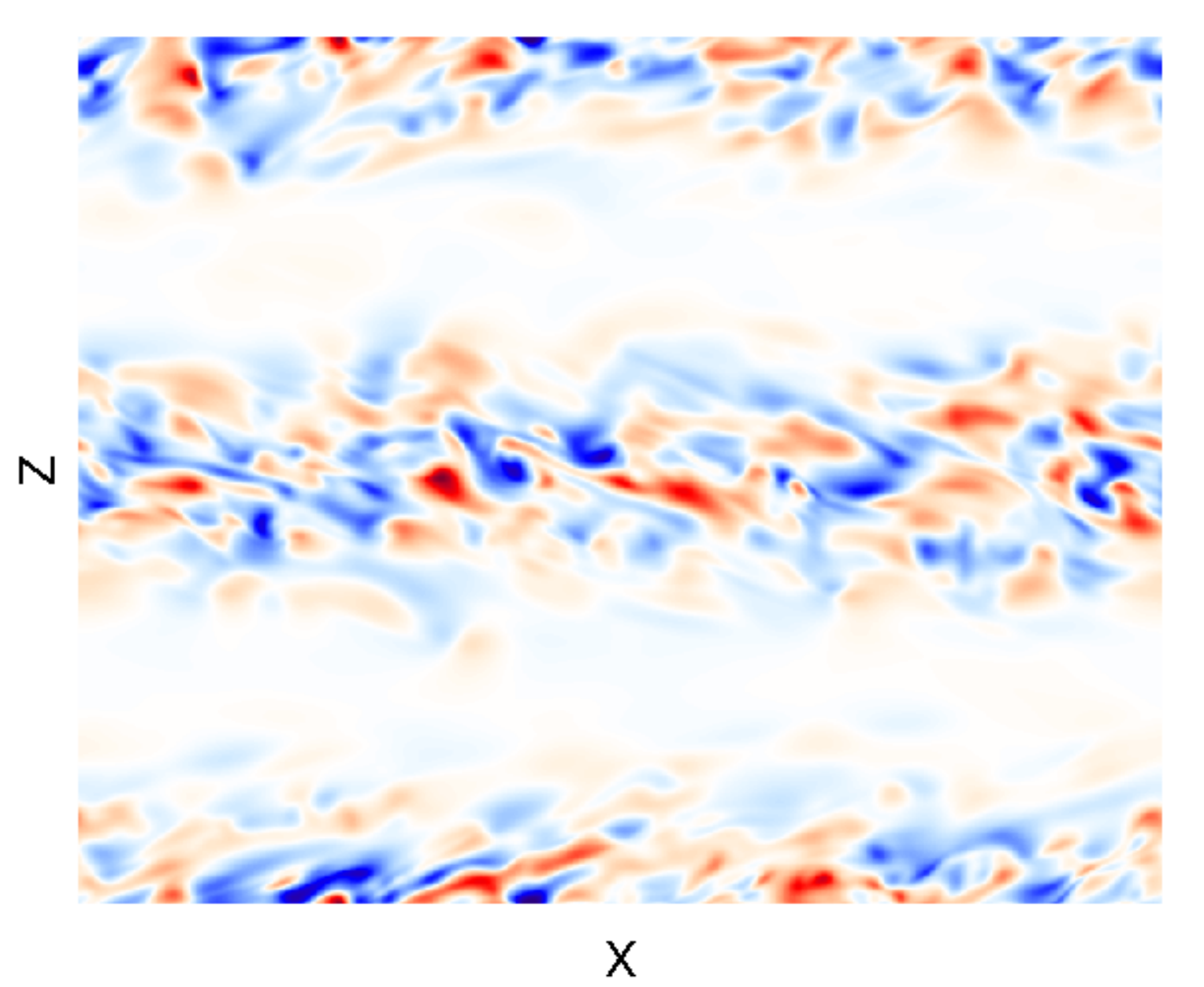}\hfill
    \includegraphics[width=.2\textwidth]{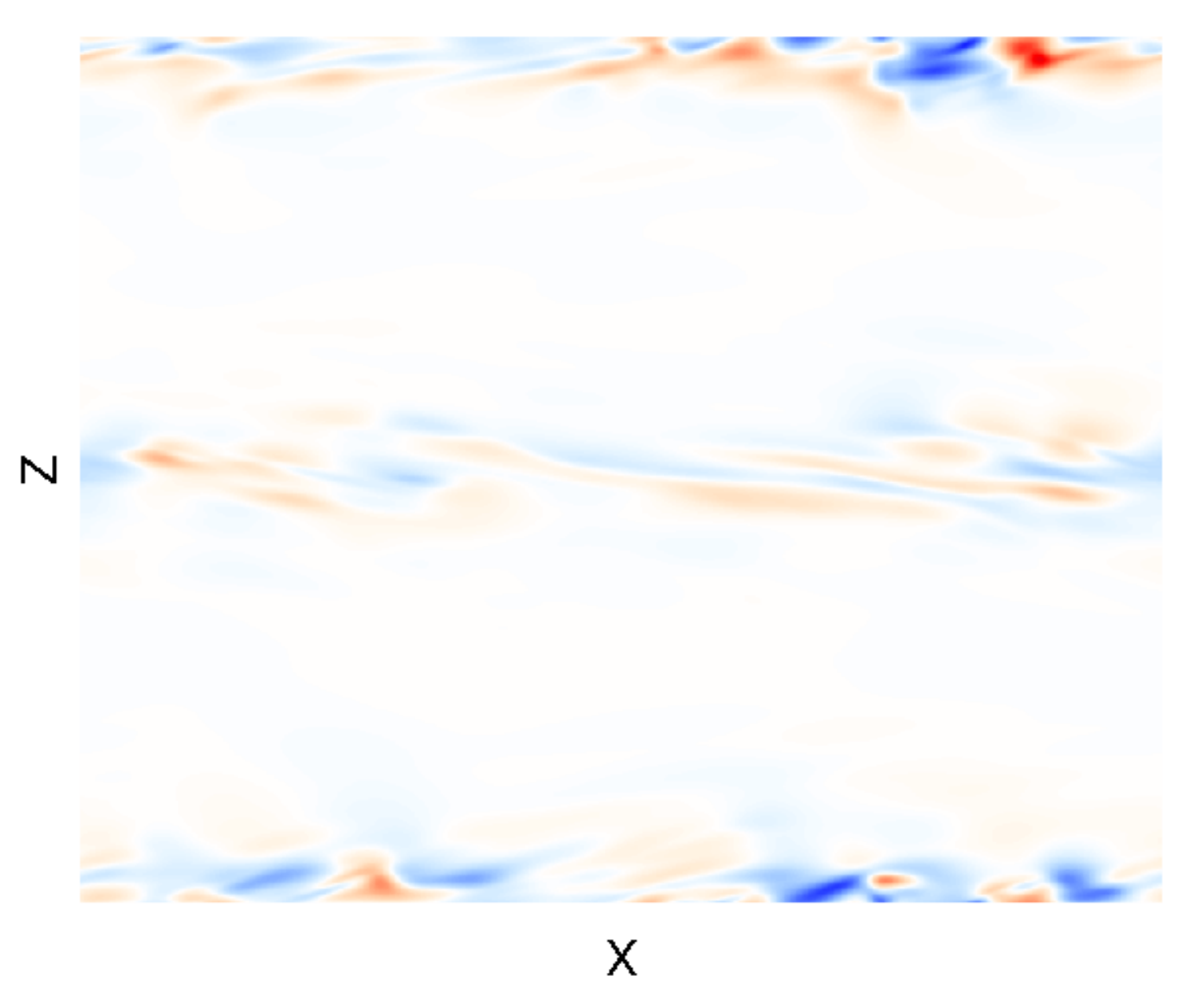}\hfill
    \includegraphics[width=.2\textwidth]{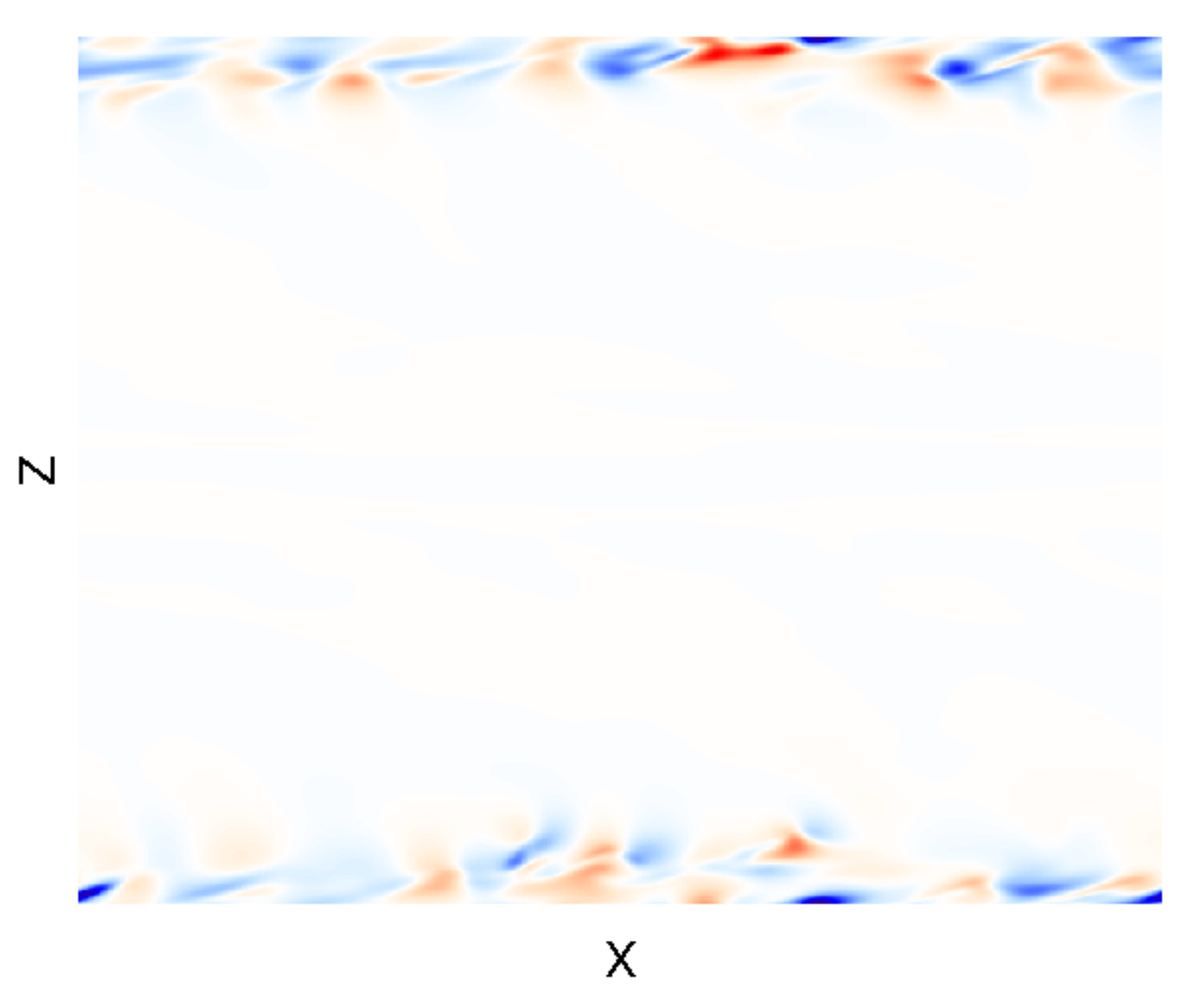}\hfill
    \includegraphics[width=.2\textwidth]{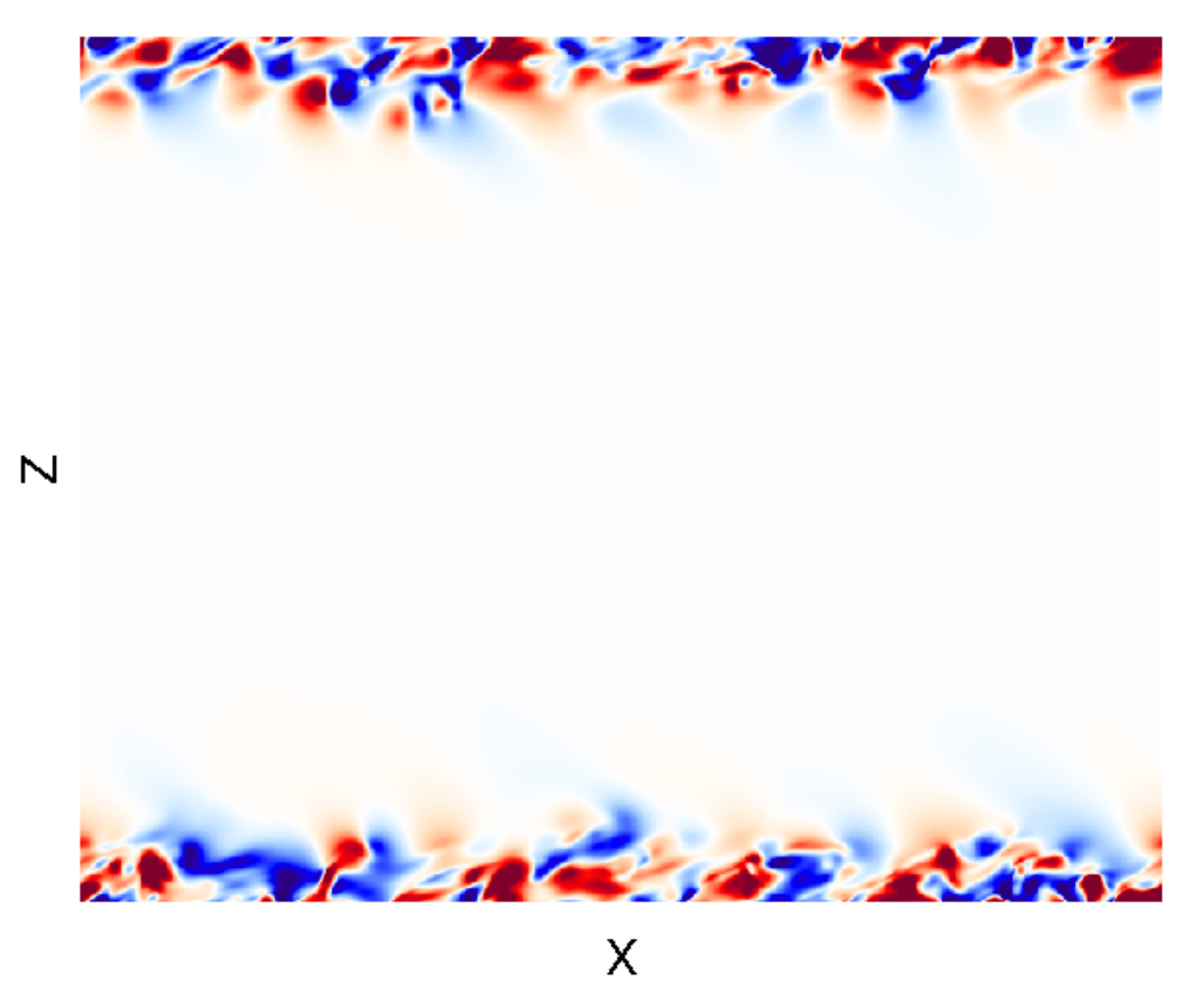}\hfill
    \includegraphics[width=.2\textwidth]{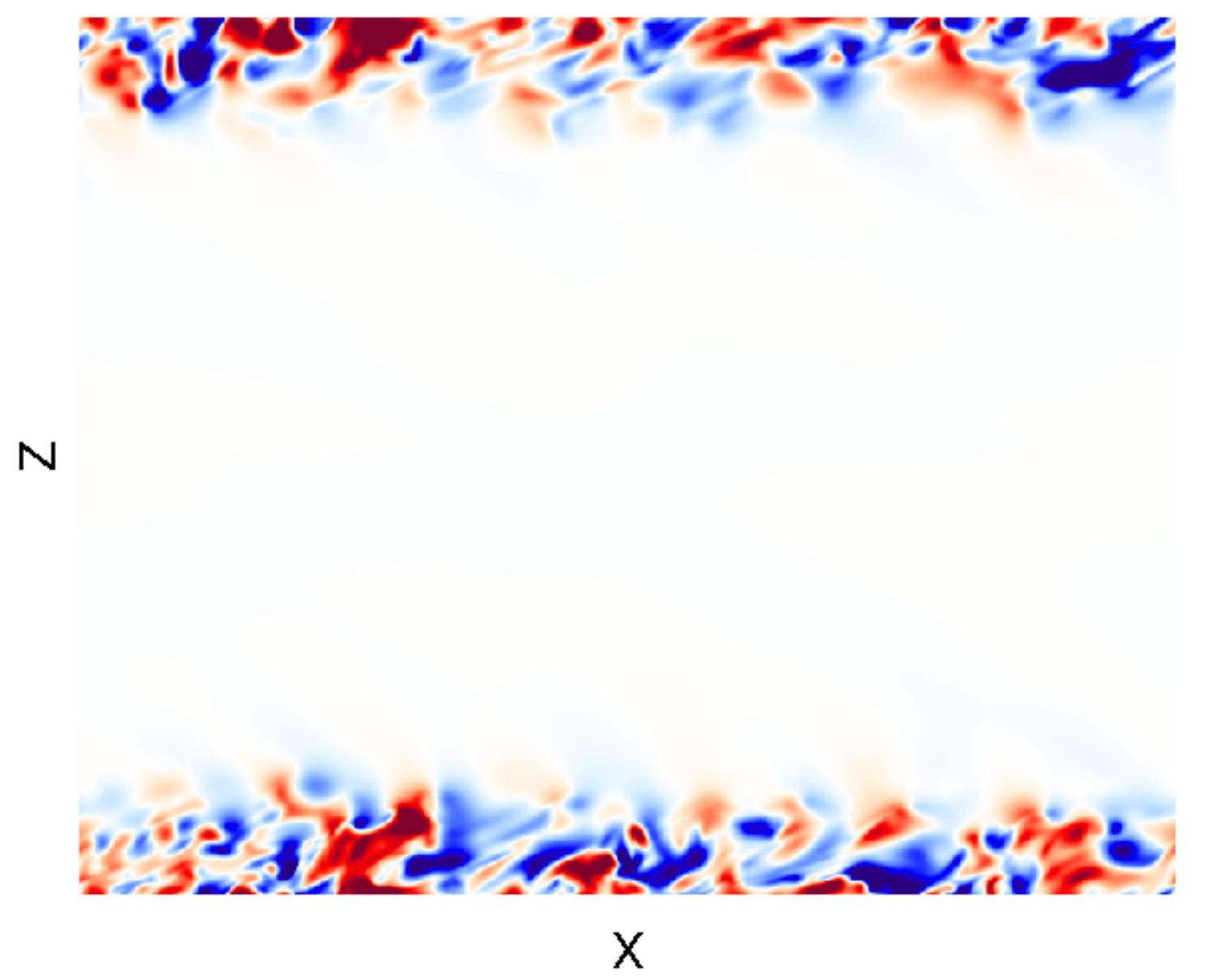}\\

    \includegraphics[width=.2\textwidth]{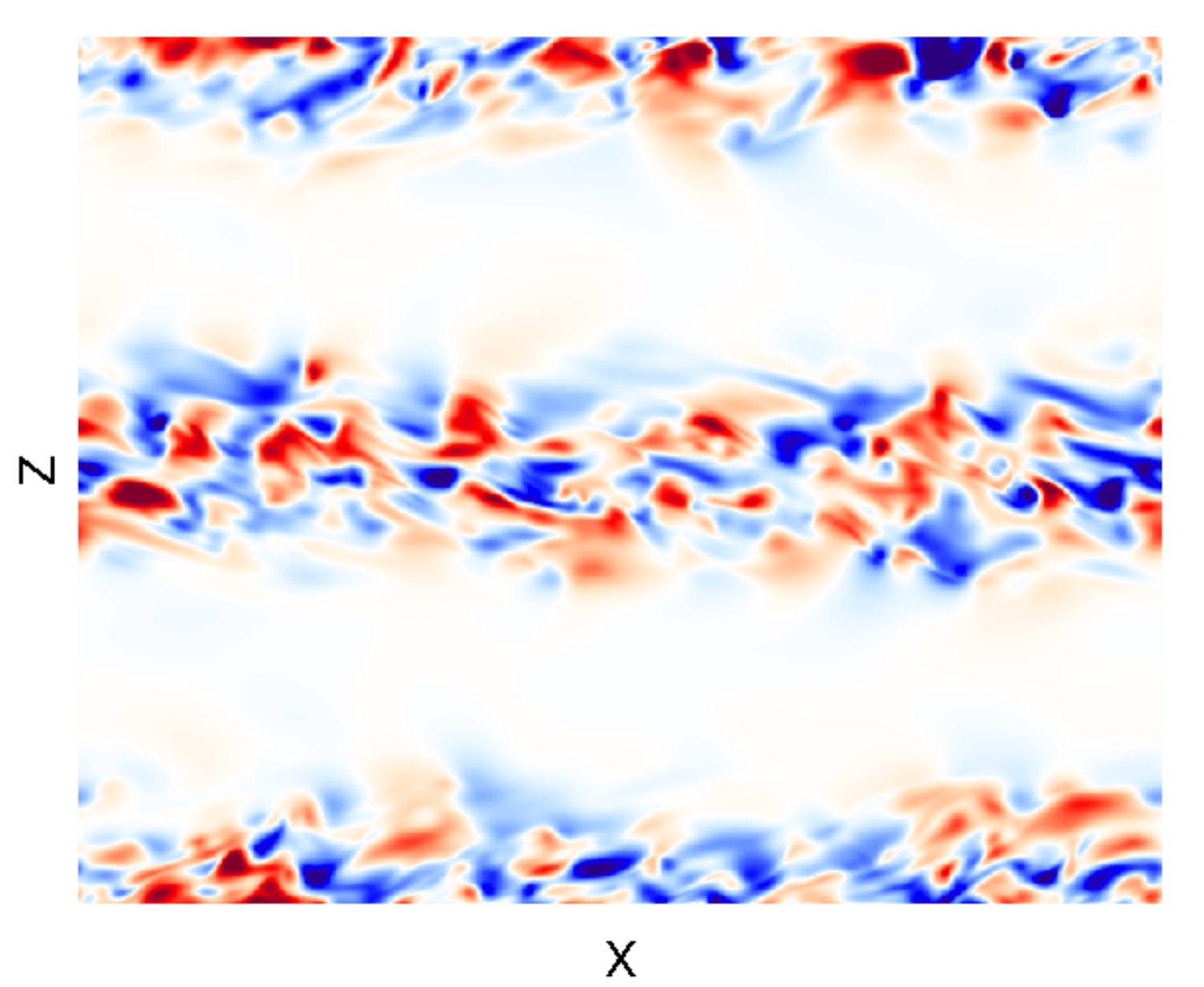}\hfill
    \includegraphics[width=.2\textwidth]{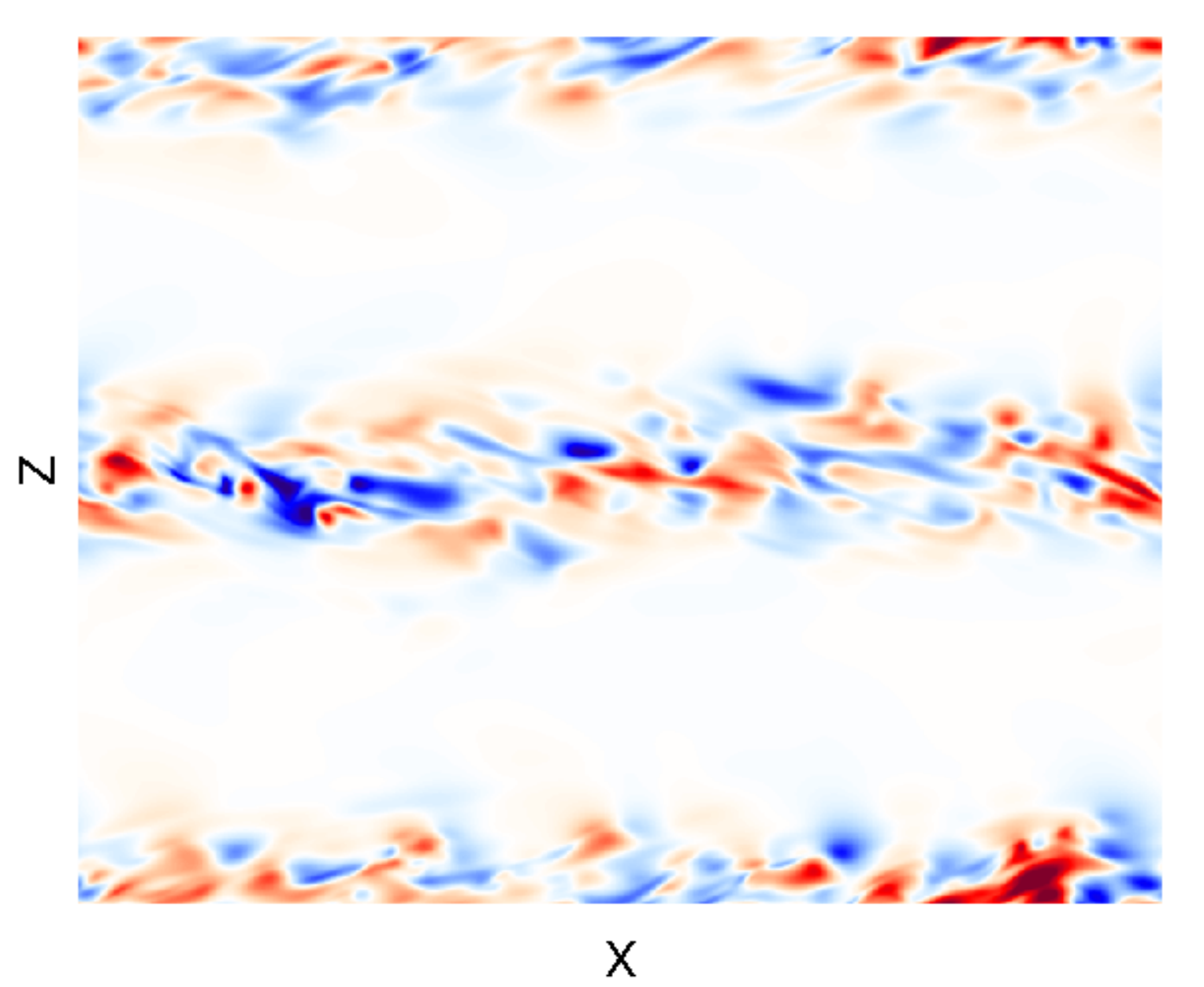}\hfill
    \includegraphics[width=.2\textwidth]{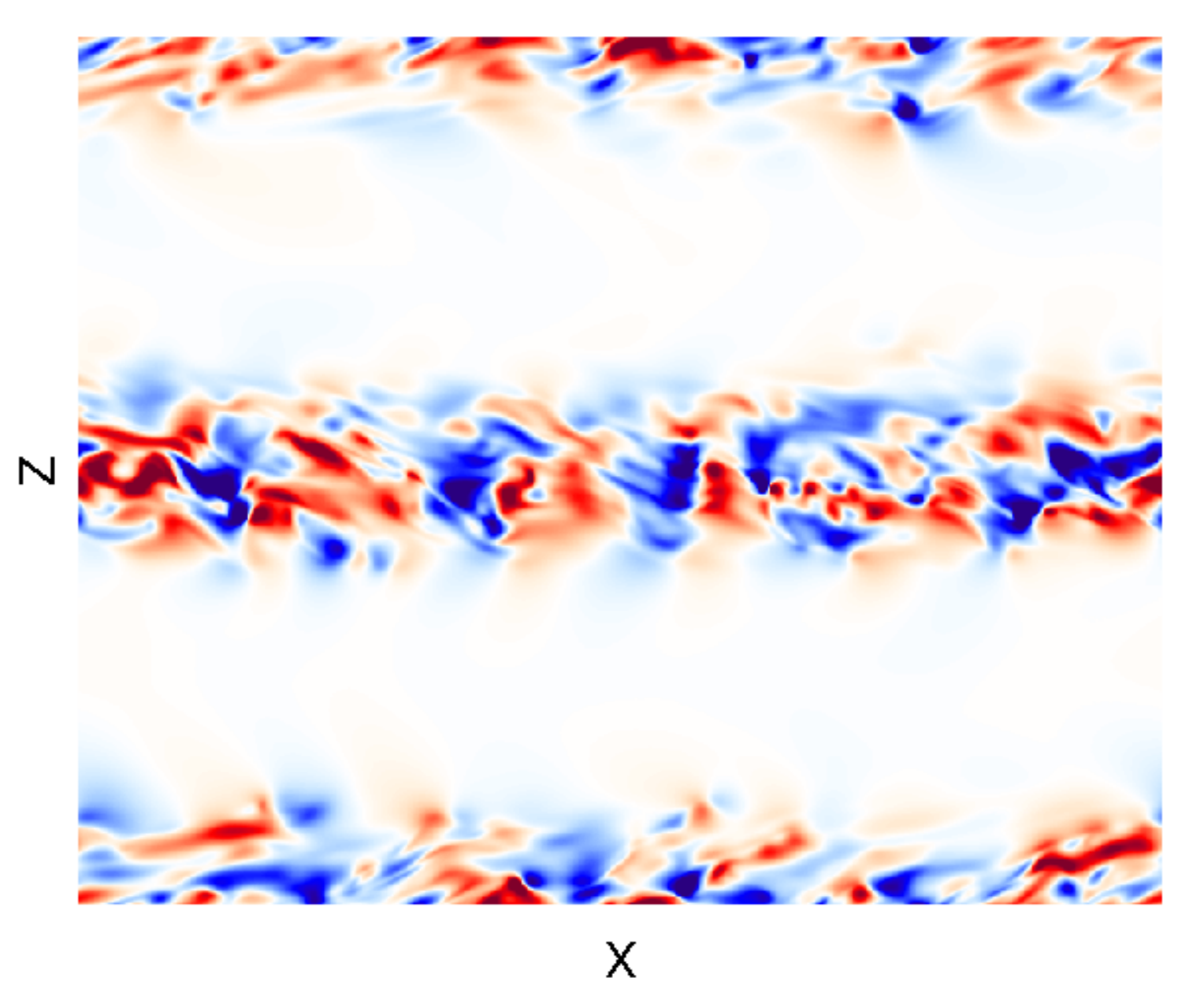}\hfill
    \includegraphics[width=.2\textwidth]{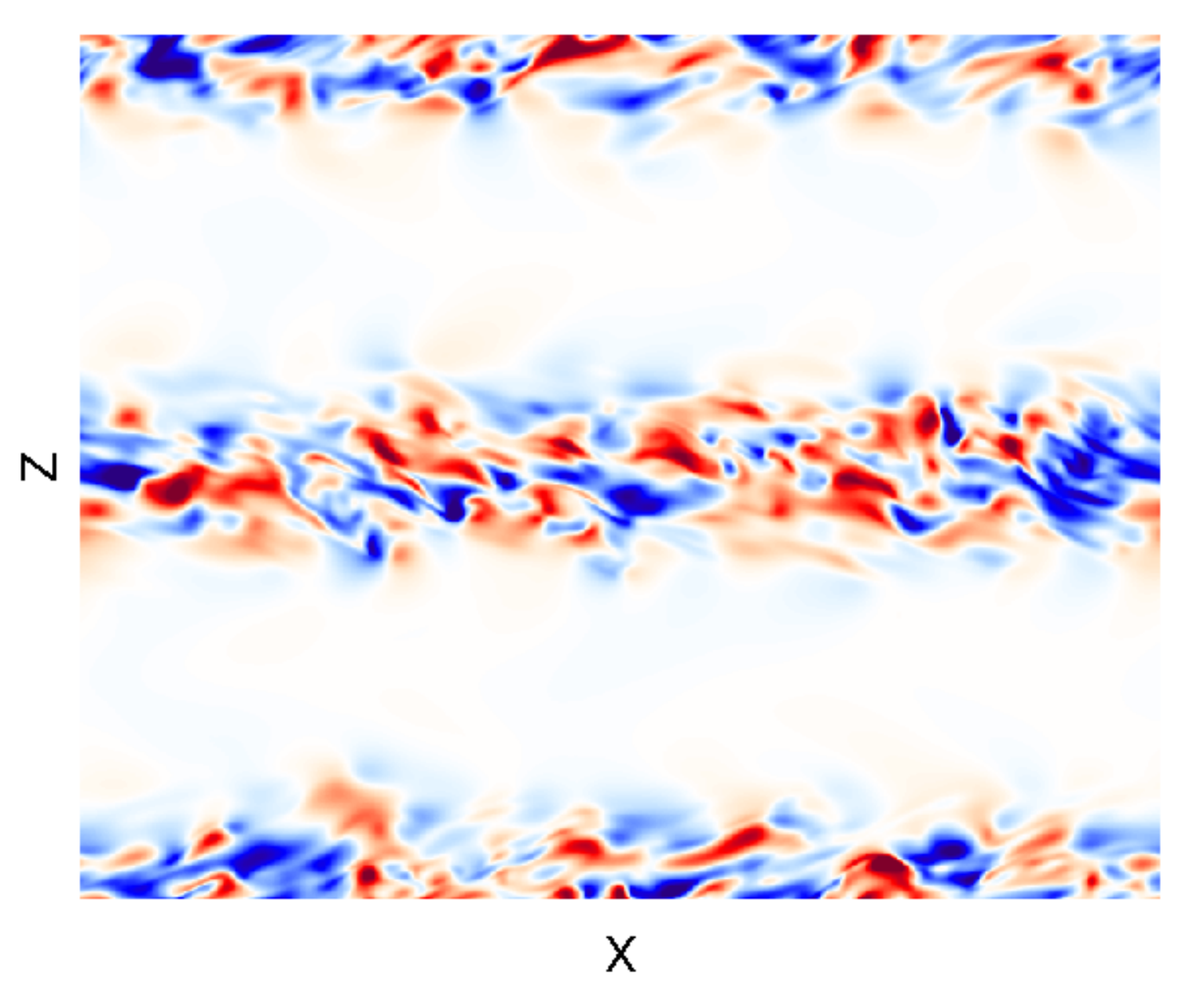}\hfill
    \includegraphics[width=.2\textwidth]{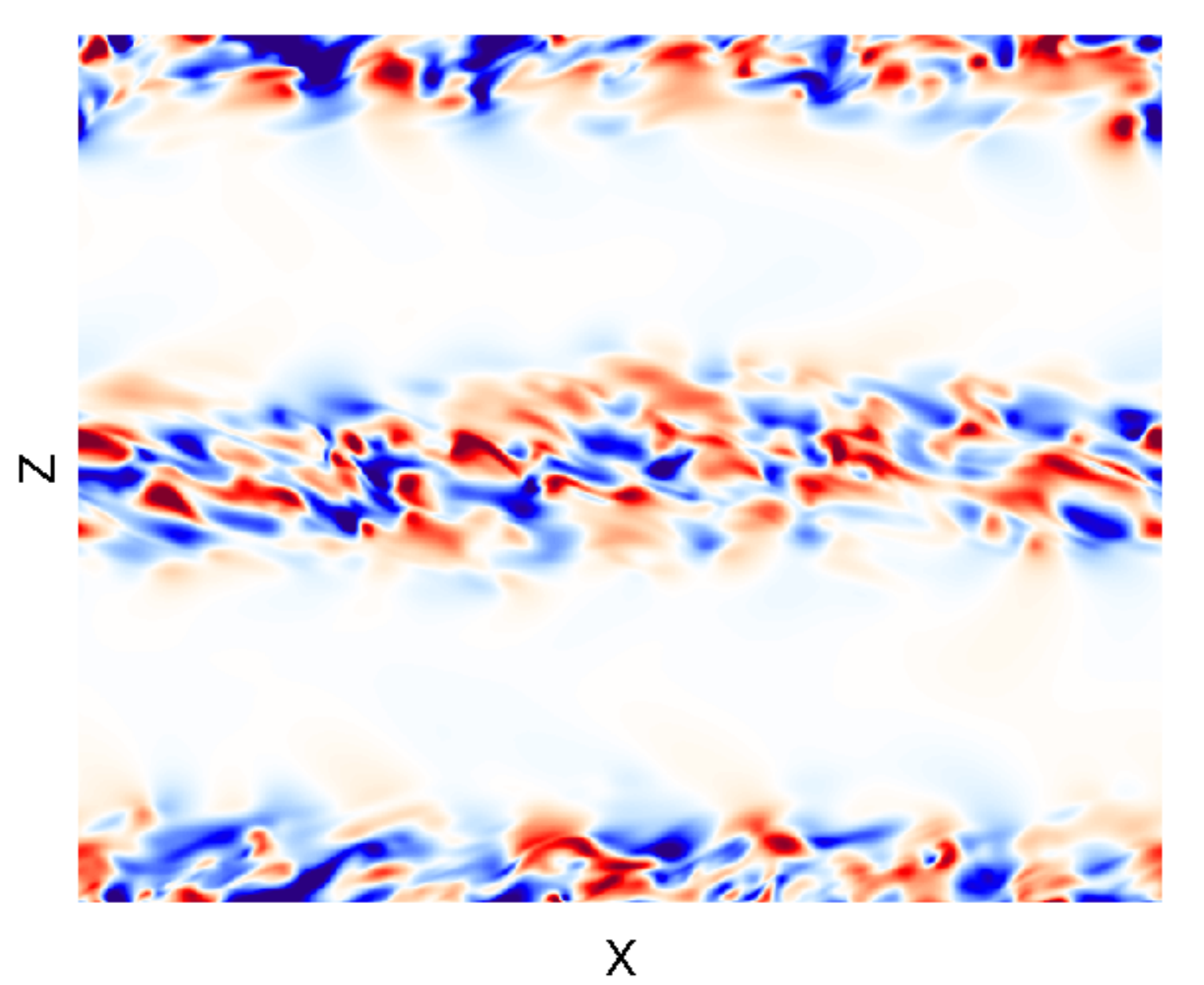}

  \caption{Snapshots of the vertical velocity field for $\Re_F=100$, $\Ri_F\Pe_F=50$ and $a=2$ taken from Simulations I (top) and II (bottom) at five different times: $t=0, \ 4, \ 11, \ 14, \ \rm{and} \ 17$ from left to right. In Simulation I, turbulence decays everywhere  up to $t \sim 11$, then rises again on the top/bottom shearing regions only. In Simulation II, turbulence is maintained in all three strong shearing regions. In all snapshots, the vertical velocity ranges from -2 (bright blue) to 2 (bright red).}
\label{fig:sv0}
\end{figure}

Figure~\ref{fig:urmsRiPe50} shows the time dependence of $u_{\rm rms}$ and $w_{\rm rms}$ for the two simulations. In Simulation II (dashed line), which was restarted from $\Ri_F\Pe_F = 30$, we see that the amplitude of the turbulence as measured by $w_{\rm rms}$ decreases by a factor of about 2 from $t = 0$ up to $t \sim 5$, but later recovers. The system eventually reaches a statistically stationary state around $t \sim 12$. In Simulation I (solid line), which was restarted from a more weakly stratified run with $\Ri_F\Pe_F = 10$, the initial drop in $w_{\rm rms}$ is much more pronounced and lasts significantly longer. The turbulence only begins to recover around $t \sim 10$, and $w_{\rm rms}$ reaches a statistically stationary state around $t \sim 20$. Interestingly, $w_{\rm rms}$ in this state is only about 1/2 of that of Simulation II. Even more curiously, $u_{\rm rms}$ does not seem to reach a statistically stationary state, but instead appears to continue growing on a much slower timescale. 

Figure~\ref{fig:sv0}, which shows snapshots of $w(x, y = 0, z)$ in both simulations at selected times between $t = 0$ and $t = 17$, provides clues as to these surprising differences. We see that the ultimate statistically stationary states achieved by Simulations I and II are different. In Simulation I (top row), the initial decay of the turbulence dramatically affects both shear layers, and the middle one never recovers. Instead, it becomes fully laminar, a situation that is allowed since the laminar solution is linearly stable at these parameters. In Simulation II on the other hand turbulence also decays initially, but ultimately survives in {\it both} middle and top/bottom shear layers. This explains why $w_{\rm rms}$ is ultimately twice as large in Simulation II than in Simulation I. Finally, the fact that the middle layer laminarizes in Simulation I also explains why $u_{\rm rms}$ continues to evolve on a slow timescale in that run. Indeed, without any turbulent stresses present, the laminar layer must achieve a viscous balance with the forcing to be in a steady state, and this adjustment takes place on a viscous timescale.

The comparison between Simulations I and II therefore provides an interesting example of the behavior of intrinsically nonlinear instabilities: they are subject to hysteresis so that their overall evolution and statistically stationary state depend sensitively on the initial conditions selected. This type of behavior is commonly discussed in the dynamical systems literature \citep[e.g. see the textbook by][]{strogatz2018}, and has been discussed more specifically in astrophysics in the contexts of thermocompositional convection \citep[][]{Moll2017} and dynamo theory \citep[e.g. see the review by][]{Tobias2011}, among others.

\begin{figure}[h]
	\includegraphics[width=.5\textwidth]{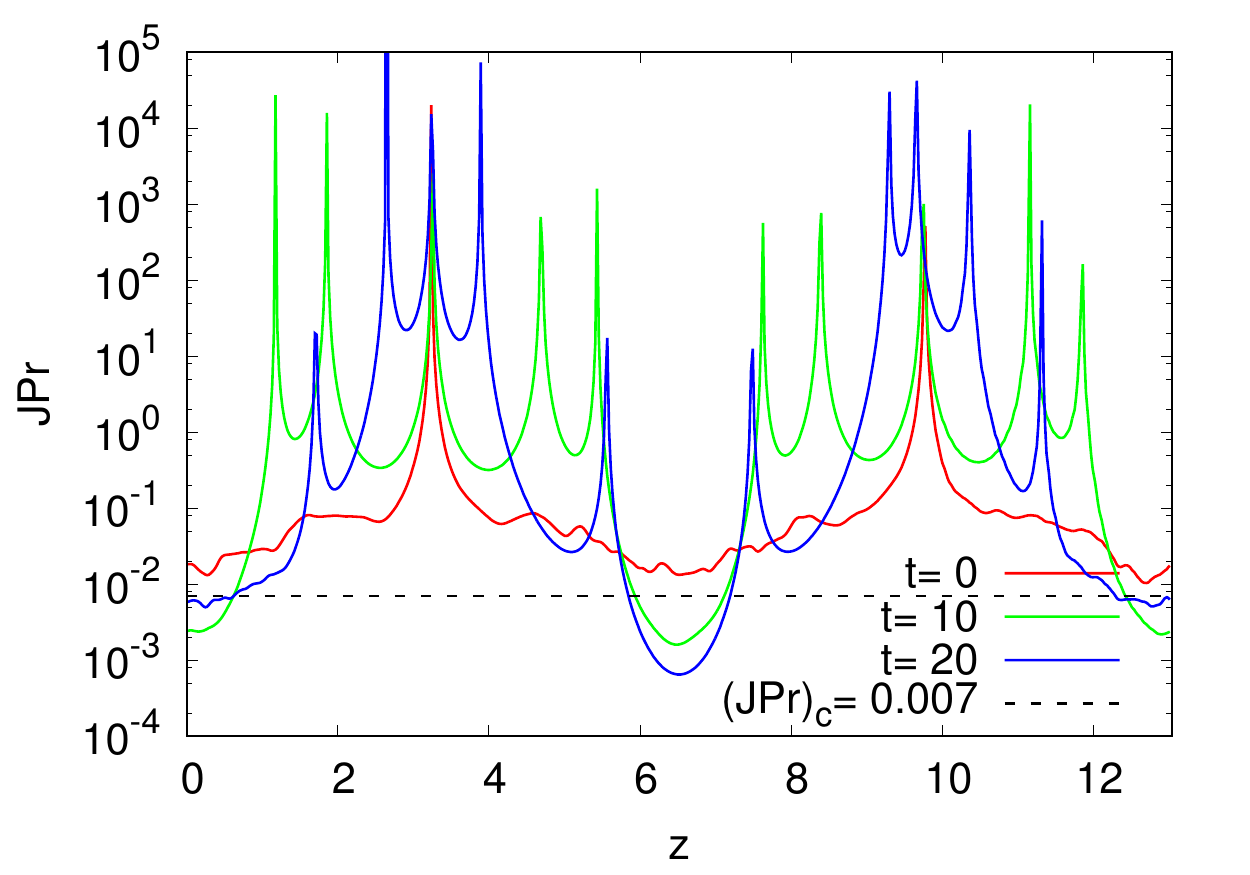}\hfill
	\includegraphics[width=.5\textwidth]{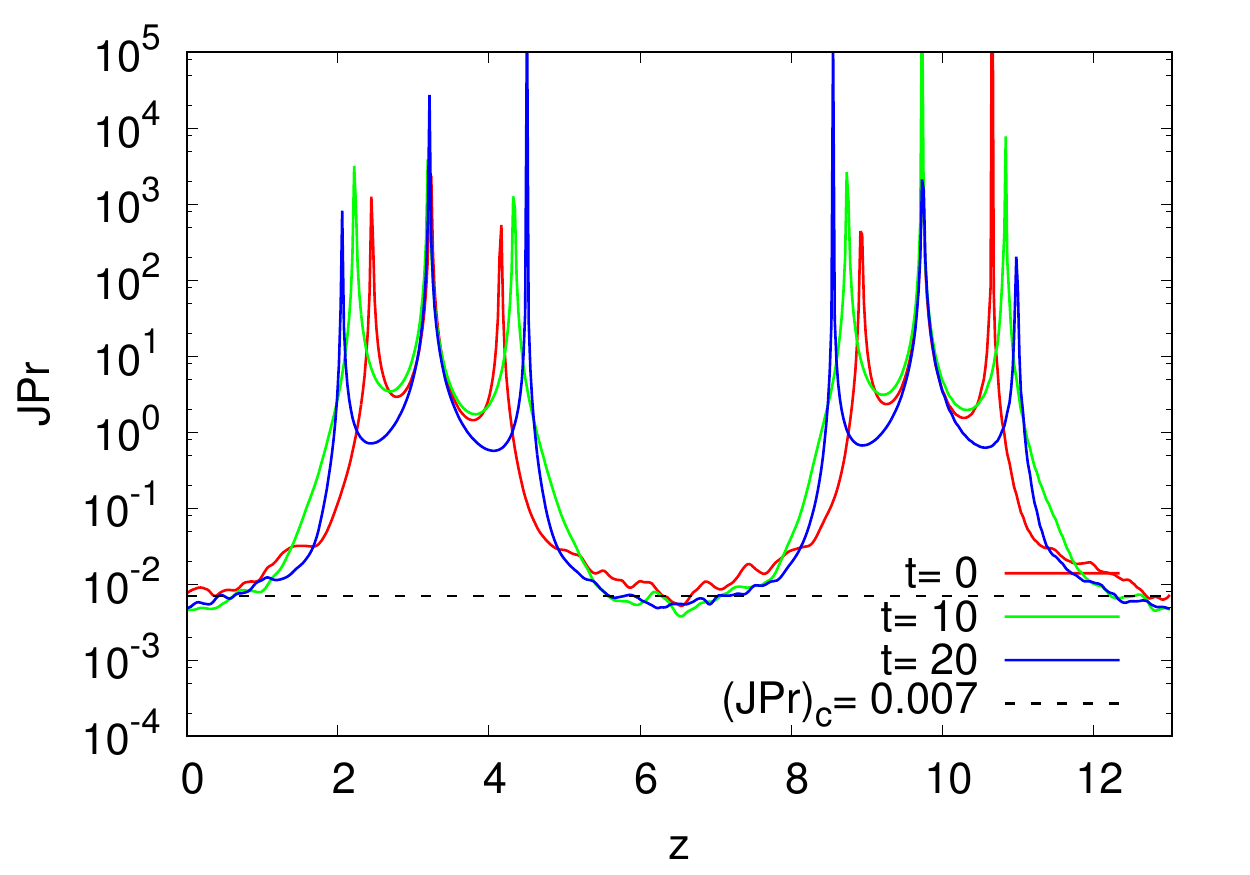}
  \caption{$J\Pr$ as a function of $z$ at three times $t= 0$, $t= 10$ and $t=20$, for Simulations I (left) and II (right). The black dashed line shows $(J\Pr)_c=0.007$ below which the shear should be unstable according to  Zahn's nonlinear criterion for instability. In Simulation I, $J\Pr$ is initially above the critical threshold everywhere, so the flow is nonlinearly stable, whereas in Simulation II $J\Pr$ remains slightly below $(J\Pr)_c$ in the center of each shear layer.}
  \label{fig:JPr_difft}
\end{figure}

To understand why the turbulence decays in both simulations in the first place, we show in Figure~\ref{fig:JPr_difft} the profiles of $J {\rm Pr}$ at $t = 0$, $10$ and $t = 20$. In Simulation I, $J {\rm Pr}$ is initially above the critical threshold $(J{\rm Pr})_c \simeq 0.007$ everywhere, so the flow is nonlinearly stable according to Zahn's criterion. This explains why the turbulent kinetic energy drops so suddenly. The forcing $F(z)$ gradually reinforces the shear in both layers, so $J{\rm Pr}$ later drops below 0.007 again in selected locations. However, by the time the shear layers satisfy Zahn's criterion once more, the remaining turbulent kinetic energy in the middle layer is too small to kick-start the nonlinear instability, and the turbulence ultimately dies out. In the top/bottom layer on the other hand the amplitude of the remaining perturbations was presumably just large enough to restart the turbulence\footnote{Which, if any, of the shear layers ultimately remains turbulent after  jumping from $\Ri_F\Pe_F= 10$ to $\Ri_F\Pe_F= 50$ is likely a stochastic process, i.e. had we restarted the simulation from a slightly different point in time, we might have found that the middle layer remains turbulent but the side one does not, or that all of them do, or that none of them do.}. In Simulation II the initial profile of $J {\rm Pr}$ is just below 0.007 in the very center of each shear layer, which explains why the turbulence does not decay as much as in Simulation I, and is able to survive in all the existing shear layers.

The comparison between Simulations I and II therefore illustrates the fact that a nonlinearly unstable flow (i.e. a flow which satisfies $J{\rm Pr} < (J{\rm Pr})_c$) is not guaranteed to be turbulent: whether it is or not will depend on its history (and presumably also on the background level of turbulence in a more realistic system). As a result, one should be careful in using Zahn's criterion for diffusive shear instabilities in the strongly stratified limit since it only provides a necessary condition for instability, but not a sufficient one. 

\begin{figure}[h]
  \centerline{\includegraphics[width=0.7\textwidth]{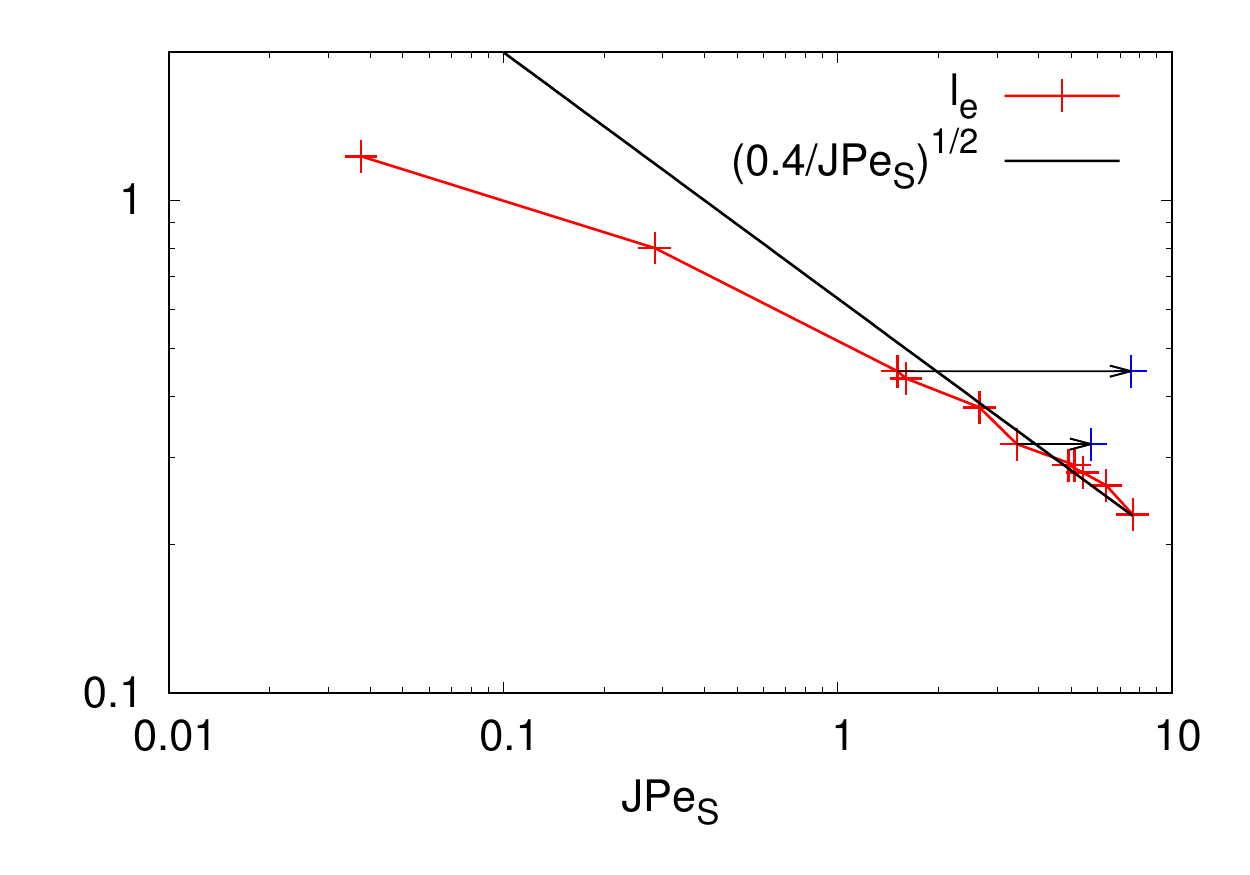}}
  \caption{Turbulent eddy lengthscale $l_e$ as a function of $J\Pe_S$ for all the DNSs presented in Table~\ref{table:0}.  The black line shows $l_{\rm Z}$ (see (\ref{eq:lZ})) with the constant of proportionality $(J{\rm Pe})_c \simeq 0.4$ selected to fit the $l_e$ data for large stratification. The two blue crosses correspond to the values of $l_e$ at the first timestep of Simulations I and II which were restarted from the endpoint of $\Ri_F\Pe_F=10$ and $\Ri_F\Pe_F=30$ simulations, respectively. The arrows clearly show the impact of abruptly changing $\Ri_F\Pe_F$. The two initial eddy lengthscales are much  larger than their corresponding $l_{\rm Z}$ which explains the initial decay of the turbulence in these simulations.}
  \label{fig:leJPes}
\end{figure}

Another way of understanding the difference between the behavior of Simulations I and II is to consider the scale of the turbulent eddies in the shear layers. \citet{Zahn1974} argued that turbulent eddies of size $l$ can only extract energy from the shear provided $J {\rm Pe}_l < O(1)$, where ${\rm Pe}_l = S l^2 / \kappa_T$ is the eddy-scale P\'eclet number (see Paper I for details). Equivalently, this implies that only eddies of size equal to or smaller than the so-called Zahn scale $l_{\rm Z}$ can be energetically self-sustained, where 
\begin{equation}
l_{\rm Z} = \sqrt{ \frac{(J{\rm Pe})_c \kappa_T}{JS}}, 
\end{equation} 
and $(J{\rm Pe})_c$ is a constant of order one and all the other quantities in this expression only are  dimensional. When expressed in the units based on the forcing,  
\begin{equation}\label{eq:lZ}
l_{\rm Z} = \sqrt{ \frac{(J{\rm Pe})_c S}{\Ri_F\Pe_F}} L_0 \ .
\end{equation}
We have therefore measured the typical eddy scale $l_e$ in each of our DNSs, to compare it to the Zahn scale. We did so by using the protocol discussed in Paper I: we define the vertical autocorrelation function of the spanwise flow at a given time $t$ as  
\begin{equation}
a_v(l,t)= \frac{1}{L_x L_y}\frac{1}{L_z-l} \int\limits_{0}^{L_x}\int\limits_{0}^{L_y}\int\limits_{0}^{L_z-l} v(x,y,z,t)v(x,y,z+l,t)dxdydz \ ,
\end{equation}
and let $l_e$ be the first zero of $a_v$, which we then average over time. The results are reported in Table~\ref{table:0}. Figure \ref{fig:leJPes} shows $l_e$ against $J{\rm Pe}_S \equiv \frac{\Ri_F\Pe_F}{S}$, measured in the statistically stationary state of each of our available simulations, as red symbols. As found in Paper I, we indeed see that the turbulent eddy scale adjusts itself to be equal to the Zahn scale,  $l_e \simeq l_{\rm Z} = \sqrt{\frac{(J{\rm Pe})_c}{J{\rm Pe}_S}}L_0$ for sufficiently large stratification (i.e. large $J{\rm Pe}_S$), as long as we fit the unknown order-unity constant $(J{\rm Pe})_c$ to be $\simeq 0.4$. As a result, the dominant turbulent eddies are just marginally unstable. 

When $\Ri_F\Pe_F$ is abruptly increased to 50 from the endpoint of a $\Ri_F\Pe_F = 10$ run (for Simulation I) or from a $\Ri_F\Pe_F = 30$ run (for Simulation II), however, $J{\rm Pe}_S$ increases accordingly but $l_e$ does not immediately adjust. As a result, the eddy scale $l_e$ at the start of Simulations I and II is now above the Zahn scale (see the blue symbols in Figure~\ref{fig:leJPes}), and the turbulence must  decay. Since $l_e$ is not far above the Zahn scale in Simulation II, not much energy is lost by the time the system adjusts to its new parameter regime. In Simulation I on the other hand $l_e$ is nearly two times larger than the Zahn scale at these new parameters, so the turbulence loses much more energy. In both runs, the typical eddy scale is ultimately forced to decrease down to the Zahn scale at $\Ri_F\Pe_F  = 50$. When this happens, the turbulence can finally be self-sustained, and its amplitude begins to increase again. 

\begin{table}[h]
\caption{Summary of the main results for all the runs. The first columns reports  $\Ri_F\Pe_F$, the second column reports the resolution used for the simulation, the third column is the measured eddy lengthscale $l_e$. The fourth column is the overshooting scale $\delta$, and its corresponding uncertainty column five. The last column is the measured shearing rate in the middle of the domain (or at $z=0$ for runs where the central region becomes stable).  Unless otherwise specified, each simulation has been restarted from the endpoint of the one on the line directly above.} 
\label{table:0}      
\centering    
\begin{threeparttable}                               
\begin{tabular}{c c c c c c}   
\hline 
\\        
${\Ri_F\Pe_F}$ &  $N_x, \ N_y, \ N_z$ & $l_e$ & $\delta$ & $\epsilon_\delta$ & $|\breve{S}|$ \\
\hline      \\                        
    0.1\tnote{a} & $384 \times 192 \times 384$& 1.23& 1.513 & $2.48 \cdot 10^{-1}$ &2.67   \\    
    \\    
    1\tnote{a} &  $384 \times 192 \times 384$&0.80 & 0.760 & $3.66 \cdot 10^{-2}$ & 3.52 \\    
    \\       
                           
    10\tnote{a} &  $720 \times 288 \times 576$&0.45 & 0.534 & $1.93 \cdot 10^{-2}$ & 6.62 \\    
    \\   
    10\tnote{b} & $720 \times 288 \times 576$&0.44 &0.483 & $1.89 \cdot 10^{-2}$ & 6.24\\
    \\
    20 & $720 \times 288 \times 576$&0.38 & 0.416 & $2.89 \cdot 10^{-2}$ & 7.53   \\    
    \\
    30 & $720 \times 288 \times 576$&0.32 & 0.398 & $4.72 \cdot 10^{-2}$ & 8.72 \\    
    \\     
   
    50\tnote{c} & $720 \times 288 \times 576$&0.29 &0.369 & $1.25 \cdot 10^{-2}$ & 9.76  \\    
    \\ 
    50\tnote{d} & $720 \times 288 \times 576$&0.29 &0.292 & $2.23 \cdot 10^{-3}$ & 10.16  \\    
    \\  
    60\tnote{e} & $720 \times 288 \times 576$&0.28 &0.390 & $4.50 \cdot 10^{-2}$  & 11.07 \\    
    \\  
    70 & $720 \times 288 \times 576$&0.264 &0.359 & $2.58 \cdot 10^{-2}$ & 11.00 \\    
    \\   
    100 & $720 \times 288 \times 576$&0.23 &0.344 & $3.49 \cdot 10^{-3}$ & 13.08\\    
                      
\hline                                             
\end{tabular}
\begin{tablenotes}
\item [a] (Simulation A) Initialized with the laminar profile from equation~(\ref{eq:laminar}).
\item [b] (Simulation B) Initialized with a weak amplitude sinusoidal mean flow.
\item [c] Simulation II
\item [d] (Simulation I) Restarted from the endpoint of the $\Ri_F\Pe_F= 10^{a}$ simulation.
\item [e] Restarted from the endpoint of the Simulation II.

\end{tablenotes}
\end{threeparttable}
\end{table}

To summarize the results of this section, we have shown that diffusive shear instabilities exhibit more complex dynamics than what had previously been discussed. In particular (1) the outcome of a linear instability analysis of the mean flow (of either the initial profile or the final profile) is not always of practical use in understanding the nonlinear evolution of the shear layer, (2) nonlinearly unstable diffusive shear flows can exhibit multiple statistically stationary states, and can either be turbulent or laminar depending on their history, and (3) Zahn's instability criterion $J{\rm Pr}<(J{\rm Pr})_c$ should not be used as a strict criterion to determine the edge of region that is mixed by the shear-driven turbulence. Instead, there is clear evidence for mixing beyond that edge due to some form of turbulent overshooting. In the following section, we now investigate this last issue more quantitatively. 

\section{Turbulence extension into theoretically laminar regions}\label{sec:extension}

In the previous section we have shown that Zahn's mixing model \citep{Zahn92} fails to account for the existence of substantial mixing beyond the edge of the region that is unstable according to (4), and that is because it ignores non-local effects. Turbulent eddies that are driven by the shear in the nominally unstable region can extend into the stable region by virtue of their finite size.

\begin{figure}[h]
\center
    \includegraphics[width=0.7\textwidth]{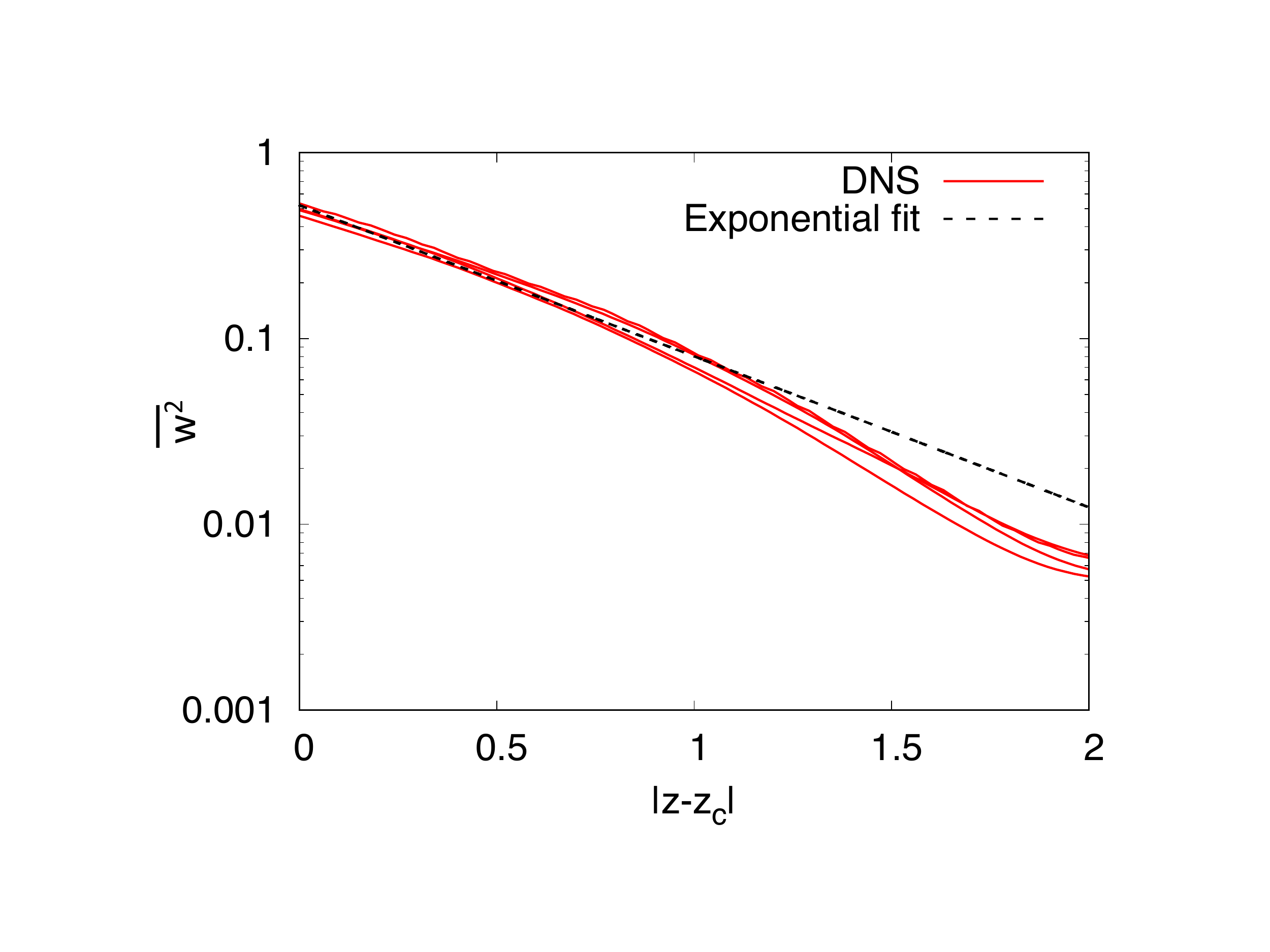}
  \caption{Illustration of the extraction of the lengthscale $\delta$ from a DNS with ${\rm Ri}_F\Pe_F = 10$. The statistically stationary state of this run contains two turbulent regions (one in the middle and one in the top/bottom layer). The black dashed line has been obtained fitting the average of the four red curves giving the value of $\delta$ reported in Table ~\ref{table:0}, while the error $\epsilon_\delta$ is the standard deviation of the individual fits from this average.} 
\label{fig:FitExp}
\end{figure}

In an attempt to characterize how far mixing can propagate into the theoretically stable region, we first look at the profile of $\overline{w^2}(z)$ in the vicinity of each critical height $z_c$, defined as the positions where $J{\rm Pr} = 0.007$. This profile looks exponential, so we assume from now on that $\overline{w^2}(z) \propto e^{-(z-z_c)/\delta}$ in the vicinity of $z_c$ but outside of the nominally turbulent region. We fit these theoretical profiles to the data, to extract the length scale $\delta$ from each available DNS. An illustration of our fitting procedure for the data from Simulation A with ${\rm Ri}_F{\rm Pe }_F = 10$ is shown in Figure~\ref{fig:FitExp}. The results are reported in Table~\ref{table:0}. 

\begin{figure}[h]
\center
    \includegraphics[width=0.7\textwidth]{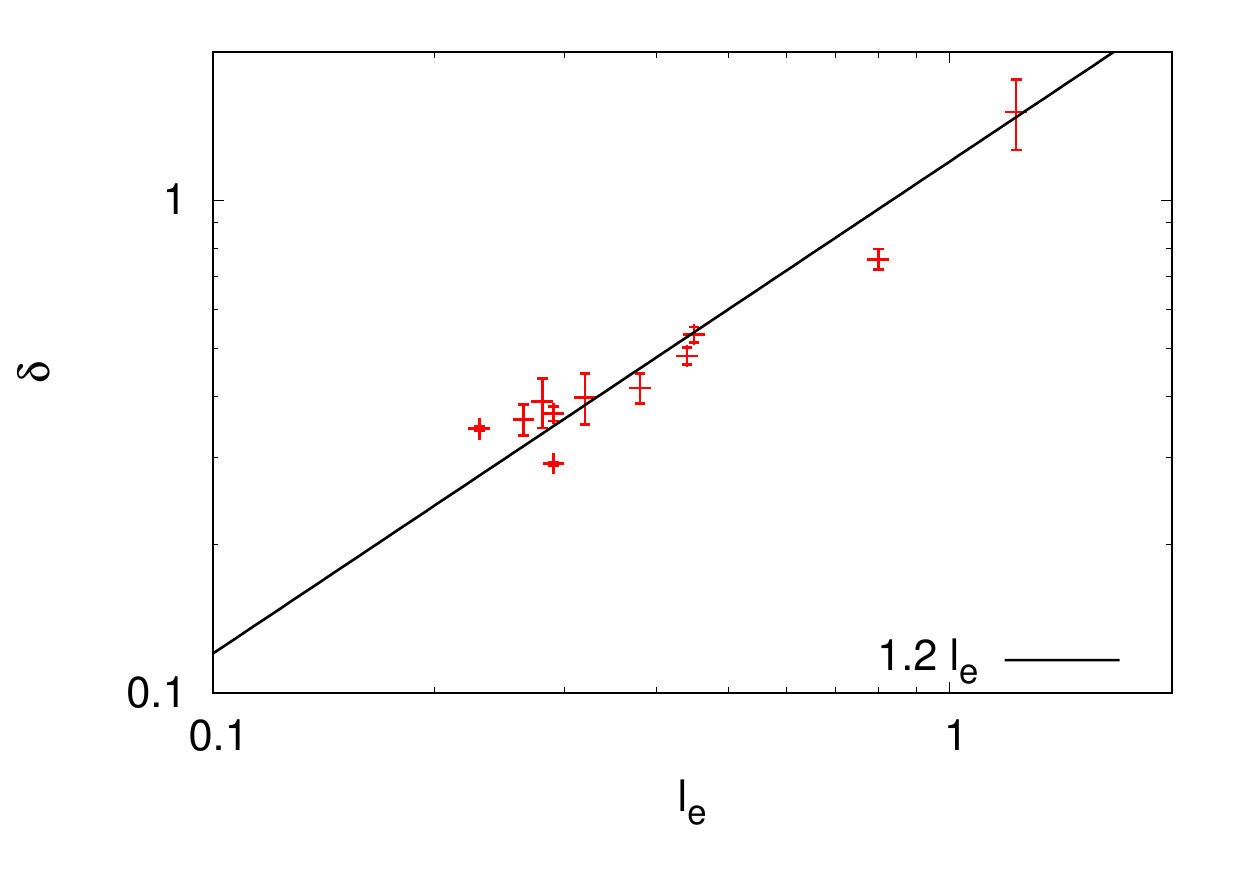}
  \caption{Overshooting lengthscale $\delta$ as a function of the measured eddy lengthscale $l_e$ for all of our simulations. The black line was fitted to the data, revealing that  $\delta \simeq 1.2 l_e$.}
\label{fig:FDiffusionLength}
\end{figure}

Note how $\delta$ decreases as ${\rm Ri}_F{\rm Pe}_F$ increases, meaning that this shear-induced overshooting is less important in more strongly stratified systems. In fact, one may naturally expect that the overshooting scale should be related to the size of the turbulent eddies in the unstable zone. To check whether this is the case, we compare $\delta$ for each simulation to the corresponding value of the turbulent eddy scale $l_e$ computed in the previous section.  Figure~\ref{fig:FDiffusionLength} presents the results. We see that across several orders of magnitude in ${\rm Ri}_F {\rm Pe}_F$, $\delta \simeq 1.2 l_e$. This confirms that turbulent eddies can overshoot from the nonlinearly unstable part of the shear layer into the adjacent stable one, and that their influence decays exponentially on a lengthscale commensurate with their actual size. Since we showed in the previous section that $l_e$ can be predicted from the Zahn scale, we now have a fairly simple way of predicting the overshooting scale as $\delta \simeq 1.2 l_{\rm Z}$. To see whether this effect is important in stars, note that dimensionally speaking, we have 

\begin{equation}
l_{\rm Z} = \sqrt{ \frac{(J{\rm Pe})_c \kappa_T}{JS}} = 10^9 \sqrt{ (J{\rm Pe})_c } \left( \frac{\kappa_T}{10^{15}{\rm cm}^2{\rm s}^{-1}} \right)^{1/2} \left( \frac{10^{-9} {\rm s}^{-1} }{N^2} \right)^{1/2} \left( \frac{S}{10^{-6} {\rm s}^{-1} } \right)^{1/2} {\rm cm}
\end{equation}
where the values selected for each of these quantities are appropriate for the outer envelope of a 30 solar mass star (where the low P\'eclet number approximation applies, see \citet{Garaud2016}), assuming that the shear is equal to 10\% of its rotation rate, and where the rotation rate is taken to be $\sim 100$km/s, which is typical of the rotation rates of such massive stars \citep{Conti1977, Penny2004}. With that estimate, $l_{\rm Z}$ appears to be small relative to the radius of the star (which is $\sim 10^{12}$cm), suggesting that turbulent mixing will decay very rapidly away from edge of the turbulent region, and that non-local effects can ultimately be neglected. However, it is worth remembering that $S$, $N$ and $\kappa_T$ can vary by several orders of magnitudes both as a function  of position within the star, and with stellar mass, so $l_{\rm Z}$ could reach values closer to $10^{10}$cm in the near surface layers where $\kappa_T$ approaches $10^{18}$cm$^2$/s  \citep[see][]{Garaud2016}. In these more weakly stratified or more diffusive systems where $l_{\rm Z}$ is much larger, the non-local overshooting effect could be more significant, and should be taken into account. 

\section{Discussion and conclusion} 
\label{sec:ccl} 

In this paper, we have continued our investigation of the validity of Zahn's model for diffusive shear instabilities, following from the work of \citet{Garaud2016} and of Paper I. This time, we tested the model against numerical experiments that were by design significantly less idealized than before, and could probe the limits of its validity in situations that are closer to what may take place in stars.

To do so, we considered a model setup in which a well-chosen body force can maintain a localized shear layer, flanked by two regions with very weak shear on each side. The laminar velocity profile this force would drive is not too different from a hyperbolic tangent \citep[as in][]{Lignal1999}, albeit in a periodic domain. With this setup, we can test in particular whether Zahn's instability criterion can correctly predict the position of the edge of the turbulent region(s) thus created, which is something that had not yet been determined in prior work. 

By looking at the linear stability of the laminar solution, we were able to identify two modes of instability, one which dominates in more weakly stratified systems and has a spatial extent (both horizontally and vertically) commensurate with the width of the shear layer (mode S, see Section~\ref{sec:stab}), and one which dominates at larger values of the stratification and, perhaps surprisingly, has much larger horizontal and vertical scales than the shear lengthscale (mode H). Modes of type H dominate when the shear layer is flanked by sufficiently large regions of no shear \citep[as in the pure hyperbolic tangent shear layer of][]{Lignal1999}, but disappear in the sinusoidal limit studied by \citet{Garaudal15}, where type-S modes  dominate instead. This reconciles the apparent discrepancies between these two papers.

More importantly, however, we also showed that linear theory turns out to be fairly irrelevant to the long-term nonlinear evolution of the shear layer. First, we found (as many have before us) that a significant region of parameter space exists that is linearly stable but nonlinearly unstable (i.e that is only unstable to finite amplitude perturbations of the right spatial form). While that region is admittedly artificially enlarged in our DNSs (where the streamwise length of the domain is limited, and therefore does not allow for the development of linearly unstable low $k_x$ modes) compared to what it would be in a real star, we still expect it to be significant in stars especially when the shear layer is relatively wide. Secondly, even within the linearly unstable regime we have found that the spatial properties of the unstable modes are not always consistent with those of the fully turbulent flow. Indeed, when the stratification is substantial, the turbulence ultimately does appear to be localized in the regions of strong shear, even though the linear instability of the mean flow at the statistically stationary state is not (see Section~\ref{sec:intermstrat}). In other words, it seems that linear theory is of little practical use in the study of turbulent mixing in diffusive shear flows. This is by contrast with other instabilities where linear theory can provide interesting insight into the nonlinear regime \citep[e.g. fingering convection in particular, see the review by][]{Garaud2018}.

We also tested Zahn's {\it nonlinear} criterion for instability \citep{Zahn1974}, $J{\rm Pr} < (J{\rm Pr})_c \simeq 0.007$, against the DNSs. We found that there are two potentially significant issues with the use of this criterion. The first is that the existence of turbulence in linearly stable strongly stratified diffusive shear flows is subject to hysteresis (as expected from the nonlinear nature of the instability). Hence, a shear layer that satisfies $J{\rm Pr} < (J{\rm Pr})_c$ is not necessarily the seat of turbulent mixing,  whether it is or not strongly depends on its evolutionary history. As such, if a shear layer within a star was historically nonlinearly stable (with $J{\rm Pr} > (J{\rm Pr})_c$), then it will remain stable even when $J{\rm Pr}$ drops below $(J{\rm Pr})_c$, and will only be destabilized when the shear grows large enough, or the stratification becomes weak enough, for the onset of linear instability. On the other hand, if the layer is initially turbulent but the shear later weakens, turbulence will continue to be sustained as long as $J{\rm Pr} < (J{\rm Pr})_c$. 

The second issue with Zahn's criterion is that while it can {\it in theory} be used to predict the location of the edge of the turbulent region, we also found the presence of mixing beyond that edge due to non-local effects that take the form of turbulent overshoot. We quantified this to show that the turbulence decays exponentially on a lengthscale $\delta$ that is of the order of the turbulent eddy scale $l_e$ (see Section 5). The eddy scale itself is well-approximated by the Zahn scale, $l_{\rm Z}$, which can range widely in size depending on the local thermal diffusivity, thermal stratification and shearing rate within the star. For the more weakly stratified systems, these non-local effects can be very significant (see Section~\ref{sec:weakstrat} for instance), and must be taken into account to avoid seriously underestimating shear-induced mixing. For the more strongly stratified systems on the other hand the non-local effects are less important, and the edge of the turbulent region is well-approximated by the location where $J{\rm Pr} = (J{\rm Pr})_c$ (albeit with the caveat discussed above regarding hysteresis). 

These findings raise the question of how one could better account for the complex dynamics of diffusive stratified shear flows. This question will be answered to the best of our ability in Part III of this series of papers. In the meantime, however, should anyone prefer a simple mixing parametrization to implement in their stellar evolution code, we now propose the following ``recipe''. At each timestep, the code should
\begin{enumerate} 
\item compute $J{\rm Pr}$ as a function of radius $r$, identify the regions where $J{\rm Pr} < 0.007$ as being unstable, as well as the locations of the edges of these regions $r_i$; 
\item wherever $J{\rm Pr} < 0.007$, compute the turbulent diffusion coefficient and the turbulent viscosity as $D_{\rm turb} \simeq \nu_{\rm turb} \simeq f_{\rm turb} \frac{C}{J} \kappa_T$, where $C \simeq 0.08$, and $f_{\rm turb}$ is a factor discussed below. Note that one could alternatively use the formula proposed in Paper I (see Equations 38 and 39), which is valid in a broader range of parameter space (in particular, it better captures the limit of weakly stratified flows). However, the user who chooses to do so should set the exponent $b = 0$ in that formula, otherwise it is not possible to capture the turbulent overshoot into the stable layer;
\item compute the Zahn scale $l_{\rm Z} \simeq \sqrt{\frac{\kappa_T}{JS}}$ (ignoring the factors of order unity) at the edges of the turbulent region $r_i$; 
\item for any radial position $r$ outside of the turbulent region, let $D_{\rm turb}(r) \sim D_{\rm turb}(r_i) \exp\left[ -|r-r_i| / l_{\rm Z} \right]$ where $r_i$ is the closest edge to $r$, and similarly for $\nu_{\rm turb}$. 
\end{enumerate}
The factor $f_{\rm turb}$ should in principle take the value $1$ in the linearly unstable regime\footnote{Assessing whether the shear flow is linearly unstable can only be done with a global linear stability analysis, which may not be practical during the run-time of a stellar evolution code. In doubt, the user should set $f_{\rm turb} = 1$; this would simply have the effect of ignoring the possibility of hysteresis altogether.} where turbulence is expected, and some value between 0 and 1 in the nonlinearly unstable regime. That factor could be made to depend on the history of the shear layer in an attempt to account for the possibility of hysteresis. It could also be made to depend on the low-level ambient kinetic energy, which could come for instance from a nearby convection zone, or from gravity waves emitted from further away. It could also be interpreted as a filling factor for the turbulence, noting that nonlinearly unstable flows are often observed to have spatial intermittency\footnote{We did not observe such intermittency in the DNSs presumably because our computational domain was too small.} \citep{Bottin1998}.
Of course, this very simplistic model is merely a recipe for mixing. In Part III of this series we provide a more physically-based closure model for turbulence that more realistically captures the dynamics of diffusive stratified shear flows.

Finally, it is worth remembering some of the rather strict assumptions that were made in our studies, which in principle constrain the applicability of our model. First, all of the DNSs performed so far have been in non-rotating systems, while most of the shear in stars originates from differential rotation. This means that the model can only be applied with confidence when the shearing rate is much larger than the rotation rate of the star. We are presently planning future work to investigate diffusive shear flows in rotating systems to address the complementary limit of strong rotation. 
Second, all of the results obtained so far were in the limit of low P\'eclet number which, as discussed by \citet{Garaud2016}, only applies when the thermal diffusivity is particularly large. It will be interesting to determine, through further DNSs, which aspects of the model remain valid and which ones do not when the P\'eclet number is large \citep[see][for preliminary analyses of this question]{PratLignieres13, PratLignieres14, Garaud2016}. Finally, many other additional effects could radically alter the predictions for mixing by diffusive shear flows, such as lateral shear \citep{TalonZahn97}, compositional gradients \citep{Maeder97, PratLignieres14}, and of course, magnetic fields \citep[e.g.][]{Acheson1978}. As such, the present work should be viewed as a valuable step forward in modeling diffusive shear flows, but should also be used with all these caveats in mind.

\acknowledgements

D.G and P.G. gratefully acknowledge funding by NSF AST-1517927. We thank J. Verhoeven for valuable discussions. The simulations were run on the Hyades supercomputer at UCSC, purchased with an NSF MRI grant. The PADDI code used was generously provided by S. Stellmach.


\begin{thebibliography}{19}
\expandafter\ifx\csname natexlab\endcsname\relax\def\natexlab#1{#1}\fi

\bibitem[{{Acheson} \& {Gibbons}(1978)}]{Acheson1978}
{Acheson}, D.~J., \& {Gibbons}, M.~P. 1978, Philosophical Transactions of the
  Royal Society of London Series A, 289, 459

\bibitem[{{Beaumont}(1981)}]{Beaumont1981}
{Beaumont}, D.~N. 1981, Journal of Fluid Mechanics, 108, 461

\bibitem[{{Bottin} {et~al.}(1998){Bottin}, {Daviaud}, {Manneville}, \&
  {Dauchot}}]{Bottin1998}
{Bottin}, S., {Daviaud}, F., {Manneville}, P., \& {Dauchot}, O. 1998, EPL
  (Europhysics Letters), 43, 171
  
\bibitem[{{Conti} \& {Ebbets}(1977)}]{Conti1977}
{Conti}, P.~S., \& {Ebbets}, D. 1977, \apj, 213, 438

\bibitem[{{Garaud}(2018)}]{Garaud2018}
{Garaud}, P. 2018, Annual Review of Fluid Mechanics, 50, 275

\bibitem[{{Garaud} {et~al.}(2017){Garaud}, {Gagnier}, \&
  {Verhoeven}}]{Garaudal17}
{Garaud}, P., {Gagnier}, D., \& {Verhoeven}, J. 2017, \apj, 837, 133

\bibitem[{{Garaud} {et~al.}(2015){Garaud}, {Gallet}, \&
  {Bischoff}}]{Garaudal15}
{Garaud}, P., {Gallet}, B., \& {Bischoff}, T. 2015, Physics of Fluids, 27,
  084104

\bibitem[{{Garaud} \& {Kulenthirarajah}(2016)}]{Garaud2016}
{Garaud}, P., \& {Kulenthirarajah}, L. 2016, \apj, 821, 49

\bibitem[{{Gotoh} {et~al.}(1983){Gotoh}, {Yamada}, \& {Mizushima}}]{Gotoh83}
{Gotoh}, K., {Yamada}, M., \& {Mizushima}, J. 1983, J. Fluid Mech., 127, 45

\bibitem[{{Green}(1974)}]{Green1974}
{Green}, J.~S.~A. 1974, Journal of Fluid Mechanics, 62, 273

\bibitem[{{Ligni{\`e}res}(1999)}]{Lign1999}
{Ligni{\`e}res}, F. 1999, \aap, 348, 933

\bibitem[{{Ligni{\`e}res} {et~al.}(1999){Ligni{\`e}res}, {Califano}, \&
  {Mangeney}}]{Lignal1999}
{Ligni{\`e}res}, F., {Califano}, F., \& {Mangeney}, A. 1999, \aap, 349, 1027

\bibitem[{{Maeder}(1997)}]{Maeder97}
{Maeder}, A. 1997, A\&A, 321, 134

\bibitem[{{Moll} {et~al.}(2017){Moll}, {Garaud}, {Mankovich}, \&
  {Fortney}}]{Moll2017}
{Moll}, R., {Garaud}, P., {Mankovich}, C., \& {Fortney}, J.~J. 2017, \apj, 849,
  24

\bibitem[{{Peltier} \& {Caulfield}(2003)}]{Peltier2003}
{Peltier}, W.~R., \& {Caulfield}, C. 2003, Annual Review of Fluid Mechanics,
  35, 135
  
\bibitem[{{Penny} {et~al.}(2004){Penny}, {Sprague}, {Seago}, \&
  {Gies}}]{Penny2004}
{Penny}, L.~R., {Sprague}, A.~J., {Seago}, G., \& {Gies}, D.~R. 2004, \apj,
  617, 1316

\bibitem[{{Prat} {et~al.}(2016){Prat}, {Guilet}, {Viallet}, \&
  {M{\"u}ller}}]{Pratal2016}
{Prat}, V., {Guilet}, J., {Viallet}, M., \& {M{\"u}ller}, E. 2016, \aap, 592,
  A59

\bibitem[{{Prat} \& {Ligni{\`e}res}(2013)}]{PratLignieres13}
{Prat}, V., \& {Ligni{\`e}res}, F. 2013, A\&A, 551, L3

\bibitem[{{Prat} \& {Ligni{\`e}res}(2014)}]{PratLignieres14}
---. 2014, aap, 566, A110

\bibitem[{Strogatz(2018)}]{strogatz2018}
Strogatz, S.~H. 2018, Nonlinear dynamics and chaos: with applications to
  physics, biology, chemistry, and engineering (CRC Press)

\bibitem[{{Talon} \& {Zahn}(1997)}]{TalonZahn97}
{Talon}, S., \& {Zahn}, J.-P. 1997, A\&A, 317, 749

\bibitem[{{Tobias} {et~al.}(2011){Tobias}, {Cattaneo}, \&
  {Brummell}}]{Tobias2011}
{Tobias}, S.~M., {Cattaneo}, F., \& {Brummell}, N.~H. 2011, \apj, 728, 153

\bibitem[{Traxler {et~al.}(2011)Traxler, Garaud, \& Stellmach}]{Traxleral2011a}
Traxler, A., Garaud, P., \& Stellmach, S. 2011, Astrophys. J., 728, L29

\bibitem[{{Zahn}(1974)}]{Zahn1974}
{Zahn}, J.-P. 1974, in IAU Symposium, Vol.~59, Stellar Instability and
  Evolution, ed. P.~{Ledoux}, A.~{Noels}, \& A.~W. {Rodgers}, 185--194

\bibitem[{{Zahn}(1992)}]{Zahn92}
{Zahn}, J.-P. 1992, A\&A, 265, 115

\end{thebibliography}


\section*{Appendix A}
\subsection*{Linear stability analysis}

This Appendix provides more detail on the  stability analysis of the laminar solution $\hat{U}(z)$ given by equation~(\ref{eq:laminar}) to infinitesimal perturbations. We assume that perturbations are two-dimensional, and therefore let the total velocity field $\tilde{\bu}+ \hat{U}(z)\be_x$, where $\tilde{\bu} = \bnabla \times(\psi \be_y)$. Linearizing equations~(\ref{eq:1})-(\ref{eq:3}) assuming velocity and temperature fluctuations are small, yields

\begin{equation}
\frac{\partial}{\partial t}\left(\nabla^2 \psi \right) + \hat{U}(z) \frac{\partial}{\partial x}\left(\nabla^2 \psi \right) - \frac{\partial \psi}{\partial x}\cdot\frac{d^2 \hat{U}(z)}{d z^2}=\Ri\frac{\partial T}{\partial x}+\frac{1}{\Re}\nabla^2\left(\nabla^2 \psi \right)
\end{equation}

\begin{equation}
\frac{\partial T}{\partial t}+\hat{U}(z)\frac{\partial T}{\partial x}+\frac{\partial \psi}{\partial x} = \frac{1}{\Pe} \nabla^2T
\end{equation}
which automatically satisfies the incompressibility condition with $\tilde{u}=-\frac{\partial \psi}{\partial z}$ and $\tilde{w}=\frac{\partial \psi}{\partial x}$.
The coefficients of these PDEs are independent of $t$ and $x$, but are periodic in $z$. Therefore, we seek solutions of the form
\begin{equation}
q(x,z,t)= e^{ikx + \lambda t} \hat{q}(z) \ ,
\end{equation}
where $q$ is either $T$ or $\psi$ , and the wavenumber $k$ is real but the growth rate $\lambda$ could be complex. The mean flow can be expressed as
\begin{equation}
\hat{U}(z)=\sum_{n= -N}^N U_{n}e^{ 2\pi i z n/L_z} \ ,
\end{equation}
where the sum has been truncated for numerical purposes. Using Floquet theory,  $\hat{q}(z)$ can be written
\begin{equation}
\hat{q}(z)=e^{if'z}\sum_{n= -N}^N q_n e^{2 \pi i z n/L_z} \ ,
\end{equation}
where $f'=2\pi f/L_z$ is real to satisfy the periodicity of the system, and $f$ is the Floquet coefficient. Substituting this ansatz into the equations of motion and temperature and using $\frac{1}{L_z}\int e^{-i\frac{2\pi z m}{L_z}} e^{i\frac{2\pi z n}{L_z}} dz= \delta_{n,m}
$, we obtain an algebraic system for the coefficients $\psi_n$ and $T_n$

\begin{multline}
-\lambda \psi_{m}\left[\left(f'+\frac{2\pi m}{L_z}\right)^2+k^2\right]+ ik\sum\limits_{n}U_n \psi_{m-n} \left[\left(\frac{2\pi n}{L_z}\right)^2-\left(\left(f'+\frac{2\pi (m-n)}{L_z}\right)^2+k^2\right)\right]\\=ik \Ri T_m +\frac{1}{\Re}\psi_{m} \left[\left(f'+\frac{2\pi m}{L_z}\right)^2+k^2\right]^2
\end{multline}
\begin{equation}
\lambda T_{m}+ ik\psi_{m}+ik\sum\limits_{n}U_n T_{m-n}=-\frac{1}{\Pe}T_m\left[\left(f'+\frac{2\pi m}{L_z}\right)^2+k^2\right].
\end{equation}

This system of $2(2N+1)$ equations can be written as a generalized eigenvalue problem of the form $\bA {\bf v}=\lambda \bB {\bf v}$, where ${\bf v}=\{\psi_{-N},...,\psi_{N},T_{-N},...,T_{N}\}$, and can be solved numerically for the eigenvalue $\lambda$ using LAPACK routines. As \citet{Green1974}, \citet{Beaumont1981}, \citet{Gotoh83} and \citet{Garaudal15}, we find that the most unstable mode have the same period in $z$ as the main flow $U(z)$, i.e. $f'=0$. We therefore restrict the presentation of our results to the case $f' = 0$. The LAPACK routine returns a number of eigenvalues $\lambda$ and eigenvectors ${\bf v}$ for each set of input parameters ($\Ri$, $\Pe$, $\Re$ and $k$). We compute the eigenvalue(s) with the largest positive real part, and report these as the growing mode(s). In some regions of parameter space there is only one growing mode, but in others there are several (Modes H and S). 

\end{document}